\newcommand{\DoPrePrint}{0} 

\ifnum\DoPrePrint=1
  \documentclass[aps,amsmath,preprint,showpacs,superscriptaddress,nofootinbib,linenumbers,floatfix]{revtex4-1}

\else
  \documentclass[aps,amsmath,reprint,showpacs,superscriptaddress,nofootinbib,floatfix]{revtex4-1}
\fi

\usepackage{graphicx}  
\usepackage{dcolumn}
\usepackage{bm}

\usepackage{xspace}
\usepackage{units}
\usepackage{color}

\begin{document}

\newcommand{\ie}{{\it i.e.}\xspace}
\newcommand{\eg}{{\it e.g.}\xspace}
\newcommand{\red}[1]{{\color{red}#1}}
\newcommand{\blue}[1]{{\color{blue}#1}}

\newcommand{\chisq}{\ensuremath{\chi^{2}}\xspace}
\newcommand{\chisqmin}{\ensuremath{\chi^{2}_{\textrm{min}}}\xspace}
\newcommand{\chisqpen}{\ensuremath{\chi^{2}_{\textrm{pen}}}\xspace}
\newcommand{\ndof}{\ensuremath{\mathrm{N_{DoF}}}\xspace}
\newcommand{\nbins}{\ensuremath{\mathrm{N_{bins}}}\xspace}
\newcommand{\units}[1]{\ensuremath{~\mbox{#1}}}

\newcommand{\clarence}[1]{\xspace\textcolor{blue}{CWret: #1}}

\newcommand{\pip}{\ensuremath{\pi^+\xspace}\xspace}
\newcommand{\pim}{\ensuremath{\pi^-\xspace}\xspace}
\newcommand{\pipm}{\ensuremath{\pi^\pm\xspace}\xspace}
\newcommand{\pin}{\ensuremath{\pi^0\xspace}\xspace}

\newcommand{\ccpip}{$\nu_\mu$CC1$\pi^{\pm}$\xspace}
\newcommand{\ccnpip}{$\nu_\mu$CC$N\pi^{\pm}$\xspace}
\newcommand{\ccpin}{$\nu_\mu$CC$1\pi^0$\xspace}
\newcommand{\ccapin}{$\bar{\nu}_\mu$CC$1\pi^0$\xspace}

\newcommand{\Pmu}{\ensuremath{p_\mu}\xspace}
\newcommand{\Tmu}{\ensuremath{\theta_\mu}\xspace}
\newcommand{\Kpi}{\ensuremath{T_\pi}\xspace}
\newcommand{\Tpi}{\ensuremath{\theta_\pi}\xspace}
\newcommand{\qq}{\ensuremath{Q^{2}}\xspace}
\newcommand{\Ev}{\ensuremath{E_{\nu}}\xspace}
\newcommand{\Wrec}{\ensuremath{W_{\mathrm{rec}}}}
\newcommand{\Ma}{\ensuremath{M_{\textrm{A}}}\xspace}

\newcommand{\Mares}{\ensuremath{M_{\mathrm{A}}^\mathrm{res}}\xspace}
\newcommand{\Normres}{\ensuremath{\mathrm{NormRes}}\xspace}
\newcommand{\nonresonepi}{NonRes$1\pi$\xspace}
\newcommand{\nonrestwopi}{NonRes$2\pi$\xspace}
\newcommand{\ThetaPi}{$\pi$-iso\xspace}

\newcommand{\GENIE}{GENIE\xspace}
\newcommand{\NUISANCE}{NUISANCE\xspace}
\newcommand{\NUANCE}{NUANCE\xspace}
\newcommand{\NOvA}{NOvA\xspace}
\newcommand{\MINERvA}{MINERvA\xspace}
\newcommand{\MINOS}{MINOS\xspace}
\newcommand{\MiniBooNE}{MiniBooNE\xspace}

\graphicspath{{./Figures/}}

\title{Tuning the \GENIE Pion Production Model with MINERvA Data}

\author{P.~Stowell}     \affiliation{\Sheffield}
\author{L.~Pickering}   \affiliation{\MSU} \affiliation{\Imperial}
\author{C.~Wilkinson}   \affiliation{\Bern}
\author{C.~Wret}        \affiliation{\Rochester} \affiliation{\Imperial}

\newcommand{\Sheffield}{University of Sheffield, Dept. of Physics and Astronomy, Sheffield, United Kingdom}
\newcommand{\MSU}{Michigan State University, Dept. of Physics and Astronomy, East Lansing, Michigan 48824, USA}
\newcommand{\Imperial}{Imperial College London, Dept. of Physics, London, United Kingdom}
\newcommand{\Bern}{University of Bern, Albert Einstein Center for Fundamental Physics, LHEP, Bern, Switzerland}


\newcommand{\Rutgers}{Rutgers, The State University of New Jersey, Piscataway, New Jersey 08854, USA}
\newcommand{\Hampton}{Hampton University, Dept. of Physics, Hampton, VA 23668, USA}
\newcommand{\Dortmund}{Institute of Physics, Dortmund University, 44221, Germany }
\newcommand{\Otterbein}{Dept. of Physics, Otterbein University, 1 South Grove Street, Westerville, OH, 43081 USA}
\newcommand{\JMU}{James Madison University, Harrisonburg, Virginia 22807, USA}
\newcommand{\Florida}{University of Florida, Dept. of Physics, Gainesville, FL 32611}
\newcommand{\UCIrvine}{Dept. of Physics and Astronomy, University of California, Irvine, Irvine, California 92697-4575, USA}
\newcommand{\CBPF}{Centro Brasileiro de Pesquisas F\'{i}sicas, Rua Dr. Xavier Sigaud 150, Urca, Rio de Janeiro, Rio de Janeiro, 22290-180, Brazil}
\newcommand{\PUCP}{Secci\'{o}n F\'{i}sica, Departamento de Ciencias, Pontificia Universidad Cat\'{o}lica del Per\'{u}, Apartado 1761, Lima, Per\'{u}}
\newcommand{\INRM}{Institute for Nuclear Research of the Russian Academy of Sciences, 117312 Moscow, Russia}
\newcommand{\Jlab}{Jefferson Lab, 12000 Jefferson Avenue, Newport News, VA 23606, USA}
\newcommand{\Pittsburgh}{Dept. of Physics and Astronomy, University of Pittsburgh, Pittsburgh, Pennsylvania 15260, USA}
\newcommand{\Guanajuato}{Campus Le\'{o}n y Campus Guanajuato, Universidad de Guanajuato, Lascurain de Retana No. 5, Colonia Centro, Guanajuato 36000, Guanajuato M\'{e}xico.}
\newcommand{\Athens}{Dept. of Physics, University of Athens, GR-15771 Athens, Greece}
\newcommand{\Tufts}{Physics Dept., Tufts University, Medford, Massachusetts 02155, USA}
\newcommand{\WM}{Dept. of Physics, College of William \& Mary, Williamsburg, Virginia 23187, USA}
\newcommand{\FNAL}{Fermi National Accelerator Laboratory, Batavia, Illinois 60510, USA}
\newcommand{\Purdue}{Dept. of Chemistry and Physics, Purdue University Calumet, Hammond, Indiana 46323, USA}
\newcommand{\MCLA}{Massachusetts College of Liberal Arts, 375 Church Street, North Adams, MA 01247}
\newcommand{\UMD}{Dept. of Physics, University of Minnesota -- Duluth, Duluth, Minnesota 55812, USA}
\newcommand{\Northwestern}{Northwestern University, Evanston, Illinois 60208}
\newcommand{\UNI}{Universidad Nacional de Ingenier\'{i}a, Apartado 31139, Lima, Per\'{u}}
\newcommand{\Rochester}{University of Rochester, Rochester, New York 14627 USA}
\newcommand{\Austin}{Dept. of Physics, University of Texas, 1 University Station, Austin, Texas 78712, USA}
\newcommand{\USM}{Departamento de F\'{i}sica, Universidad T\'{e}cnica Federico Santa Mar\'{i}a, Avenida Espa\~{n}a 1680 Casilla 110-V, Valpara\'{i}so, Chile}
\newcommand{\Geneva}{University of Geneva, 1211 Geneva 4, Switzerland}
\newcommand{\Chicago}{Enrico Fermi Institute, University of Chicago, Chicago, IL 60637 USA}
\newcommand{\hired}{}
\newcommand{\OregonState}{Dept. of Physics, Oregon State University, Corvallis, Oregon 97331, USA}
\newcommand{\oxford}{Oxford University, Dept. of Physics, Oxford, United Kingdom}
\newcommand{\umiss}{University of Mississippi, Oxford, Mississippi 38677, USA}
\newcommand{\upenn}{Dept. of Physics and Astronomy, University of Pennsylvania, Philadelphia, PA 19104}
\newcommand{\AMU}{AMU Campus, Aligarh, Uttar Pradesh 202001, India}
\newcommand{\wroclaw}{University of Wroclaw, plac Uniwersytecki 1, 50-137 Wrocław, Poland}
\newcommand{\Mohali}{IISER, Mohali, Knowledge city, Sector 81, Manauli PO 140306}
\newcommand{\chrismarshallThanks}{now at Lawrence Berkeley National Laboratory, Berkeley, CA 94720, USA}
\newcommand{\joelmousseauThanks}{now at University of Michigan, Ann Arbor, MI 48109, USA}
\newcommand{\jwolcottThanks}{now at Tufts University, Medford, MA 02155, USA}


\author{F.~Akbar}                         \affiliation{\AMU}
\author{D. A.~Andrade}                    \affiliation{\Guanajuato}
\author{M. V.~Ascencio}                   \affiliation{\PUCP}
\author{L.~Bellantoni}                    \affiliation{\FNAL}
\author{A.~Bercellie}                     \affiliation{\Rochester}
\author{M.~Betancourt}                    \affiliation{\FNAL}
\author{A.~Bodek}                         \affiliation{\Rochester}
\author{A.~Bravar}                        \affiliation{\Geneva}
\author{H.~Budd}                          \affiliation{\Rochester}
\author{G.~Caceres}                       \affiliation{\CBPF}
\author{T.~Cai}                           \affiliation{\Rochester}
\author{M. F.~Carneiro}                   \affiliation{\OregonState}
\author{J.~Chaves}                        \affiliation{\upenn}
\author{H.~da~Motta}                      \affiliation{\CBPF}
\author{S. A.~Dytman}                     \affiliation{\Pittsburgh}
\author{G.A.~D\'{i}az~}                   \affiliation{\Rochester}  \affiliation{\PUCP}
\author{J.~Felix}                         \affiliation{\Guanajuato}
\author{L.~Fields}                        \affiliation{\FNAL}  \affiliation{\Northwestern}
\author{A.~Filkins}                       \affiliation{\WM}
\author{R.~Fine}                          \affiliation{\Rochester}
\author{N.~Fiza}                          \affiliation{\Mohali}
\author{R.~Galindo}                       \affiliation{\USM}
\author{H.~Gallagher}                     \affiliation{\Tufts}
\author{A.~Ghosh}                         \affiliation{\USM}  \affiliation{\CBPF}
\author{R.~Gran}                          \affiliation{\UMD}
\author{D. A.~Harris}                     \affiliation{\FNAL}
\author{S.~Henry}                         \affiliation{\Rochester}
\author{S.~Jena}                          \affiliation{\Mohali}
\author{D.~Jena}                          \affiliation{\FNAL}
\author{J.~Kleykamp}                      \affiliation{\Rochester}
\author{M.~Kordosky}                      \affiliation{\WM}
\author{D.~Last}                          \affiliation{\upenn}
\author{T.~Le}                            \affiliation{\Tufts}  \affiliation{\Rutgers}
\author{X.-G.~Lu}                         \affiliation{\oxford}
\author{E.~Maher}                         \affiliation{\MCLA}
\author{S.~Manly}                         \affiliation{\Rochester}
\author{W. A.~Mann}                       \affiliation{\Tufts}
\author{C. M.~Marshall}\thanks{\chrismarshallThanks}  \affiliation{\Rochester}
\author{K. S.~McFarland}                  \affiliation{\Rochester}  \affiliation{\FNAL}
\author{B.~Messerly}                      \affiliation{\Pittsburgh}
\author{J.~Miller}                        \affiliation{\USM}
\author{J. G.~Morf\'{i}n}                 \affiliation{\FNAL}
\author{J.~Mousseau}\thanks{\joelmousseauThanks}  \affiliation{\Florida}
\author{D.~Naples}                        \affiliation{\Pittsburgh}
\author{J. K.~Nelson}                     \affiliation{\WM}
\author{C.~Nguyen}                        \affiliation{\Florida}
\author{A.~Norrick}                       \affiliation{\WM}
\author{Nuruzzaman}                       \affiliation{\Rutgers}  \affiliation{\USM}
\author{V.~Paolone}                       \affiliation{\Pittsburgh}
\author{G. N.~Perdue}                     \affiliation{\FNAL}  \affiliation{\Rochester}
\author{M. A.~Ram\'{i}rez}                \affiliation{\Guanajuato}
\author{R. D.~Ransome}                    \affiliation{\Rutgers}
\author{H.~Ray}                           \affiliation{\Florida}
\author{D.~Rimal}                         \affiliation{\Florida}
\author{P. A.~Rodrigues}                  \affiliation{\umiss}  \affiliation{\Rochester}
\author{D.~Ruterbories}                   \affiliation{\Rochester}
\author{H.~Schellman}                     \affiliation{\OregonState}  \affiliation{\Northwestern}
\author{C. J.~Solano~Salinas}             \affiliation{\UNI}
\author{H.~Su}                            \affiliation{\Pittsburgh}
\author{M.~Sultana}                       \affiliation{\Rochester}
\author{V. S.~Syrotenko}                  \affiliation{\Tufts}
\author{E.~Valencia}                      \affiliation{\WM}  \affiliation{\Guanajuato}
\author{J.~Wolcott}\thanks{\jwolcottThanks}  \affiliation{\Rochester}
\author{M.~Wospakrik}                     \affiliation{\Florida}
\author{B.~Yaeggy}                        \affiliation{\USM}
\author{L.~Zazueta}                       \affiliation{\WM}
\author{D.~Zhang}                         \affiliation{\WM}

\collaboration{The MINER$\nu$A Collaboration}\ \noaffiliation

\date{\today}

\begin{abstract}
Faced with unresolved tensions between neutrino interaction measurements at few-GeV neutrino energies, current experiments are forced to accept large systematic uncertainties to cover discrepancies between their data and model predictions.  In this paper, the widely used pion production model in \GENIE is compared to four \MINERvA charged current pion production measurements using \NUISANCE.  Tunings, \ie, adjustments of model parameters, to help match \GENIE to \MINERvA and older bubble chamber data are presented here.  We find that scattering off nuclear targets as measured in \MINERvA is not in good agreement with expectations based upon scattering off nucleon (hydrogen or deuterium) targets in existing bubble chamber data.  An additional ad hoc correction for the low-\qq region, where collective nuclear effects are expected to be large, is presented.  While these tunings and corrections improve the agreement of \GENIE with the data, the modeling is imperfect.  The development of these tunings within the \NUISANCE framework allows for straightforward extensions to other neutrino event generators and models, and allows omitting and including new data sets as they become available.
\end{abstract}

\maketitle

\section{Introduction}

In recent years, experimental groups have started to publish neutrino interaction cross-section measurements on nuclear targets in terms of measurable final state particle content, instead of inferred initial interaction channels.  This avoids the problem of correcting for complex nuclear effects to make a measurement in terms of the initial interaction channels.  For example, events with only a single pion can be produced by the decay of hadronic resonances formed at the primary neutrino interaction, followed by loss of a nucleon from the resonance's decay as a result of final state interactions (FSI) within the nuclear medium.  Such events can also be produced by other sequences of interactions, such as a deep inelastic collision where only a single pion is produced after FSI.  A measurement of charged current events with one identified pion in the final state is a benchmark for models, independent of the details of how each model assesses any particular interaction channel's contributions to that final state.  The limitation of giving results in terms of final state particle content is that FSI are important, and result in the contribution of many different interaction channels into a specific final state.

There are tensions between published data from the T2K, \MiniBooNE, and \MINERvA experiments~\cite{Mosel:2017nzk, Alvarez-Ruso:2017oui, Gonzalez-Jimenez:2017fea, Wilkinson:2016wmz, Mahn:2018mai}.  These tensions exist in the charged current production of both zero and one pion final states, and a model has yet to emerge that can reliably simulate all experiments at once.  This is troubling, as current and future neutrino oscillation experiments require a cross section model which is predictive across the range of energies covered by these experiments and for a variety of targets.

The differences in neutrino fluxes, scattering targets, available phase space and signal definitions between experiments make it difficult to diagnose the exact causes of disagreement within the global data set.  In particular, as results must be averaged over the neutrino flux distribution of each experiment, it is difficult to disentangle the energy dependence of an observed deficiency in a particular model, and decide how uncertainties should be propagated in neutrino energy.  Tensions between measurements from a single experiment can uncover fundamental problems with a model which should be addressed, before considering the more difficult issue of developing, or empirically tuning a model which fits data from multiple experiments.  In this work, we employ published \MINERvA pion production data.  The cross-section measurements utilized in this effort have not been reanalyzed or modified in any way.

\NUISANCE~\cite{Stowell:2016jfr} was developed to provide the neutrino scattering community with a flexible framework in which various neutrino interaction generators can be validated and empirically tuned to data.  Its structure allows for generator tunings to be easily adapted to account for changes in the underlying model or data.  In this work, the default pion production model in the \GENIE~\cite{Andreopoulos:2009rq, Andreopoulos:2015wxa} neutrino interaction generator is tuned to \MINERvA data.   Although more sophisticated pion production models exist (e.g. ~\cite{Kabirnezhad:2017jmf, Leitner:2008ue, Nieves:2011pp, Hernandez:2016yfb}), \GENIE is widely used by the neutrino scattering community, and its model uncertainties have a central importance to the field.  Although the work is only directly applicable to one generator, the methods developed in this paper are easily adaptable to different generators.  All the data and methods are publicly available and integrated into the open source \NUISANCE framework, facilitating similar studies using other generators and models.

In Section~\ref{sec:datasets-used}, the data are reviewed and the goodness-of-fit test statistic is defined for the tuning process.  Section~\ref{sec:genie-pion} describes the default \GENIE pion production model, and reviews comparisons of this model to data.  In Section~\ref{sec:genie-systs}, the parameter reweighting package in \GENIE is discussed along with the specific parameters tuned therewith. We also discuss other corrections to the \GENIE model made to improve agreement with bubble chamber data~\cite{Rodrigues:2016xjj, Wilkinson:2014yfa}.  In Section~\ref{sec:genie-tuning}, we tune additional systematic parameters in \GENIE to improve agreement with the \MINERvA data in combination with the bubble chamber data.  In Section~\ref{sec:q2-corrections}, additional low-\qq ad hoc corrections are added to the model to resolve observed tensions, motivated by the need for similar corrections observed at both \MINOS~\cite{Adamson:2014pgc} and \MiniBooNE~\cite{AguilarArevalo:2010bm}.  Finally, in Section~\ref{sec:conclusions} we present our conclusions.

\section{Data included in the fits}
\label{sec:datasets-used}
We tune to four of \MINERvA's published charged current pion production measurements taken on a polystyrene scintillator target: \ccpip~\cite{Eberly:2014mra}, \ccnpip~\cite{McGivern:2016bwh}, \ccpin~\cite{Altinok:2017xua} and \ccapin~\cite{McGivern:2016bwh}, summarized in Table~\ref{tab:datareleases}\footnote{In ``\ccnpip", the $N$ indicates one or more identified pions and does not refer to a nucleon.}. The \MINERvA detector~\cite{Aliaga:2014wtf} does not determine the polarity of charged pions.  The fraction of \pim in \ccpip sample is small ($\sim$ 1\%).  Furthermore, the \ccpip and \ccnpip signal definition allows for any number of neutral pions.  Approximately 3\% of the \MINERvA \ccpip signal events have at least one neutral pion in the final state.  All four analyses include signal definition cuts on the true ``reconstructed'' mass of the hadronic system assuming the struck nucleon is at rest, \Wrec, and the true neutrino energy \Ev.

\begin{table*}[hbtp]
\centering
{\renewcommand{\arraystretch}{1.2}
\begin{tabular}{ccccc}
\hline\hline
Channel                 & \ccpip~\cite{Eberly:2014mra}                                                                                              & \ccnpip~\cite{McGivern:2016bwh}   
                                                                          &\ccpin~\cite{Altinok:2017xua}
                                                                                                   & \ccapin~\cite{McGivern:2016bwh} \\
\hline\hline
\nbins$\;{p_\mu}$       & 8                      & 9                      & 8                      & 9 \\
\nbins$\; {\theta_\mu}$ & 9                      & 9                      & 9                      & 9 \\
\nbins$\; {T_\pi}$      & 7                      & 7                      & 7                      & 7 \\
\nbins$\; {\theta_\pi}$ & 14                     & 14                     & 11                     & 11 \\
\nbins total            & 38                     & 39                     & 35                     & 36 \\
\hline\hline
Signal definition       & 1\pipm, $\geq$ 0\pin  & $>0$\pipm, $\geq$ 0\pin & 1\pin, 0\pipm          & 1\pin , 0\pipm \\
                        & 1$\mu^{-}$            & 1$\mu^{-}$              & 1$\mu^{-}$             & 1$\mu^{+}$ \\

                        & \Wrec$<1.4\units{GeV}$ & \Wrec$<1.8\units{GeV}$ & \Wrec$<1.8\units{GeV}$ & \Wrec$<1.8\units{GeV}$ \\
                        & ---                    & ---                    &  $\theta_\mu<25^\circ$ & ---\\
\hline \hline
\end{tabular}}
\caption{Summary of the measurements used in this analysis.  \Wrec~is the true reconstructed hadronic mass assuming the struck nucleon is at rest.  None of the measurements veto on activity other than the $\mu$ and $\pi$ in their signal definition, and all selections require $1.5 < E_\nu < 20\units{GeV}$.
}
\label{tab:datareleases}
\end{table*}

The kinematic variable distributions used in this work are the momentum and angle of the outgoing muon with respect to the incoming neutrino beam, \Pmu and \Tmu, and the kinetic energy and angle of the outgoing pion with respect to the incoming neutrino beam, \Kpi and \Tpi.  In the \ccnpip channel, where there is at least one \pipm in the final state, there is one entry in the distributions of \Tpi and \Kpi for each \pipm in an event.  The data are reported as efficiency corrected results unfolded to true kinematic variables, which may introduce model dependence. This is notably problematic in regions of low efficiency---present in the charged pion channels at $\Tpi\sim90^{\circ}$, $\Kpi<50\units{MeV}$ and $\Kpi>350\units{MeV}$, where the signal efficiency is zero~\cite{Eberly:2014mra}.  The pion selection cuts, not present in the signal definition, remove about 50\% of signal events, with little dependence upon the muon variables, but a clear impact on the shape of the pion kinematic variables.

The published cross-sections  are one dimensional with correlations provided between the bins within each distribution.  No correlations are provided between measurements of different final states, or between different one-dimensional projections of the same measurement.  These correlations are expected to be large, coming predominantly from flux and detector uncertainties. Additionally, the \ccpip event sample is a subset ($\sim$64\%) of the \ccnpip event sample, and including both channels introduces a statistical correlation.  Not assessing correlations between the distributions, while common practice in this field, is a limitation when tuning models to multiple data sets.  It introduces a bias in the $\chi^{2}$ statistic that is difficult to quantify, and requires imposing ad hoc uncertainties~\cite{Wilkinson:2016wmz} as the test-statistic is not expected to follow a $\chi^2$ distribution for the given degrees of freedom.

The covariance matrices contain a flux-dominated normalization component which we expect to be fully correlated across all distributions.  To account for the correlated uncertainty, we use the full covariance matrix, $M_{ij}$, for the \Pmu distribution and shape-only covariance matrices, $S_{ij}$, for the other three distributions in each of the topologies.  Whilst any distribution could set the normalization constraint, the shape of the \Pmu distribution for each channel  was chosen since it was found to be relatively insensitive to model variations and had good shape agreement with the data. The joint \chisq is therefore defined as the sum of the full \Pmu \chisq and shape-only \Tmu, \Kpi and \Tpi \chisq's:

\begin{align}
\chi^{2} & = \sum_{ij}^{N_{p_\mu}} \Delta_{i} (M^{-1})_{ij} \Delta_{j} \notag\\
         & + \sum_{kij}^{N_{k}}  \Delta^{S}_{k,i} (S^{-1})_{ij} \Delta^{S}_{k,j}
\end{align}

\noindent where $i$ and $j$ are bin indices,

\begin{align}
\Delta_{i} & = d_{p_\mu,i} - m_{p_\mu,i} \\
\Delta^{S}_{k,i} & = d_{k,i} - \left( m_{k,i} \times \frac{\sum_{j} d_{k,j}}{\sum_{j} m_{k,j}} \right ),
\end{align}

\noindent and $d_{k,i}$ and $m_{k,i}$ are the data and MC values respectively for the $i^{th}$ bin in the $k^{th}$ distribution.
The shape-only covariance matrices are provided in the public data release for the \ccpip and \ccpin measurements, and the method of Ref.~\cite{Katori:2008zz} (section 10.6.3) was used to extract them for the \ccnpip and \ccapin channels.

\section{Pion Production in \GENIE}
\label{sec:genie-pion}
This analysis begins with version 2.12.6 of \GENIE, which is close to what is used by \MINERvA, T2K, \NOvA and MicroBooNE.  We use the Smith-Moniz relativistic Fermi gas (RFG) model~\cite{Smith:1972xh}  with an added high momentum tail as per Bodek and Ritchie~\cite{Bodek:1980ar}.  The Valencia random phase approximation screening~\cite{Nieves:2004wx} is applied as a weight to quasielastic events.  The two-particle two-hole process is simulated using the Valencia model~\cite{Nieves:2011pp,Gran:2013kda}.  \MINERvA currently uses a modification of v2.8.4~\cite{Altinok:2017xua,Patrick:2008moo,Mislivec:2017qfz} with an increased rate for the Valencia two-particle two-hole process; that modification is not used here.  An important difference in single pion production between v2.8.x and v2.12.x is the angular distributions of single pion events in the Rein-Sehgal model, discussed below.  A sample of 2.5 million events were generated using the \MINERvA flux predictions~\cite{minerva_flux}, a polystyrene target and the official \GENIE 2.12.6 splines~\cite{geniesplines}.  

To simulate pion production, \GENIE uses the Rein-Sehgal (RS) model~\cite{Rein:1980wg} with a hadronic invariant mass cut of $W\leq 1.7\units{GeV}$.  Of the 18 resonances in the RS model, the $\Delta(1600)$ and $N(1990)$ were not included due to their unclear experimental status at the time of implementation.  Resonance-resonance and resonance-nonresonance interference terms are not included.  Lepton mass terms are only included in calculating phase space limits and are neglected when calculating the cross sections. A discussion of the limitations of this simplification can be found in Ref.~\cite{Graczyk:2008zz}.  In earlier versions---including v2.8.4---the pion-nucleon distribution was isotropic in the resonance rest frame, but was changed in 2.12.x. Here we use the non-isotropic model as our default and reweight to the isotropic distribution, explained later. The RS nonresonant background is not used by \GENIE; rather, a deep inelastic scattering (DIS) model is extended to cover that invariant mass region. The DIS model uses the Bodek-Yang parametrization~\cite{Bodek:2002ps}, and the AGKY model to describe hadronization~\cite{Yang:2009zx}.  In the AGKY model, the KNO model~\cite{Koba:1972ng} is used for $W\leq2.3\units{GeV}$ and PYTHIA~\cite{Sjostrand:2007gs} is used for $W\geq3.0\units{GeV}$, with a smooth transition in between the two, implemented by randomly selecting the results of one model or the other for each event.

In addition to pion production on a single nucleon, it is also possible for a neutrino to produce a pion by scattering coherently off the nucleus.  \GENIE uses the Rein-Sehgal coherent pion production model~\cite{rein1983coherent, rein2007coherent}, including the effect of lepton masses in the cross-section calculation. \MINERvA has found that the RS coherent pion production model needs to be suppressed by $\sim50\%$ at \Kpi$<500\units{MeV}$ to agree with data~\cite{Mislivec:2017qfz}.  This correction also moves the shape of the \Kpi spectrum closer to the predictions of the Berger-Sehgal coherent model~\cite{BergerSehgalCoherent}.  The \ccpip channel has a small contribution from coherent production in the lowest \qq bins but the inclusion of this suppression has only a small effect on the MC predictions.  To maintain a model similar to that currently being used by \MINERvA, this suppression is included in the analysis presented in Sections~\ref{sec:genie-systs} and onwards.

The ``hA Intranuke'' effective cascade model~\cite{intranuke} is used to model pion and nucleon FSI.  In this model, the effect of intranuclear scattering is parameterized as a single cascade step applied to each particle emanating from the primary interaction.  This model steps hadrons through a nucleus of radius $r \sim A^{1/3}$ and a nuclear density function derived from electron scattering data.  The hadron's mean free path is determined from tabulated hadron-proton and hadron-neutron cross sections~\cite{SAID}.   The probability to interact with the nucleus is high; it is \eg $\sim$73\% for a pion from an \Ev = 3 GeV quasielastic event in carbon.  When a FSI occurs, the possible interactions (absorption, pion production, knockout, charge exchange, elastic scatter) are chosen according to their proportions for iron.

\begin{table*}[hbtp]
\centering
{\renewcommand{\arraystretch}{1.2}
\begin{tabular}{ccccccc}
\hline\hline
Distribution & Channel & \nbins & Default & ANL/BNL & FrAbs Tune & FrInel Tune \\ 
\hline\hline
\Pmu (Rate)  & \ccpip  & 8       & 19.1    & 13.8    & 12.0       & 12.3  \\ 
             & \ccnpip & 9       & 35.4    & 19.5    & 26.1       & 26.8  \\ 
             & \ccpin  & 8       & 11.1    & 19.6    & 19.0       & 19.3  \\ 
             & \ccapin & 9       & 7.4     & 6.4     & 6.2        & 6.3   \\ 
\hline\hline
\Tmu (Shape) & \ccpip  & 9       & 7.1     & 12.4    & 7.5        & 7.4   \\ 
             & \ccnpip & 9       & 4.5     & 10.4    & 4.0        & 4.1   \\
             & \ccpin  & 9       & 35.1    & 71.5    & 44.5       & 45.6  \\ 
             & \ccapin & 9       & 9.3     & 14.0    & 10.2       & 10.3  \\ 
\hline\hline
\Kpi (Shape) & \ccpip  & 7       & 2.9     & 2.6     & 2.5        & 2.3   \\ 
             & \ccnpip & 7       & 39.8    & 34.7    & 31.2       & 29.4  \\
             & \ccpin  & 7       & 28.3    & 31.4    & 30.9       & 29.9  \\ 
             & \ccapin & 7       & 19.3    & 17.9    & 16.6       & 16.0  \\ 
\hline\hline
\Tpi (Shape) & \ccpip  & 14      & 25.4    & 26.5    & 13.0       & 12.6  \\ 
             & \ccnpip & 14      & 11.7    & 11.1    & 6.9        & 6.2   \\ 
             & \ccpin  & 11      & 13.5    & 15.0    & 8.3        & 8.9   \\ 
             & \ccapin & 11      & 5.7     & 5.9     & 3.4        & 3.5   \\ 
\hline\hline
Total \chisq &         & 148     & 275.6   & 312.7  & 242.3      & 240.7 \\
\hline\hline
\end{tabular}}
\caption{Channel by channel contributions to the \chisq at different stages of the tuning process.
\label{tab:genienomchi2} }
\end{table*}

\begin{figure*}[htbp]
\centering
\includegraphics[width=0.24\textwidth]{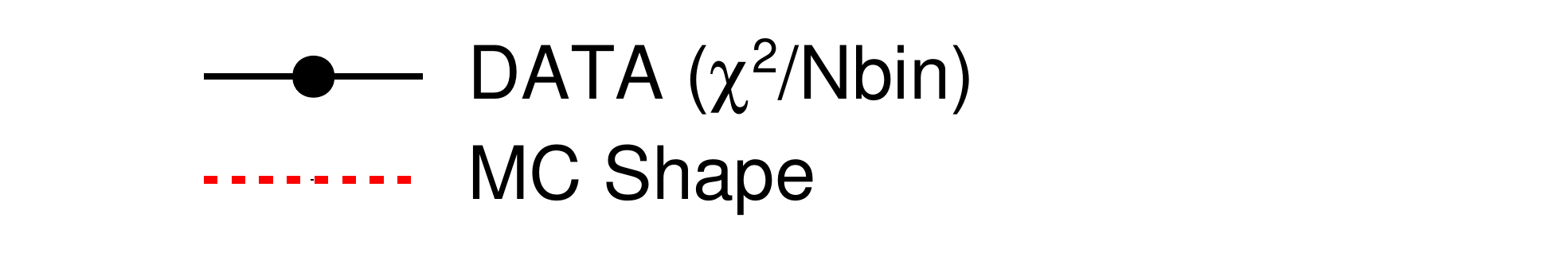}
\includegraphics[width=0.24\textwidth]{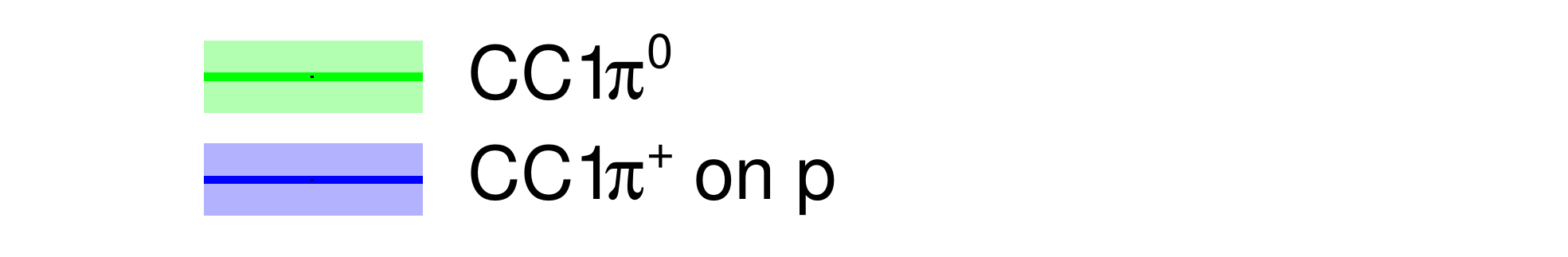}
\includegraphics[width=0.24\textwidth]{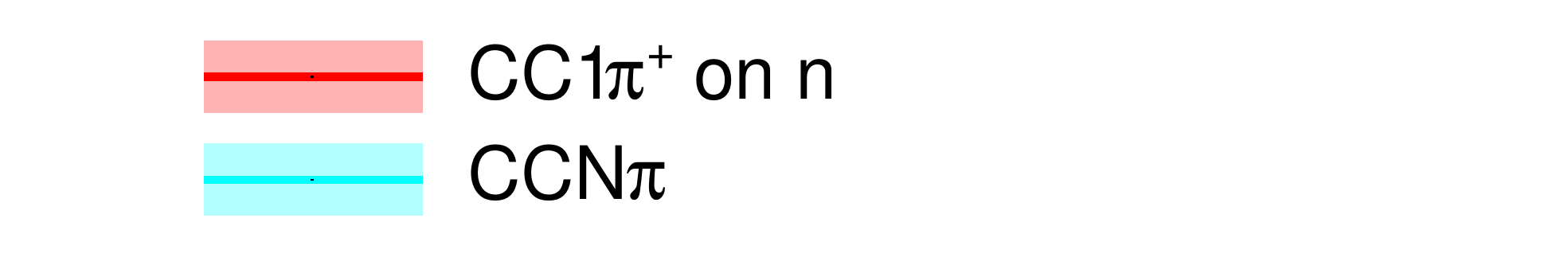}
\includegraphics[width=0.24\textwidth]{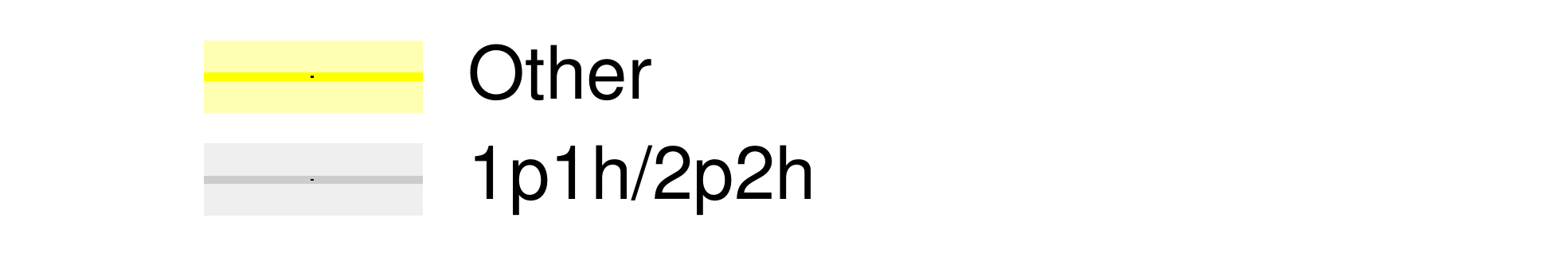}
\includegraphics[width=0.24\textwidth,trim={0mm 2mm 10mm 5mm}, clip]{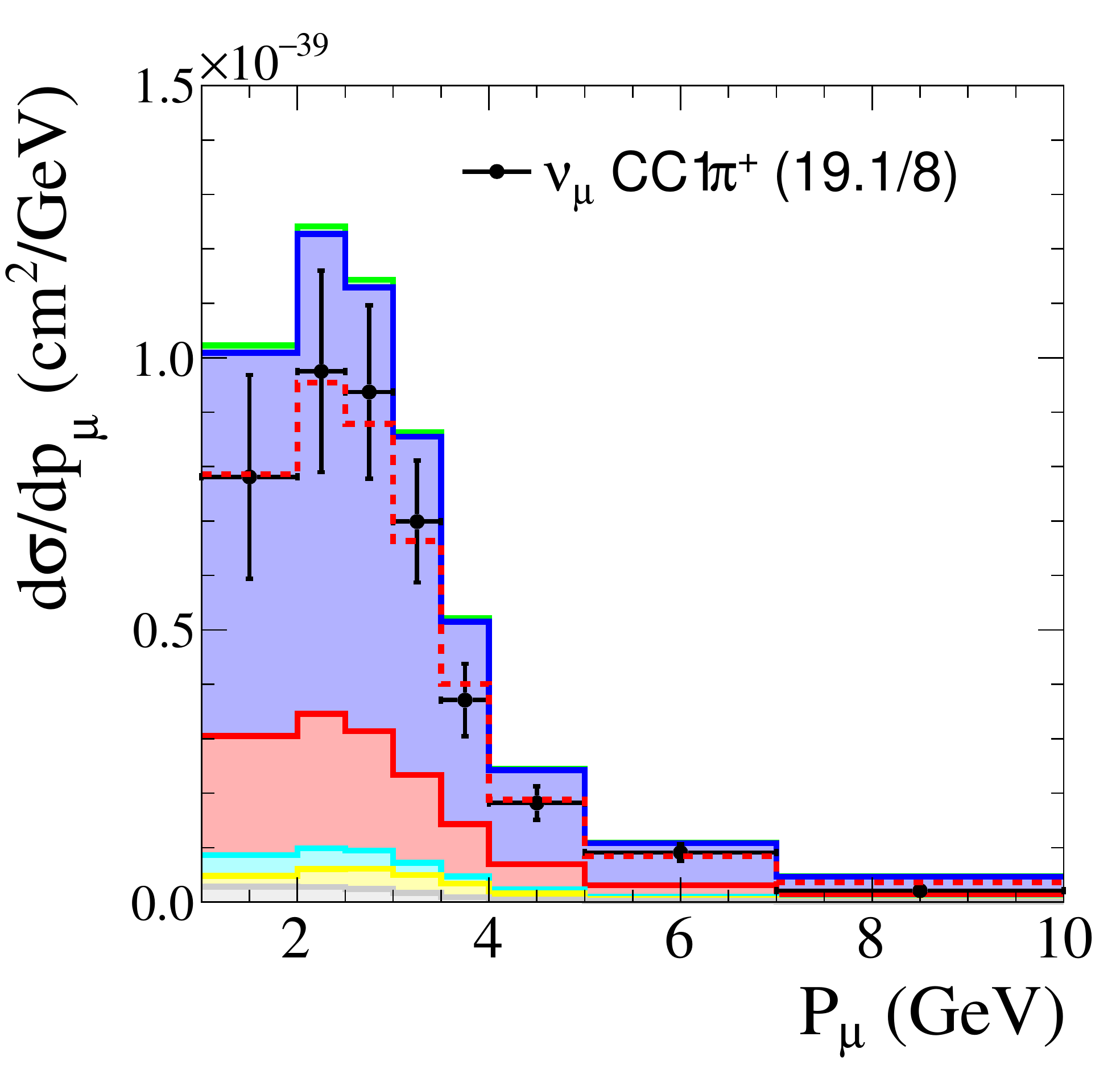}
\includegraphics[width=0.24\textwidth,trim={0mm 2mm 10mm 5mm}, clip]{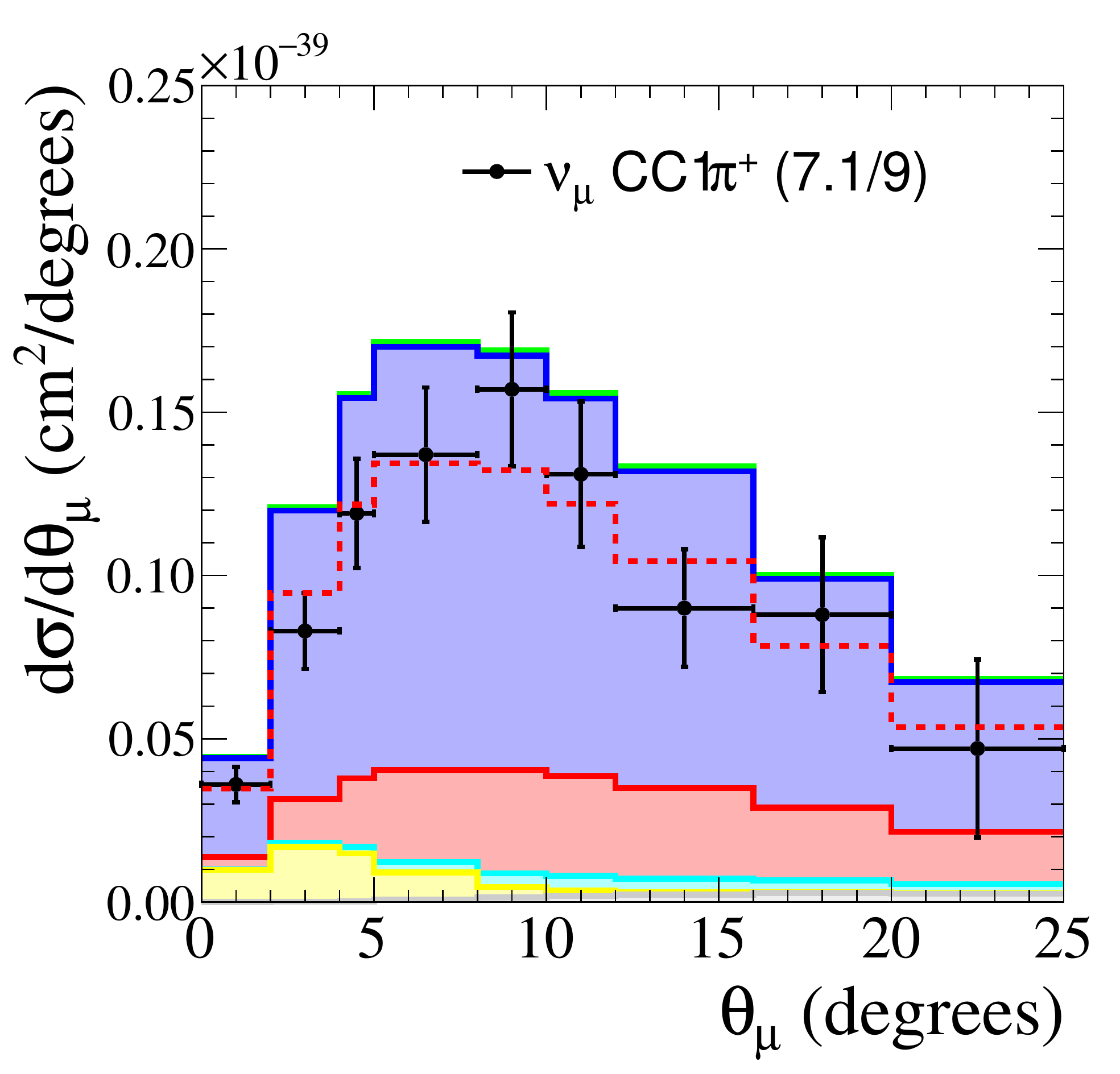}
\includegraphics[width=0.24\textwidth,trim={0mm 2mm 10mm 5mm}, clip]{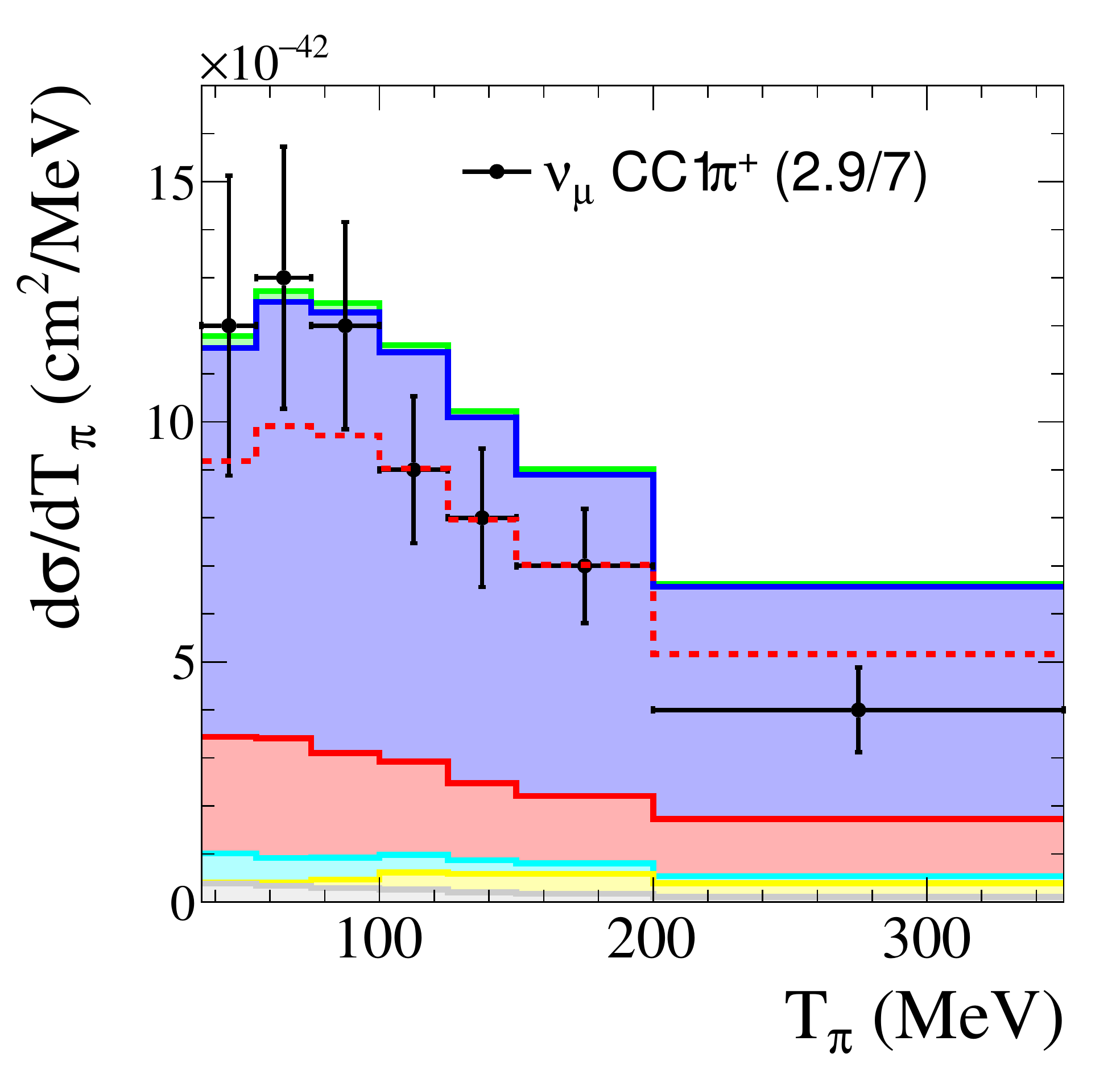}
\includegraphics[width=0.24\textwidth,trim={0mm 2mm 10mm 5mm}, clip]{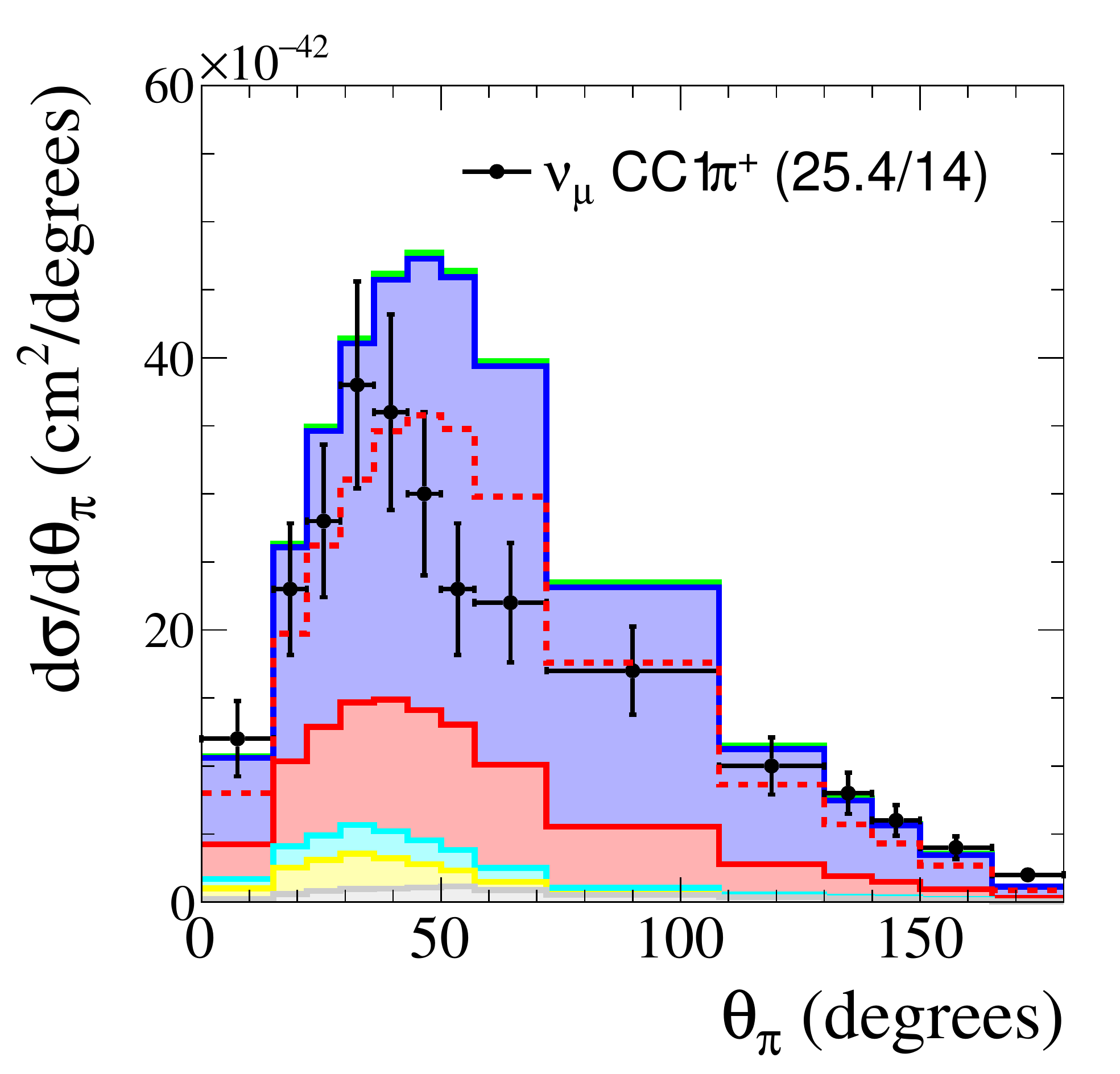} \\
\includegraphics[width=0.24\textwidth,trim={0mm 2mm 10mm 5mm}, clip]{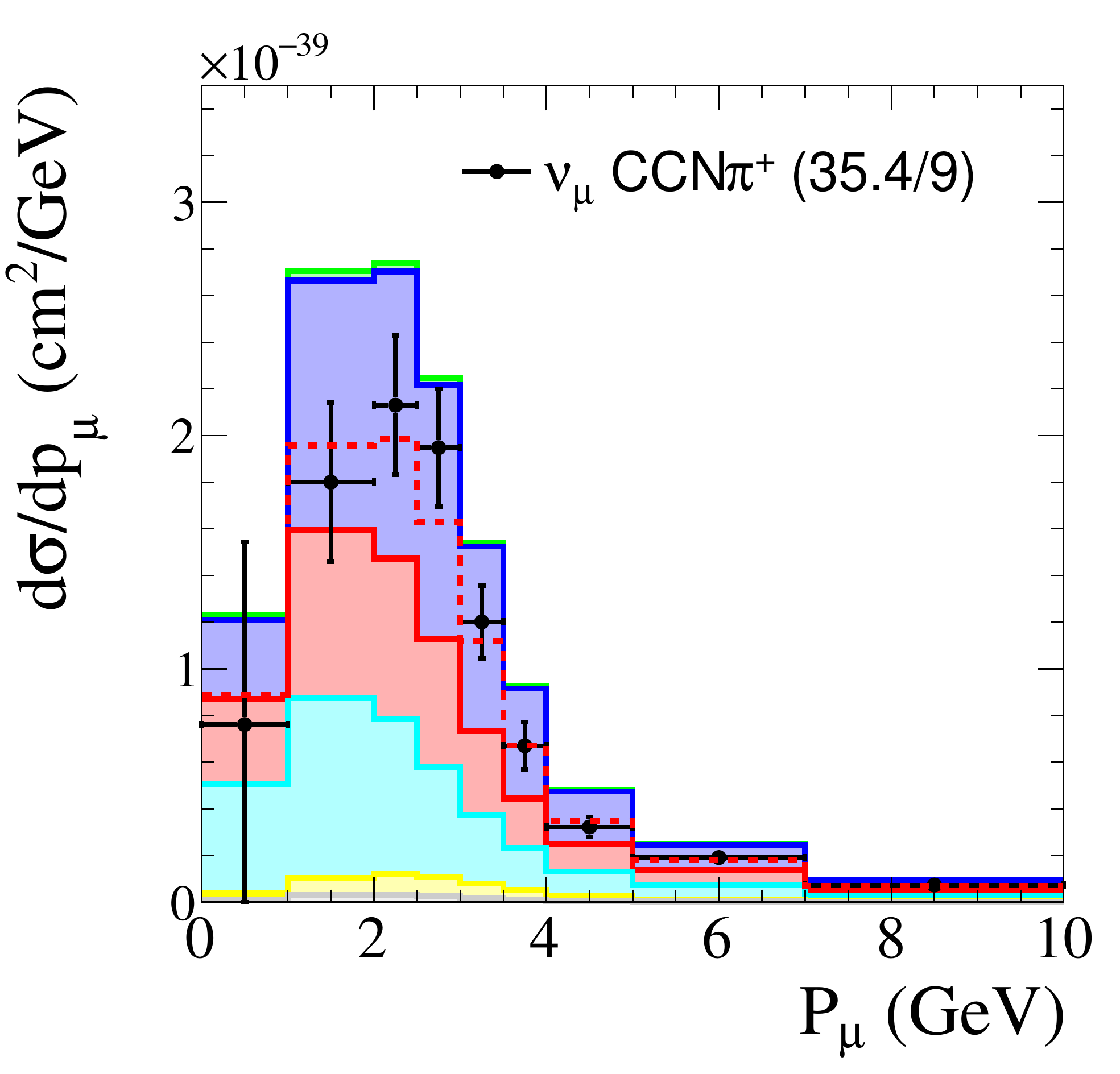}
\includegraphics[width=0.24\textwidth,trim={0mm 2mm 10mm 5mm}, clip]{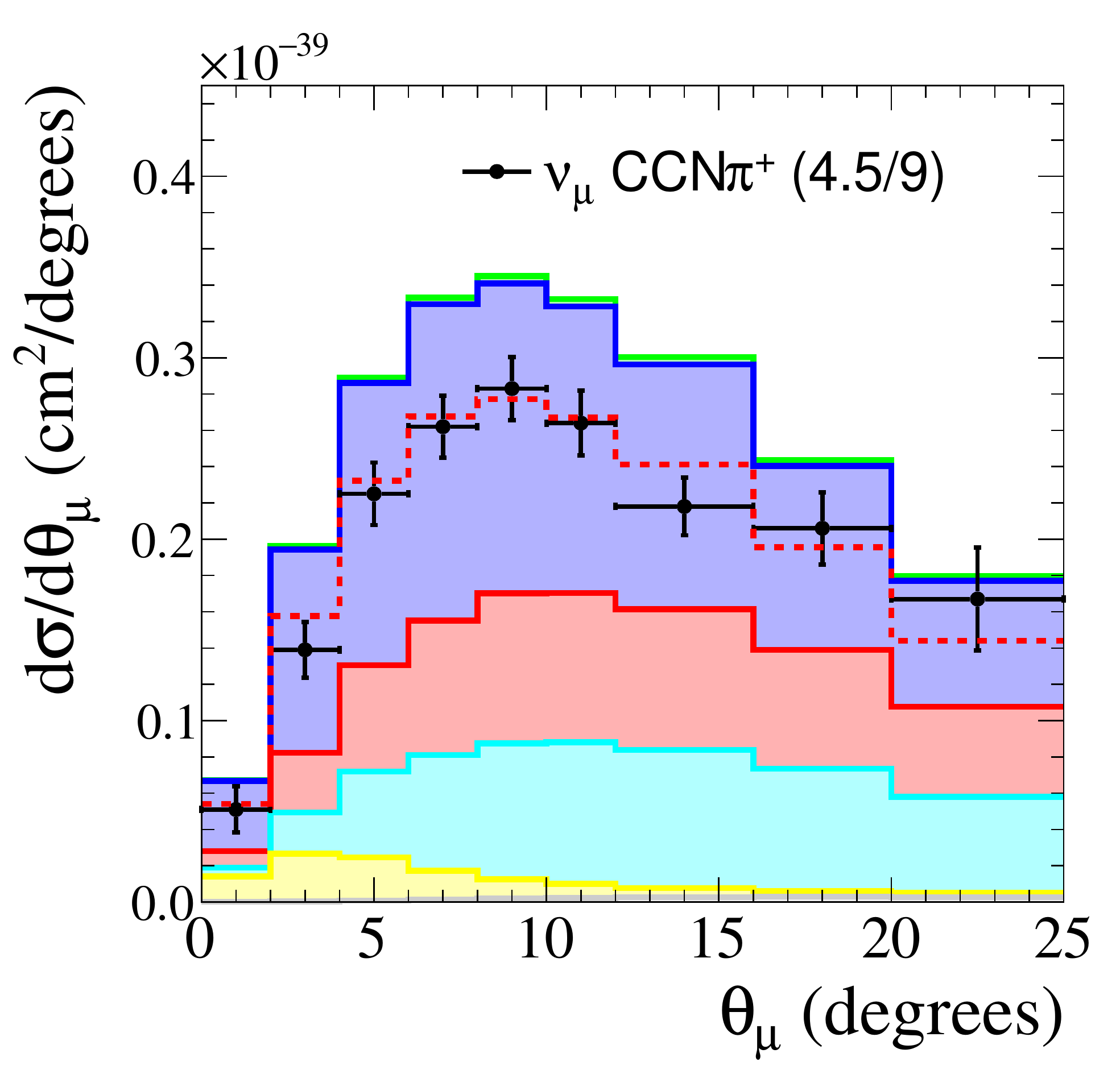}
\includegraphics[width=0.24\textwidth,trim={0mm 2mm 10mm 5mm}, clip]{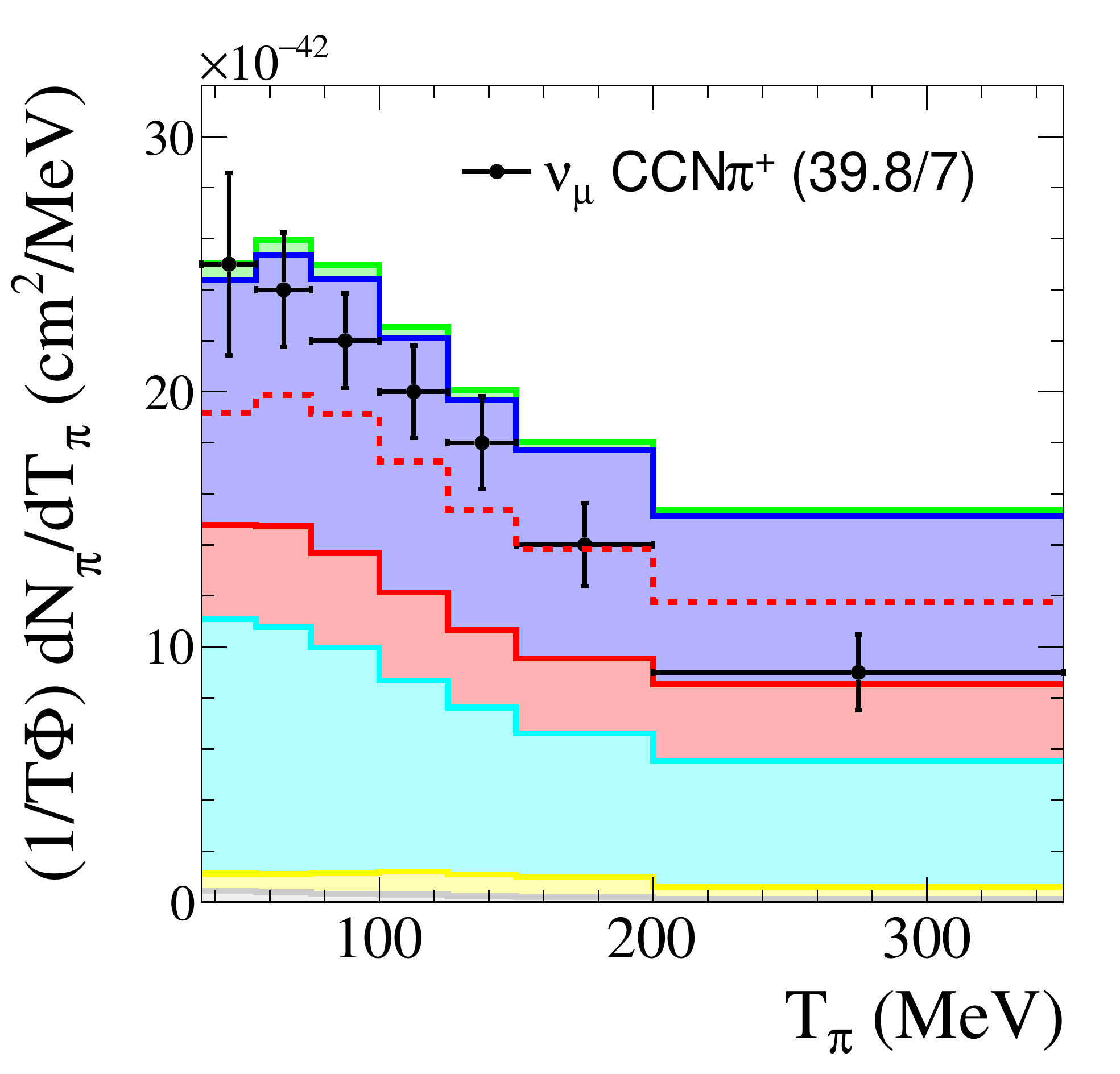}
\includegraphics[width=0.24\textwidth,trim={0mm 2mm 10mm 5mm}, clip]{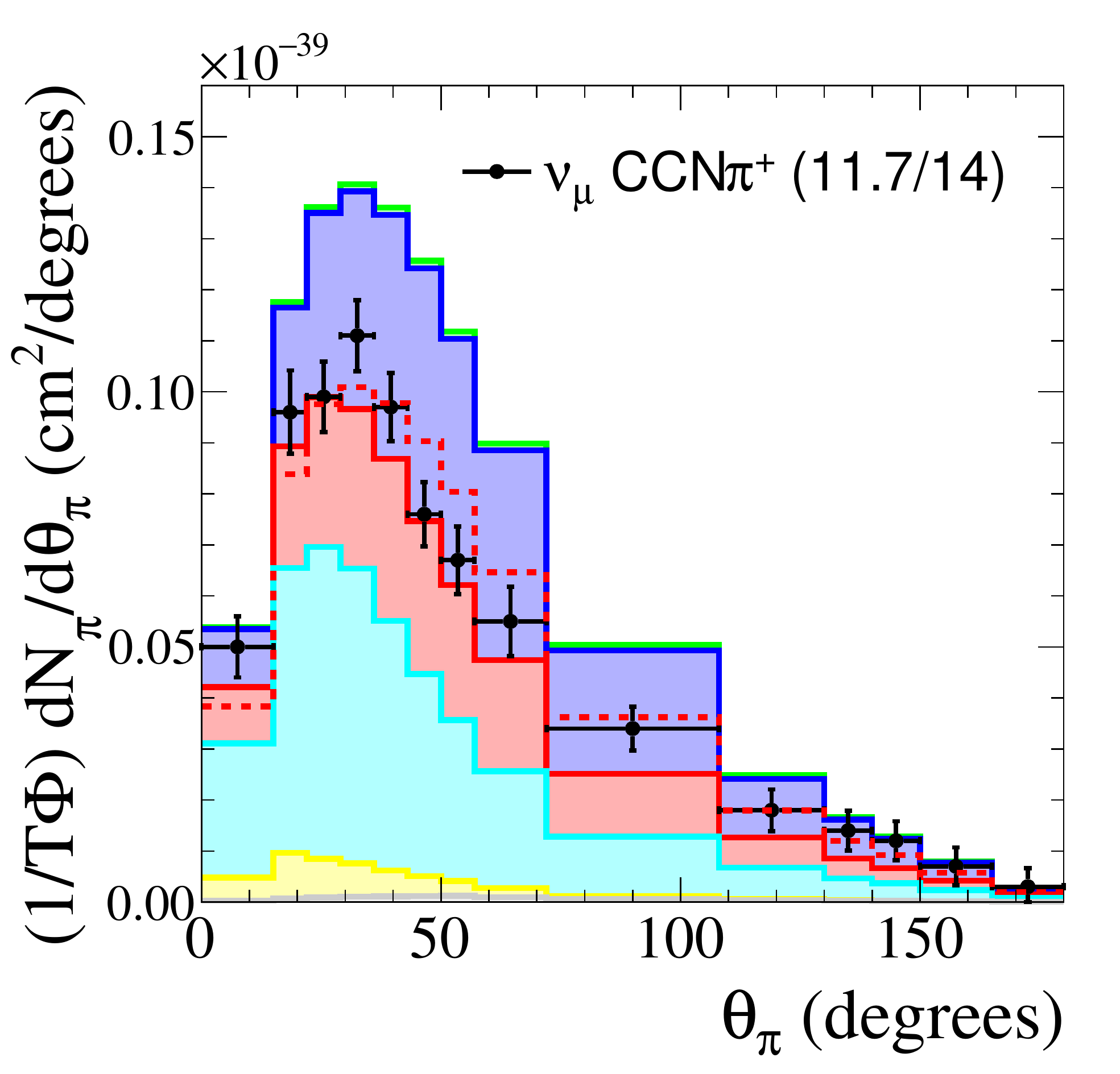} \\
\includegraphics[width=0.24\textwidth,trim={0mm 2mm 10mm 5mm}, clip]{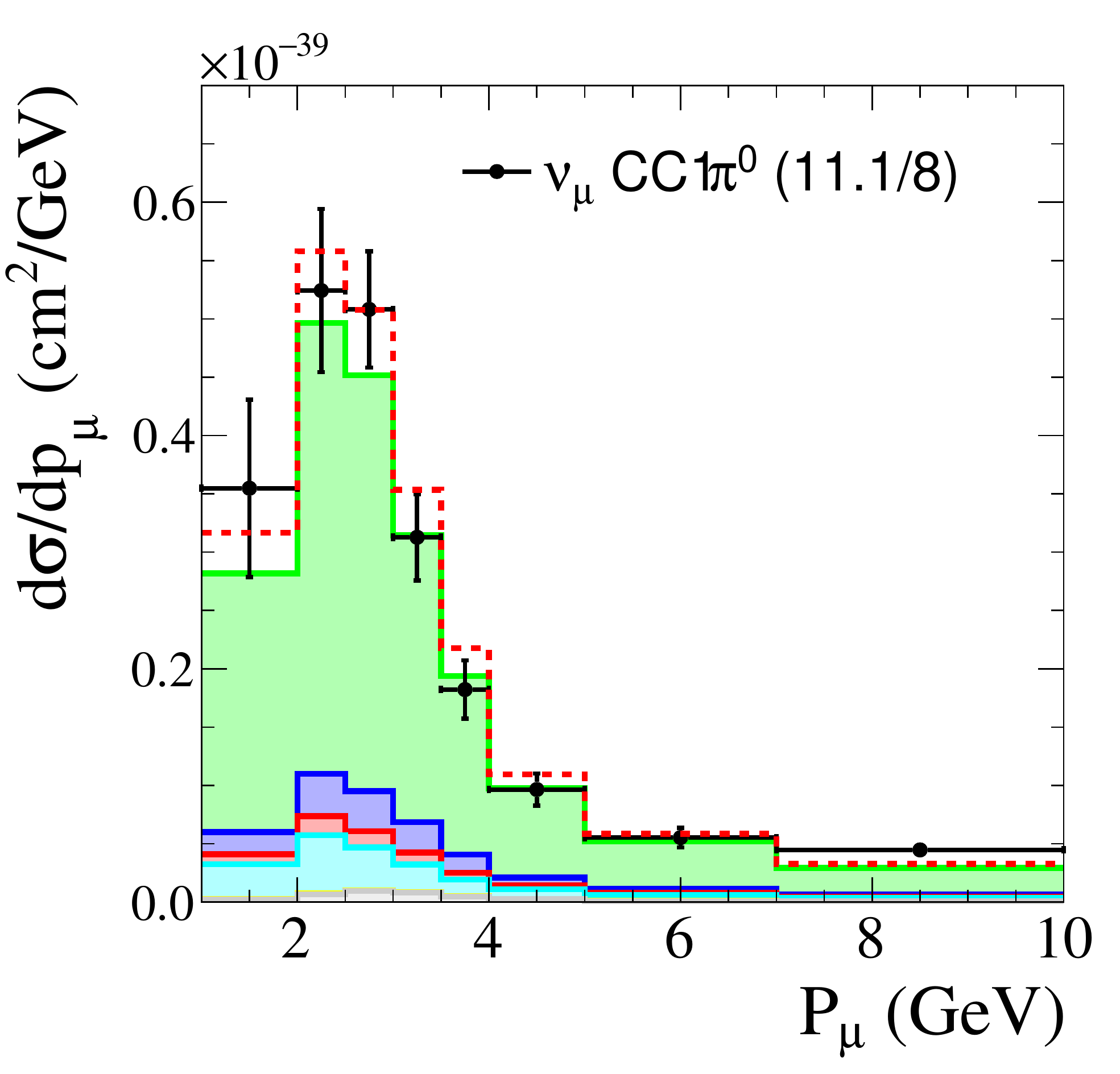}
\includegraphics[width=0.24\textwidth,trim={0mm 2mm 10mm 5mm}, clip]{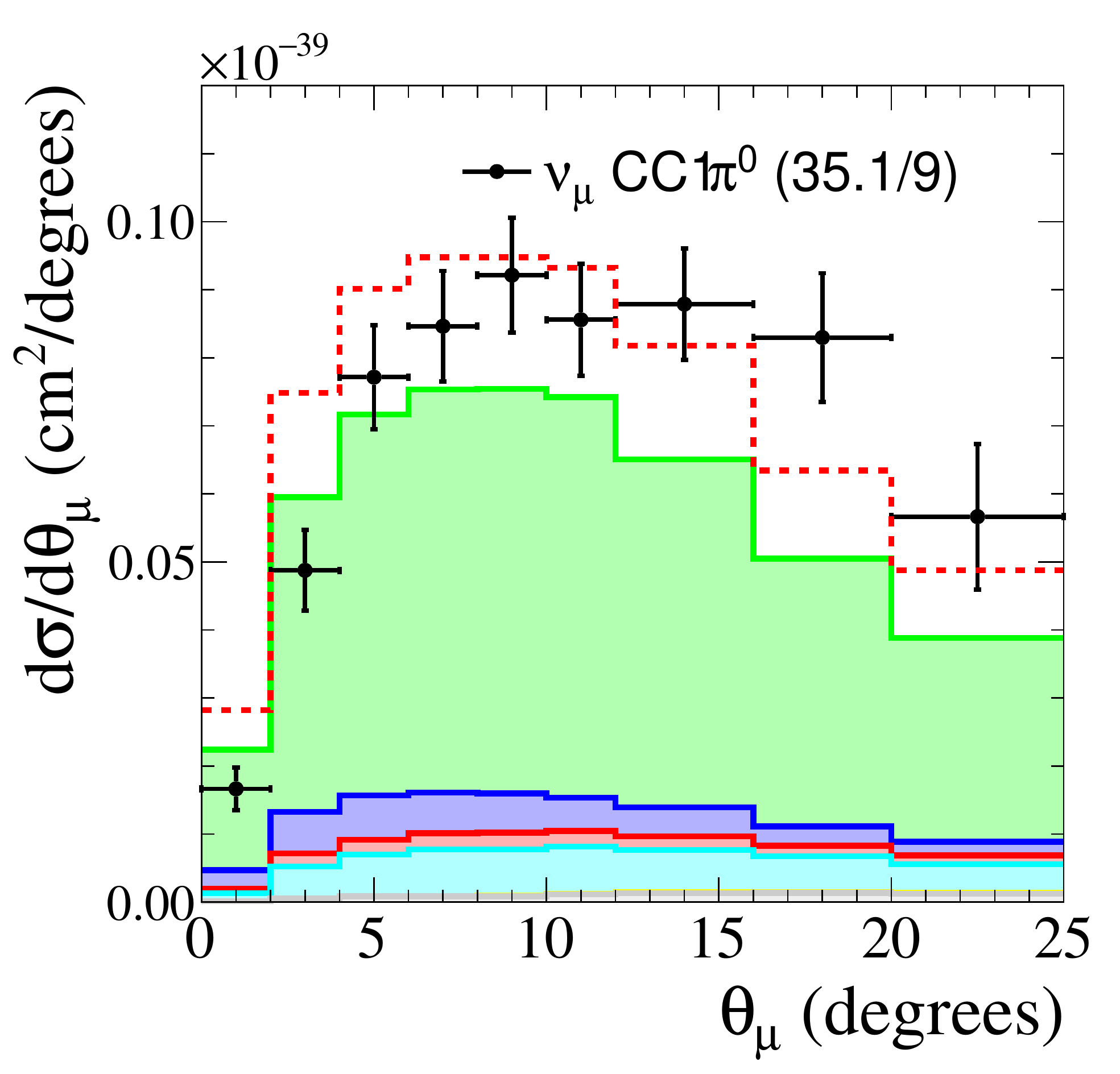}
\includegraphics[width=0.24\textwidth,trim={0mm 2mm 10mm 5mm}, clip]{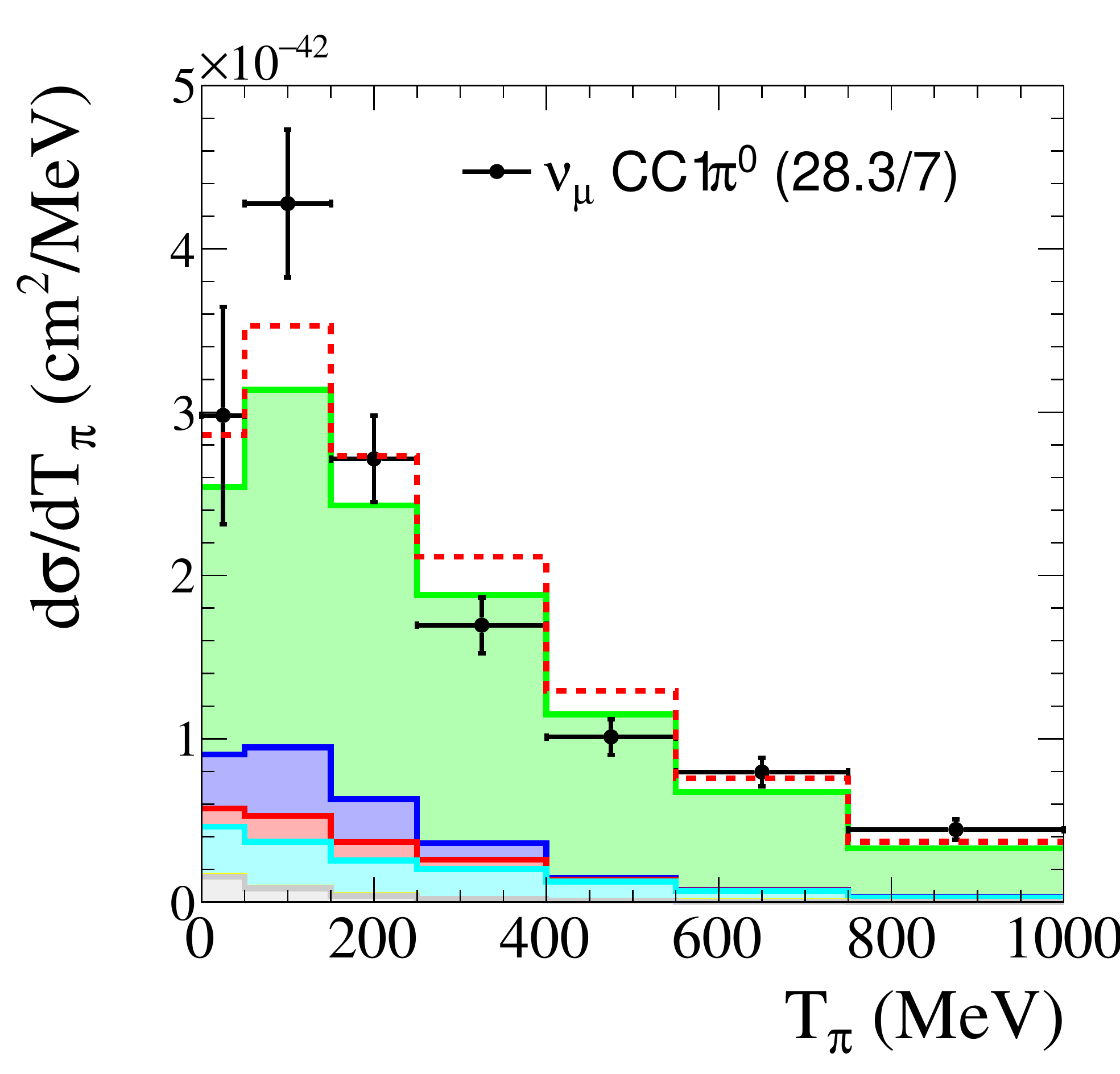}
\includegraphics[width=0.24\textwidth,trim={0mm 2mm 10mm 5mm}, clip]{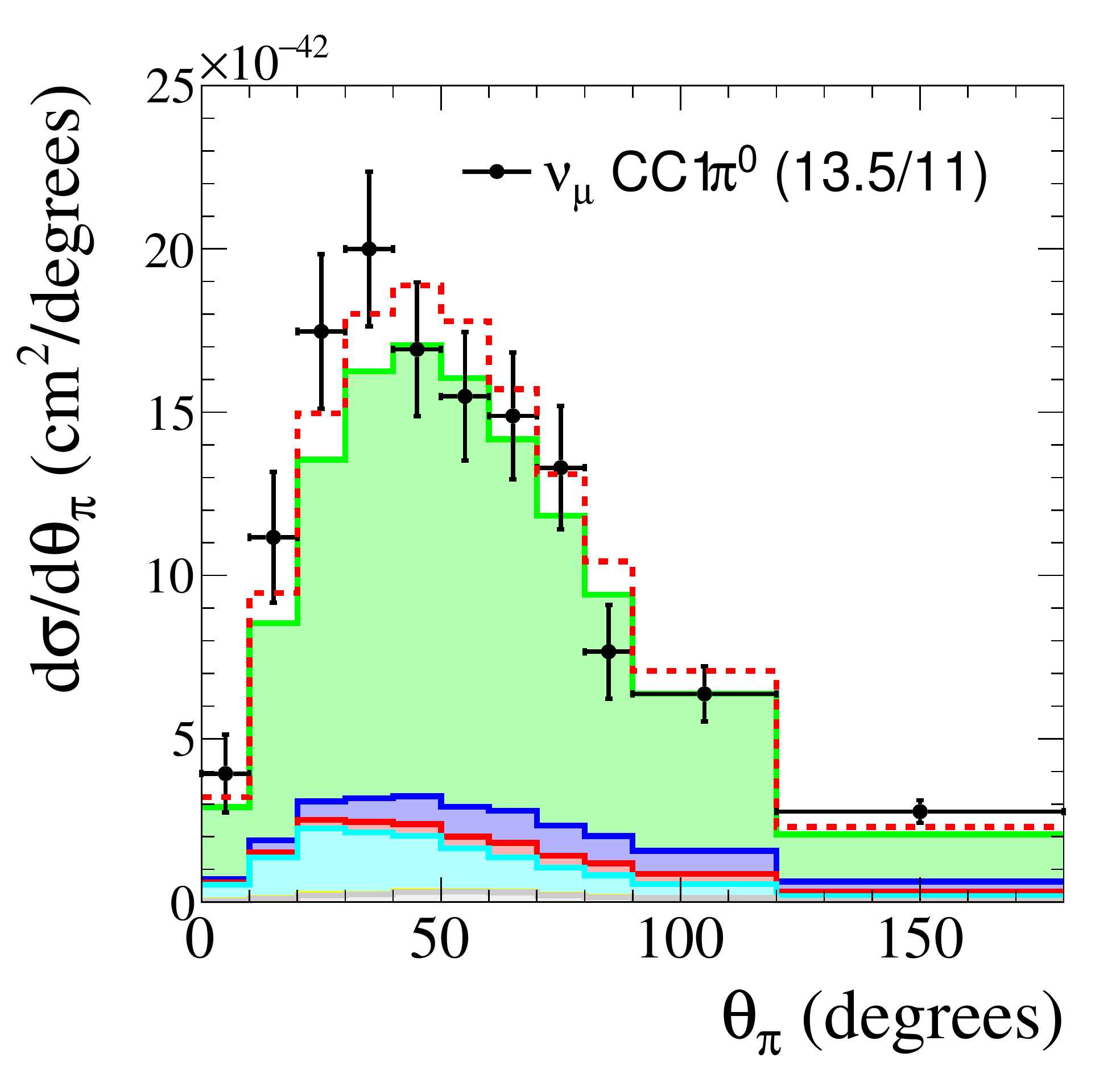} \\
\includegraphics[width=0.24\textwidth,trim={0mm 2mm 10mm 5mm}, clip]{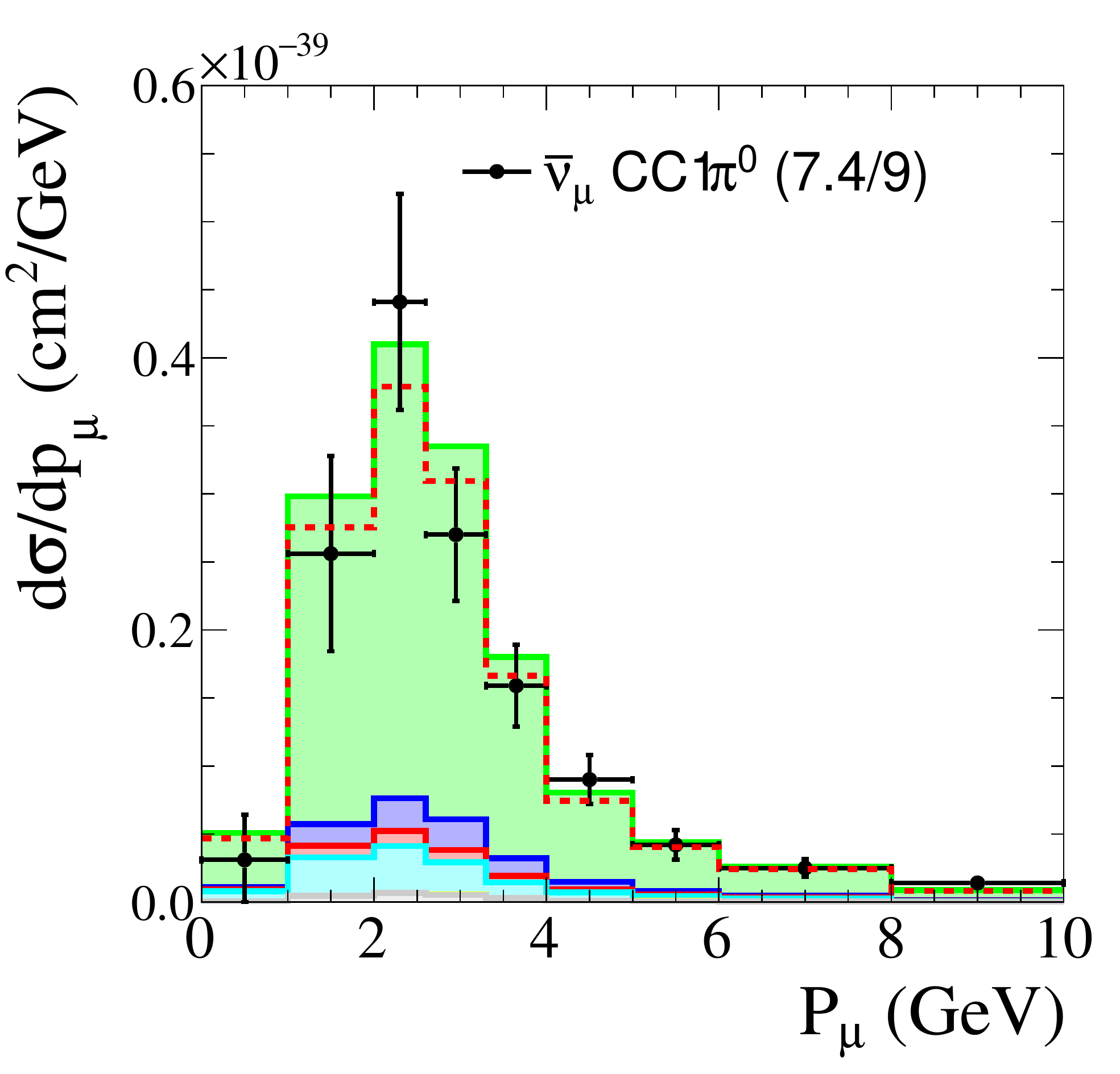}
\includegraphics[width=0.24\textwidth,trim={0mm 2mm 10mm 5mm}, clip]{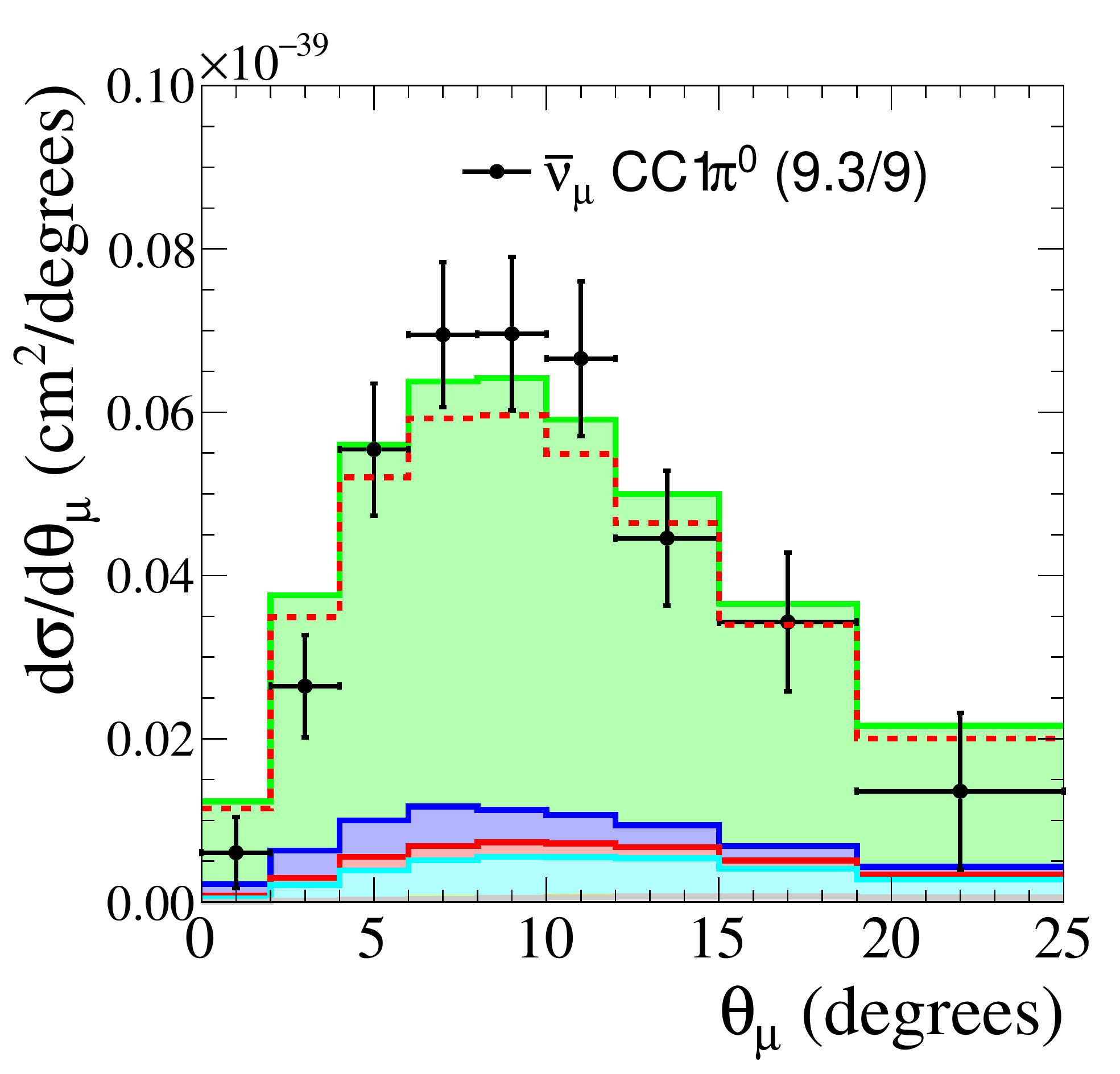}
\includegraphics[width=0.24\textwidth,trim={0mm 2mm 10mm 5mm}, clip]{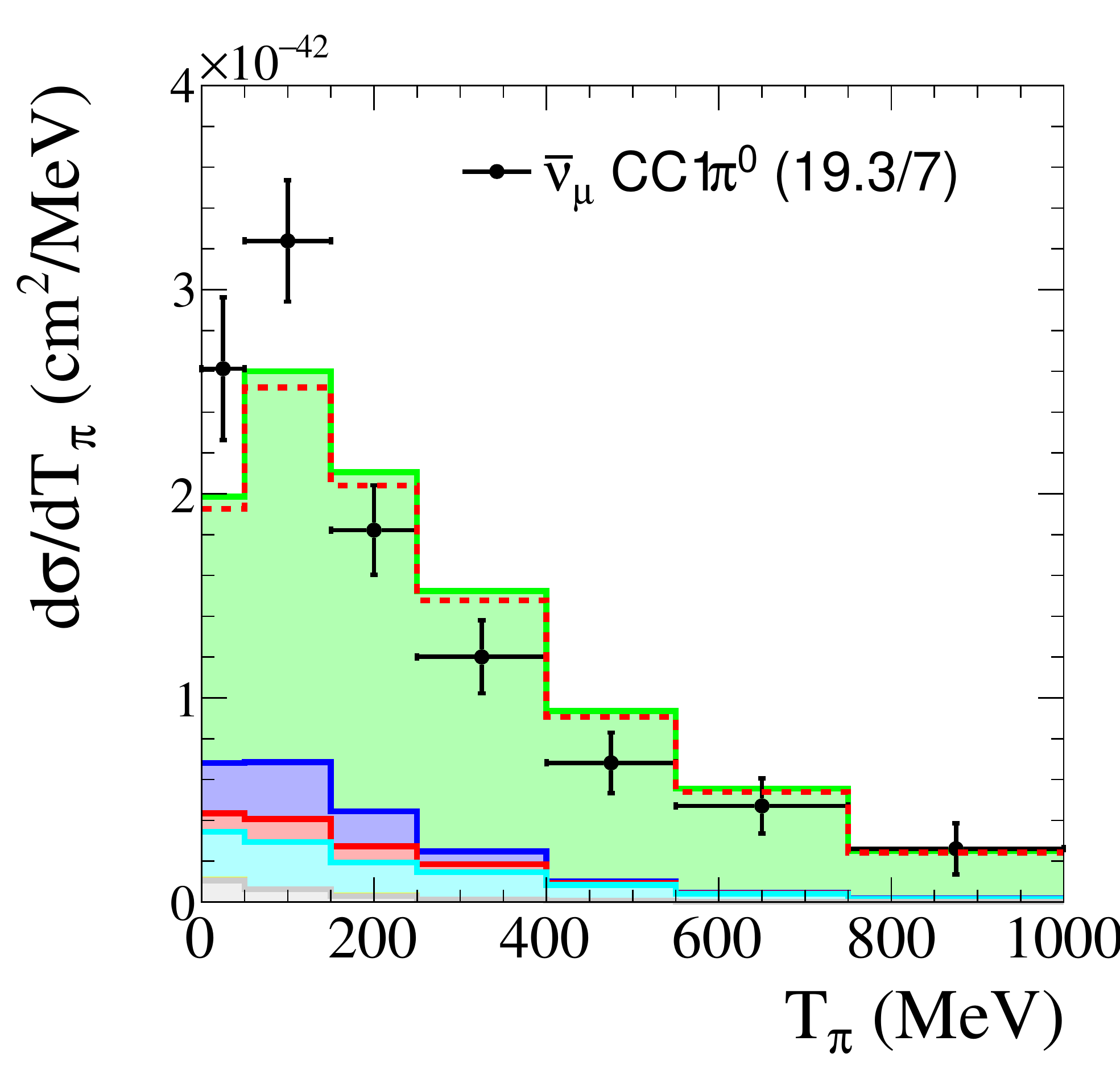}
\includegraphics[width=0.24\textwidth,trim={0mm 2mm 10mm 5mm}, clip]{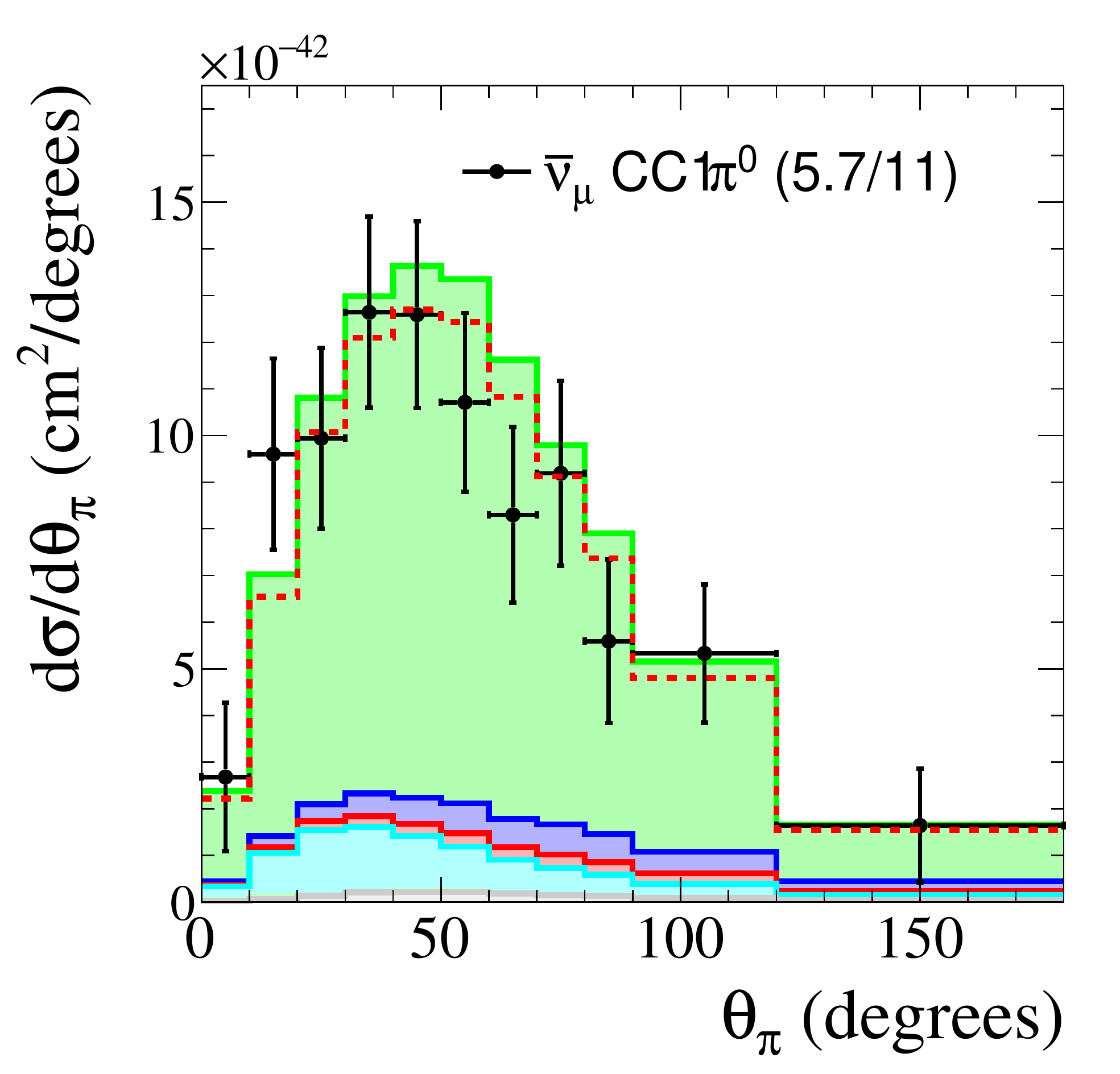}
\caption{\GENIE 2.12.6 Default model predictions compared to \MINERvA data.  Colors correspond to particle content at the nucleon interaction.  ``Other'' is dominated by coherent pion production.  ``MC Shape'' shows the total MC prediction after it has been normalized to match the total data normalization.  In the case of the shape-only distributions (\Tmu, \Kpi, \Tpi) the shape-only \chisq/\nbins values are shown.  All cross sections are per nucleon.
\label{fig:genienomcomp}}
\end{figure*}

Default \GENIE predictions separated by nucleon level interaction channels for the \MINERvA data are shown in Fig.~\ref{fig:genienomcomp}.  The shape of the \Pmu distributions agree well with the data for all four measurements.  However, the model overestimates the cross section for \pipm production and as a result the \chisq for the \ccpip and \ccnpip, given in the fourth column (``Default'') of Table~\ref{tab:genienomchi2}, are large.  The model overestimates \Tmu below $<5^\circ$ in the \pin channels, although it does correctly predict the shape of the \Tmu distribution in the \pipm channels.  The model underestimates the production rate at large \Tmu in \ccpin.  The shape of the \Kpi distribution is in larger disagreement for \ccnpip data than for \ccpip.  Since the \ccnpip distributions summed over all identified \pipm, redistributing kinetic energy between \pipm in events with more than one \pipm could resolve some of this tension.  The \pin channels are under-predicted at low \Kpi.  Finally, \GENIE predictions are too high in magnitude at $\Tpi \approx 50^\circ$ in both the \ccpip and \ccnpip channels, and the prediction has the wrong shape in the \ccpip channel.  Comparisons using the transport theory based GiBUU model~\cite{Buss:2011mx} show similar shape disagreements despite GiBUU's use of an advanced semiclassical cascade model to simulate FSI~\cite{Mosel:2017nzk}.

\begin{figure*}[hbtp]
\centering
\includegraphics[width=0.48\textwidth,trim={12mm 2mm 10mm 8mm}, clip]{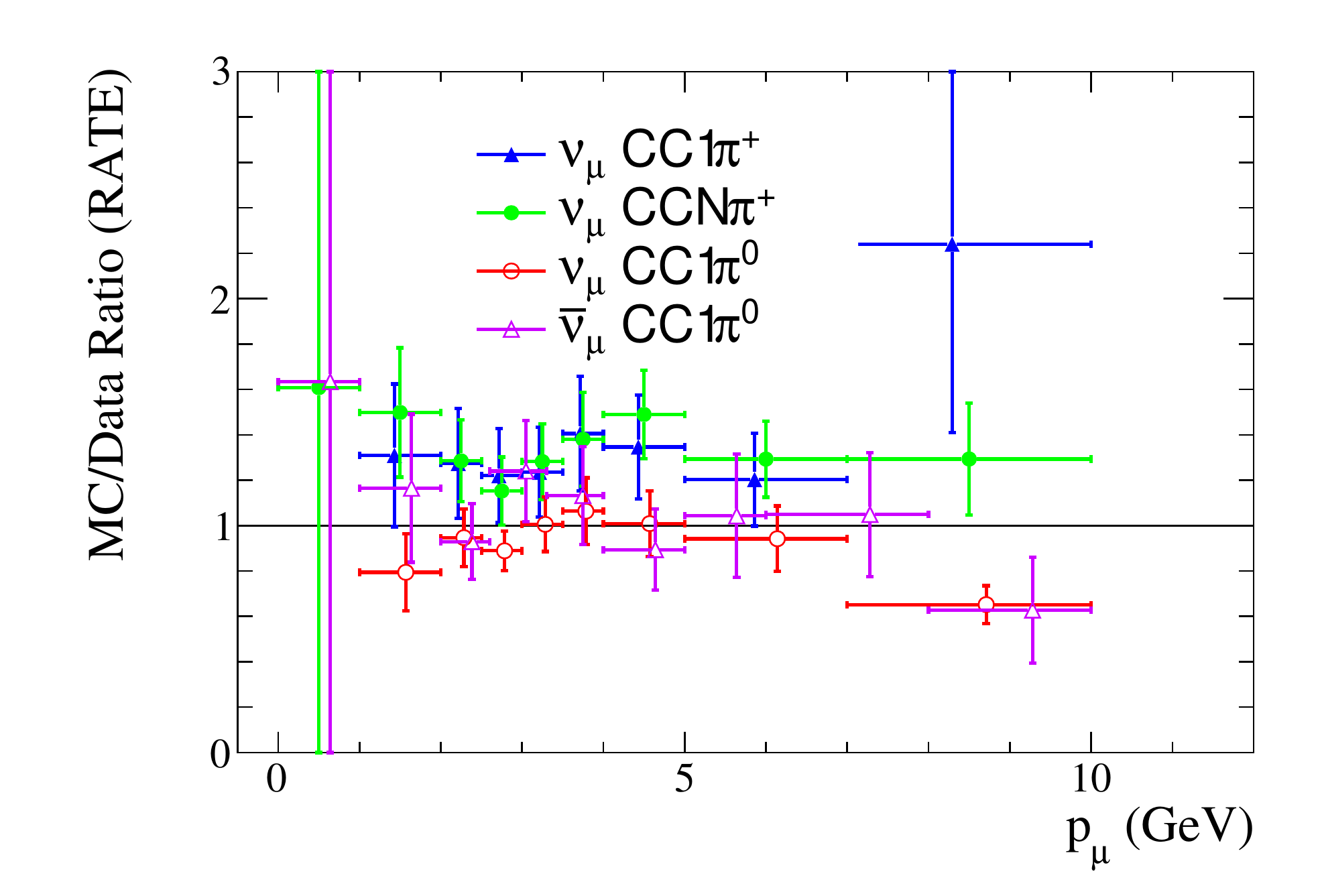}
\includegraphics[width=0.48\textwidth,trim={12mm 2mm 10mm 8mm}, clip]{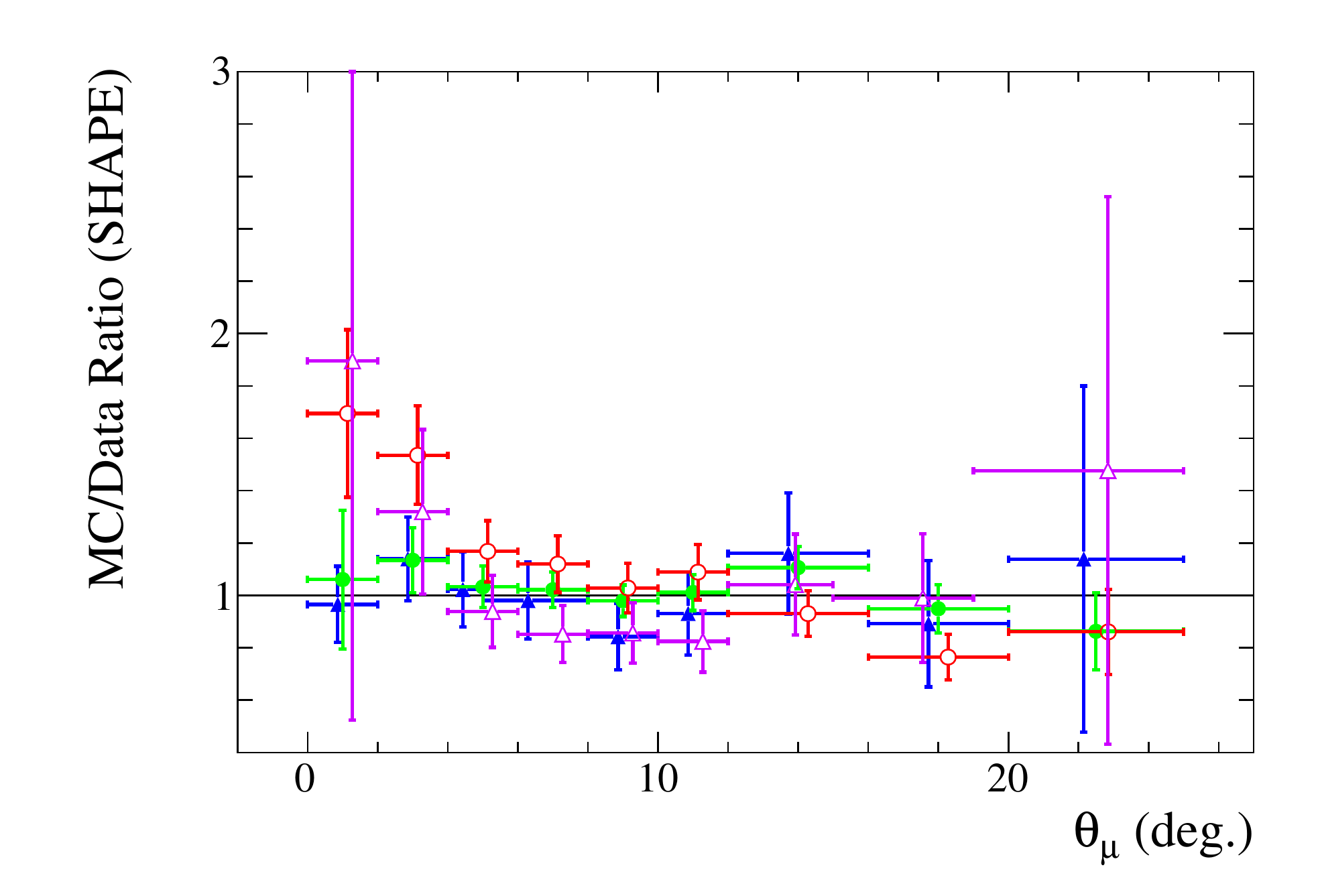}
\includegraphics[width=0.48\textwidth,trim={12mm 2mm 10mm 8mm}, clip]{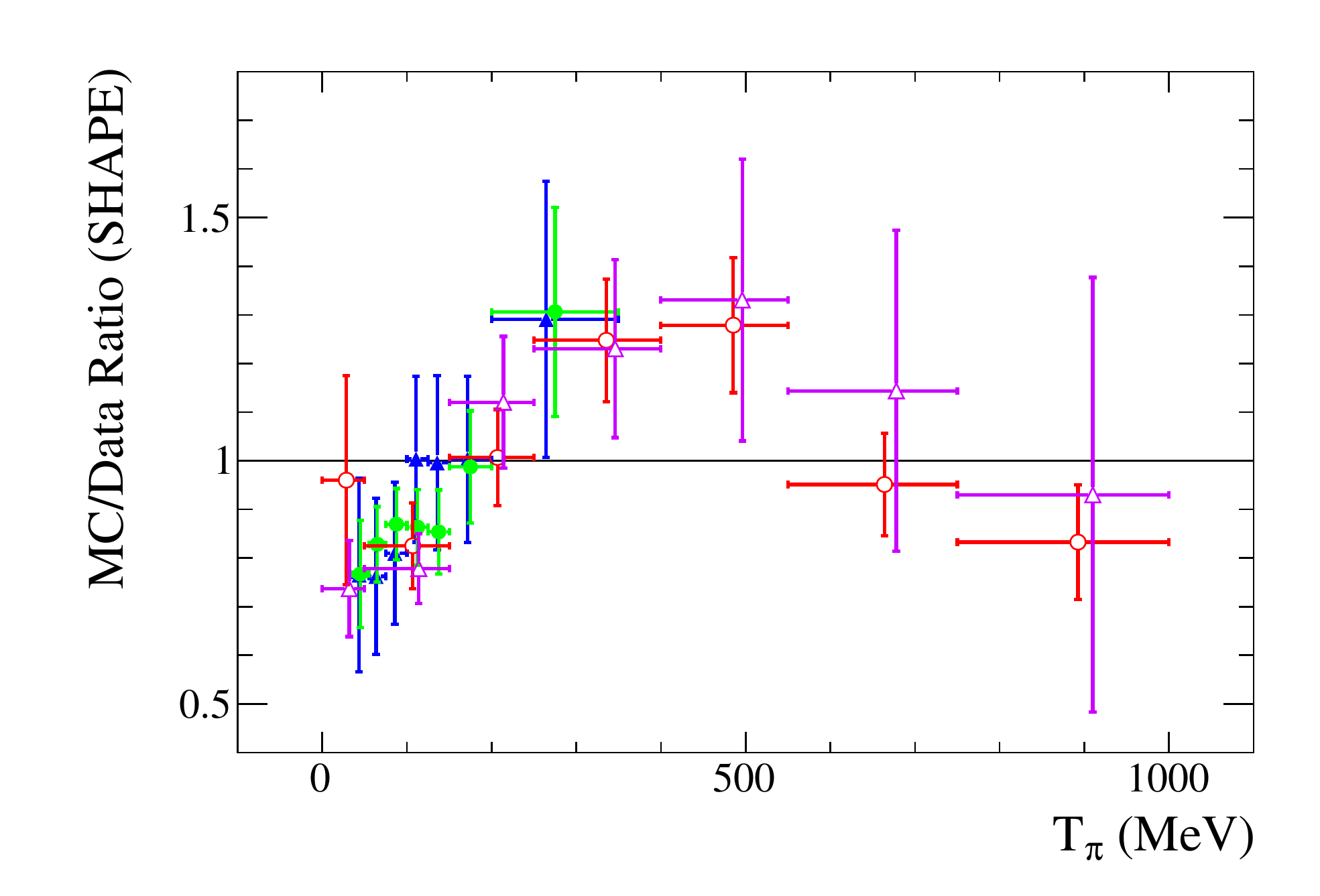}
\includegraphics[width=0.48\textwidth,trim={12mm 2mm 10mm 8mm}, clip]{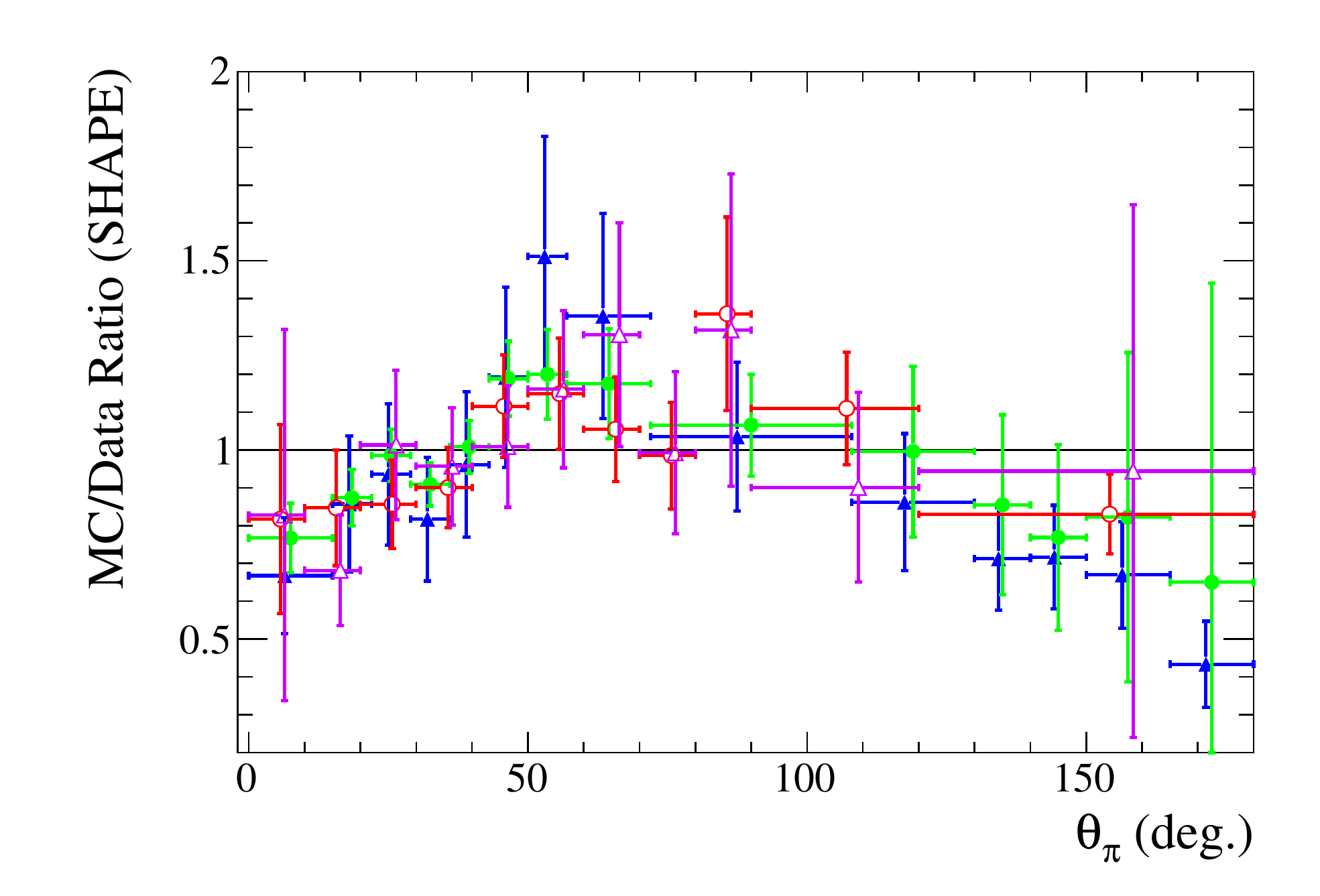}
\caption{MC/data ratios for the default \GENIE predictions. The \Pmu distribution provides a rate comparison in the \chisq calculation; the other distributions are treated as shape-only, \ie, the MC is normalized to match the data and the uncertainties are from the shape-only covariance matrix.
\label{fig:genienomratios}}
\end{figure*}

Each of the measurements are shown as MC/data ratio distributions in Fig.~\ref{fig:genienomratios}.  Similar comparisons between the \MiniBooNE and \MINERvA experiments are found in Ref.~\cite{Mahn:2018mai}. The shape-only data sets ($\theta_\mu$, $\theta_\pi$, $T_\pi$) were normalized to match the data before the ratio was taken and the error bars in Fig.~\ref{fig:genienomratios} reflect the extracted shape-only uncertainties on the data, so that the distributions reflect their contributions to the total $\chi^2$.

\section{Tunable parameters in the \GENIE Model}
\label{sec:genie-systs}
The \GENIE event generator allows assessment of systematic uncertainties through the \GENIE reweighting package.  A large number of event weighting ``dials'' are included to allow model uncertainties to be evaluated.  The dials adjusted in this note are summarized in Table~\ref{tab:geniedials} and are chosen because of their connection to the kinematic variables and interaction modes studied herein.

\begin{table*}[htbp]
\centering
{\renewcommand{\arraystretch}{1.2}
\begin{tabular}{ c  c  c  }
\hline\hline
Parameter                                         & Default Value                         & \GENIE parameter name \\
\hline\hline
CC Resonant Axial Mass (\Mares)                   & $1.12 \pm 0.22\units{GeV}$            & \texttt{MaCCRES}  \vspace{0.3cm} \\
CC Resonant Normalization (\Normres)              & $100 \pm 20\units{\%}$                & \texttt{NormCCRES} \vspace{0.3cm} \\
CC1$\pi$ Nonresonant Normalization (\nonresonepi) & $100 \pm 50\units{\%}$                & \texttt{NonRESBGvnCC1pi} \\
                                                  &                                       & \texttt{NonRESBGvpCC1pi} \\
                                                  &                                       & \texttt{NonRESBGvbarnCC1pi} \\
                                                  &                                       & \texttt{NonRESBGvbarpCC1pi} \vspace{0.3cm} \\
CC2$\pi$ Nonresonant Normalization (\nonrestwopi) & $100 \pm 50\units{\%}$                & \texttt{NonRESBGvnCC2pi} \\
                                                  &                                       & \texttt{NonRESBGvpCC1pi} \\
                                                  &                                       & \texttt{NonRESBGvbarnCC1pi} \\
                                                  &                                       & \texttt{NonRESBGvbarpCC1pi} \vspace{0.3cm}\\
Pion Angular Emission (\ThetaPi)                  & 0 (RS)                                & \texttt{Theta\_Delta2Npi} \\
Pion Absorption FSI Fraction (FrAbs)              & $100 \pm 30\units{\%}$                & \texttt{FrAbs\_pi} \vspace{0.3cm} \\
Pion Inelastic FSI Fraction (FrInel)              & $100 \pm 40\units{\%}$                & \texttt{FrInel\_pi} \vspace{0.3cm} \\
\hline\hline
\end{tabular}}
\caption{Summary of the \GENIE dials optimized in this note, their default values and the uncertainties recommended by the \GENIE collaboration. We do not use the defaults for \Mares, \Normres and \nonresonepi and instead impose central values and uncertainties from tunings to ANL and BNL data as described in the text.
\label{tab:geniedials}}
\end{table*}

\begin{figure}
\centering
\includegraphics[width=0.48\textwidth,trim={0mm 0mm 25mm 0mm}, clip]{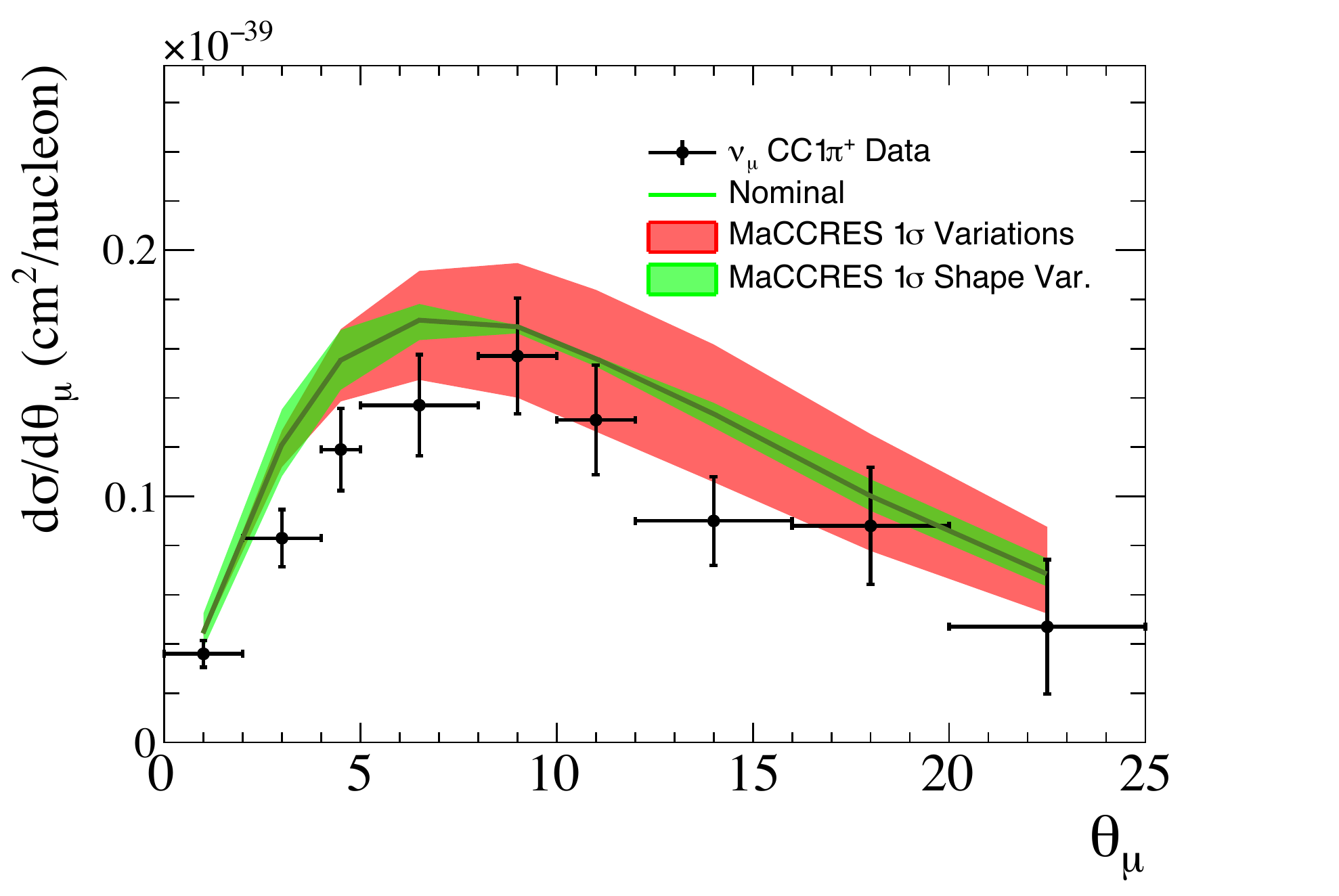}
\caption{The effect varying the \Mares dial on the default \GENIE prediction for \Tmu.  The red bands show the variation to the total rate and shape.  The green bands are obtained by normalizing the reweighted curves to the default predicted rate to highlight the smaller effect the dial has on the shape of the distributions.
\label{fig:madialeffect}}
\end{figure}

Experiments often use variations in the charged-current resonant axial mass, \Mares, as a systematic uncertainty which varies both the normalization and $Q^{2}$ shape of resonant interactions along with variations in a total resonant cross-section normalization dial, \Normres.  Variations in \Normres approximates the behaviour of varying $F_{\mathrm{A}}(0)$ in the axial form factor in the Rein-Sehgal model.  Since low \Tmu correlates with low \qq, variations in \Mares have the largest effect on the shape of the muon angular distributions as shown in Fig. \ref{fig:madialeffect}, and have a small effect on the \Tpi spectrum.

\begin{figure}[hbtp]
\centering
\includegraphics[width=0.48\textwidth,trim={3mm 3mm 25mm 5mm}, clip]{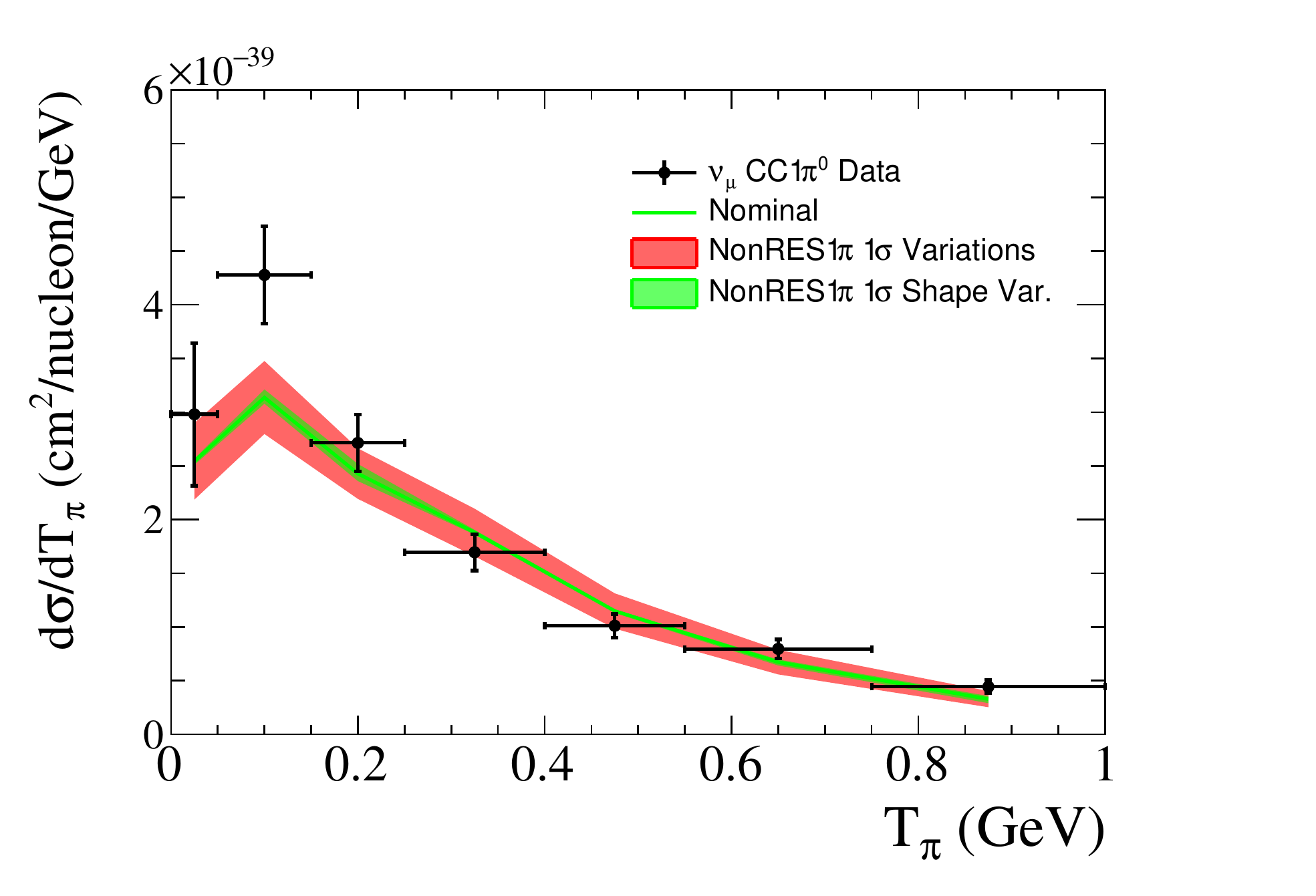}
\includegraphics[width=0.48\textwidth,trim={3mm 3mm 25mm 5mm}, clip]{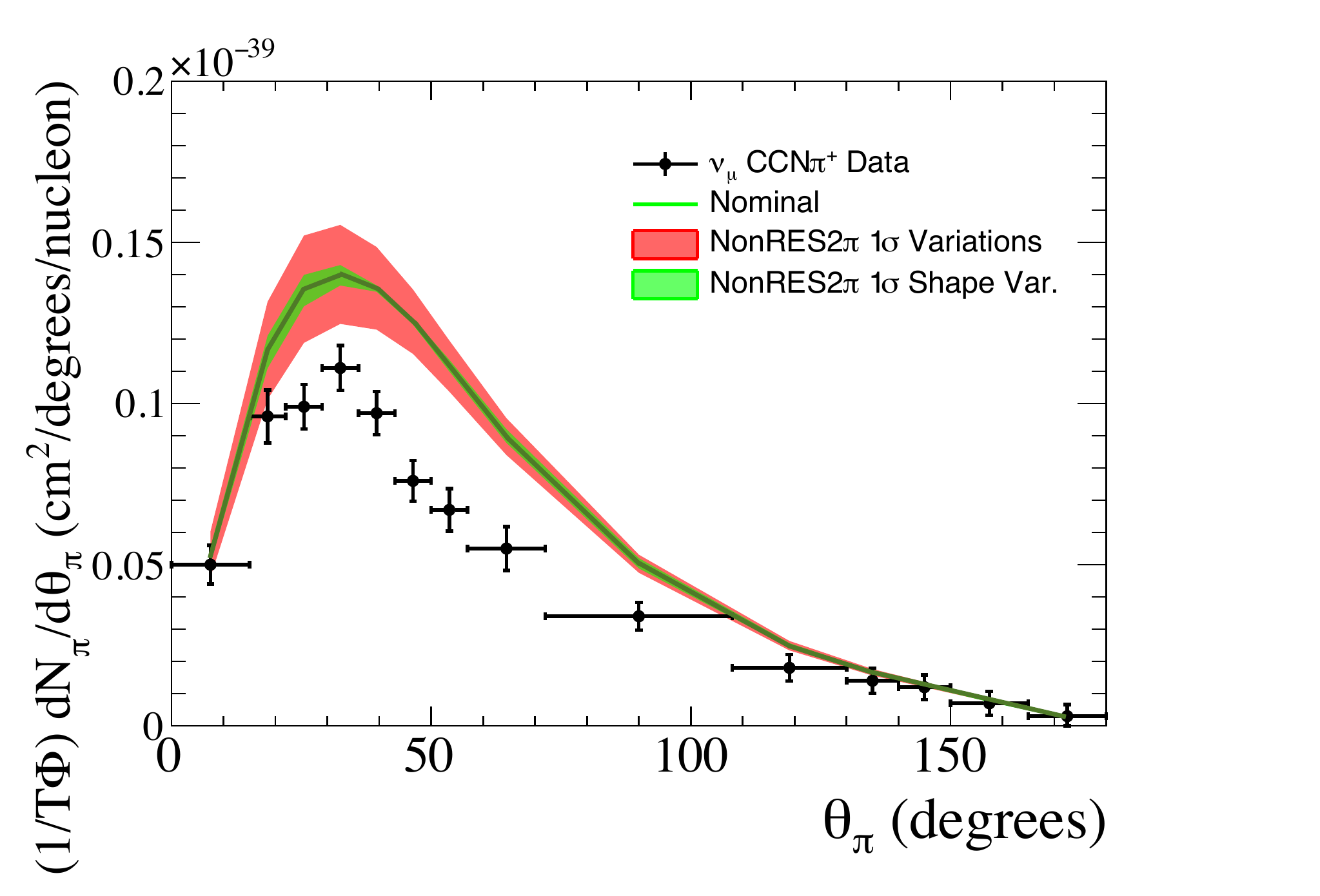}
\caption{The effect of varying the \nonresonepi dial on the default \GENIE prediction for \Kpi (top) and of varying the \nonrestwopi dial on the prediction for \Tpi (bottom).
\label{fig:nonreseffect}}
\end{figure}

Dials are available to vary the normalization of the nonresonant $1\pi$  production channels in \GENIE (\eg  \texttt{NonRESBGvnCC1pi}, \texttt{NonRESBGvpCC1pi}) but each dial introduces similar modifications to the predictions.  To reduce the number of free parameters in the fit described in Section~\ref{sec:genie-tuning}, these dials were grouped into a single background scaling for nonresonant $1\pi$ production, \nonresonepi, following the approach in Ref.~\cite{Rodrigues:2016xjj, Wilkinson:2014yfa}.  A similar treatment was also applied to nonresonant $2\pi$ production, \nonrestwopi, with the neutrino and antineutrino related parameters assumed to be 100\% correlated in both cases. 
The effects of varying the nonresonant contributions are shown in Fig.~\ref{fig:nonreseffect}. Variations in the \nonrestwopi dial introduce a large change in normalization for the \ccnpip channel and has a minor effect in the other single pion channels as the fraction of multi-$\pi$ events is small.

\begin{figure}
\centering
\includegraphics[width=0.4\textwidth,trim={8mm 5mm 8mm 5mm}, clip]{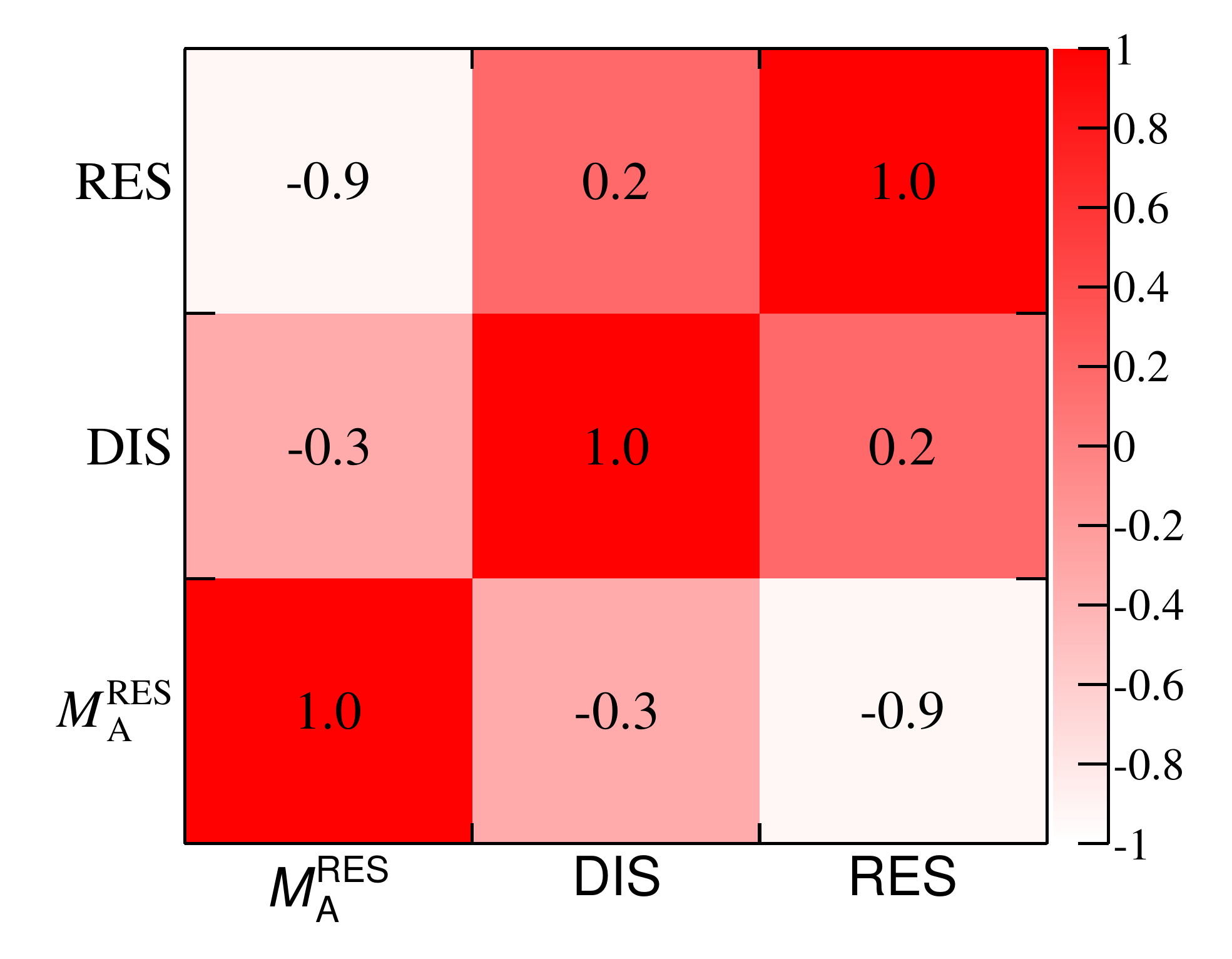}
\caption{Correlation matrix from tuning \GENIE to reproduce the ANL/BNL pion production measurements included in our \chisq penalty term. \label{fig:callumtunecor} }.
\end{figure}

Reanalysis of data from ANL and BNL bubble chambers has provided a tuning of \GENIE's single pion production model on free nucleons. The work showed that a small shift in \Mares was required to model the low-\qq region and a large suppression of the nonresonant $\pi$ production ($-54\%$) was required to match the observed cross sections of \pip and \pin production. The reanalysis used the measured ratios of the rates of single $\pi$ production to CCQE measurements to cancel errors in the flux.  We note that by using CCQE data multiple times, they introduce hidden correlations which may have a small effect on the postfit uncertainties.  However, as the single pion statistical uncertainties at ANL~\cite{ANL_pion} and BNL~\cite{BNL_pion} were magnitudes higher than the CCQE statistical uncertainty~\cite{ANL_CCQE,BNL_CCQE}, the effect was neglected in that work, and is also neglected here.
The resulting parameter tunes shown in Table~\ref{tab:callumtuneres} and Fig.~\ref{fig:callumtunecor} have been partially adopted by \MINERvA and \NOvA which both apply the nonresonant rescaling of 43\% but leave the other parameters unchanged.

\begin{table}[ht]
\centering
{\renewcommand{\arraystretch}{1.2}
\begin{tabular}{c c c }
\hline\hline
Parameter         & \GENIE default & ANL/BNL tune \\
\hline\hline
\Mares [GeV]      & $1.12\pm0.22$ & $0.94\pm0.05$ \\
\Normres [\%]     & $100\pm20$    & $115\pm7$ \\
\nonresonepi [\%] & $100\pm50$    &  $43\pm4$ \\
\hline\hline
\end{tabular}}
\caption{Prefit and best fit central values and uncertainties from tuning \GENIE to the ANL/BNL pion production measurements.  The prefit uncertainties are those recommended by the \GENIE collaboration.  The tuned values are used as penalty terms with supplied covariance matrix of Fig~\ref{fig:callumtunecor}.
\label{tab:callumtuneres}}
\end{table}

\begin{figure*}[hbpt]
\centering
\includegraphics[width=0.24\textwidth]{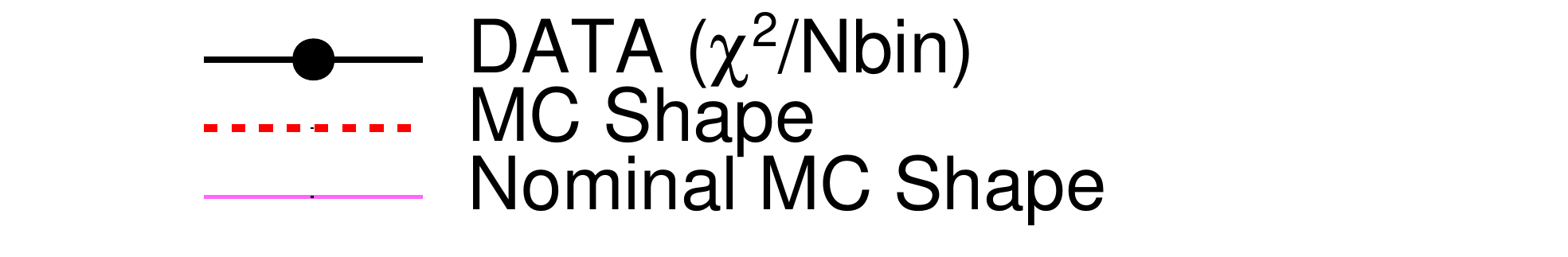}
\includegraphics[width=0.24\textwidth]{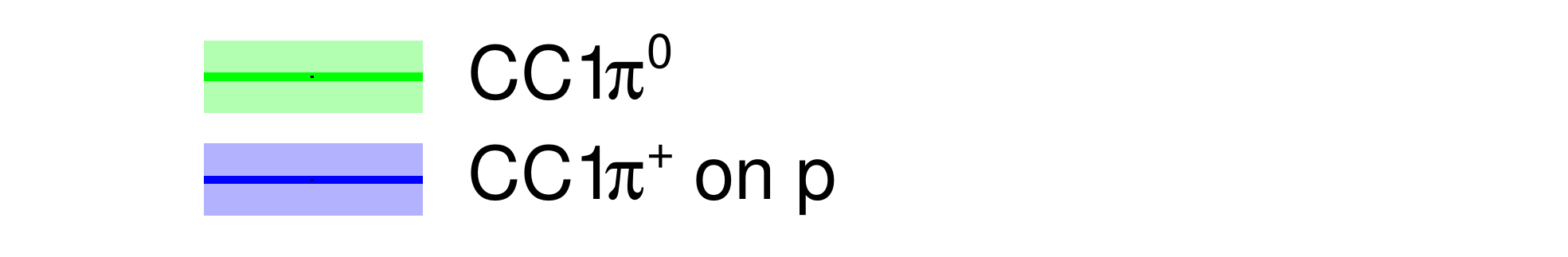}
\includegraphics[width=0.24\textwidth]{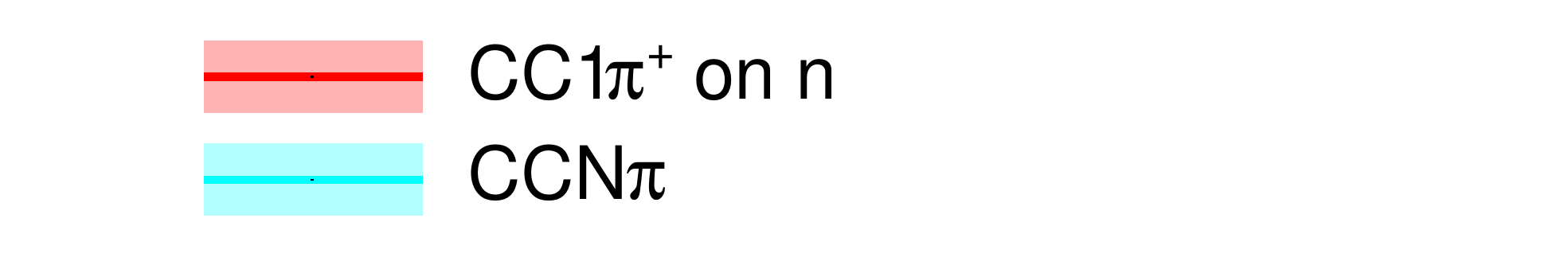}
\includegraphics[width=0.24\textwidth]{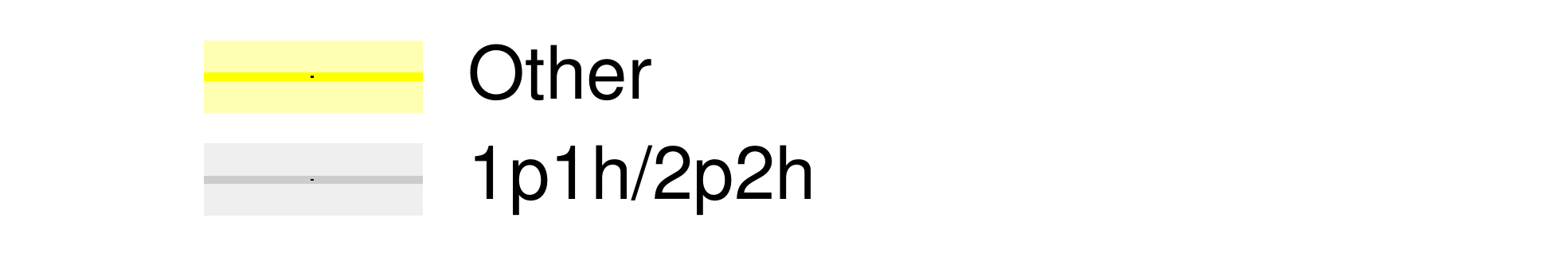}\\
\includegraphics[width=0.24\textwidth,trim={0mm 2mm 10mm 5mm}, clip]{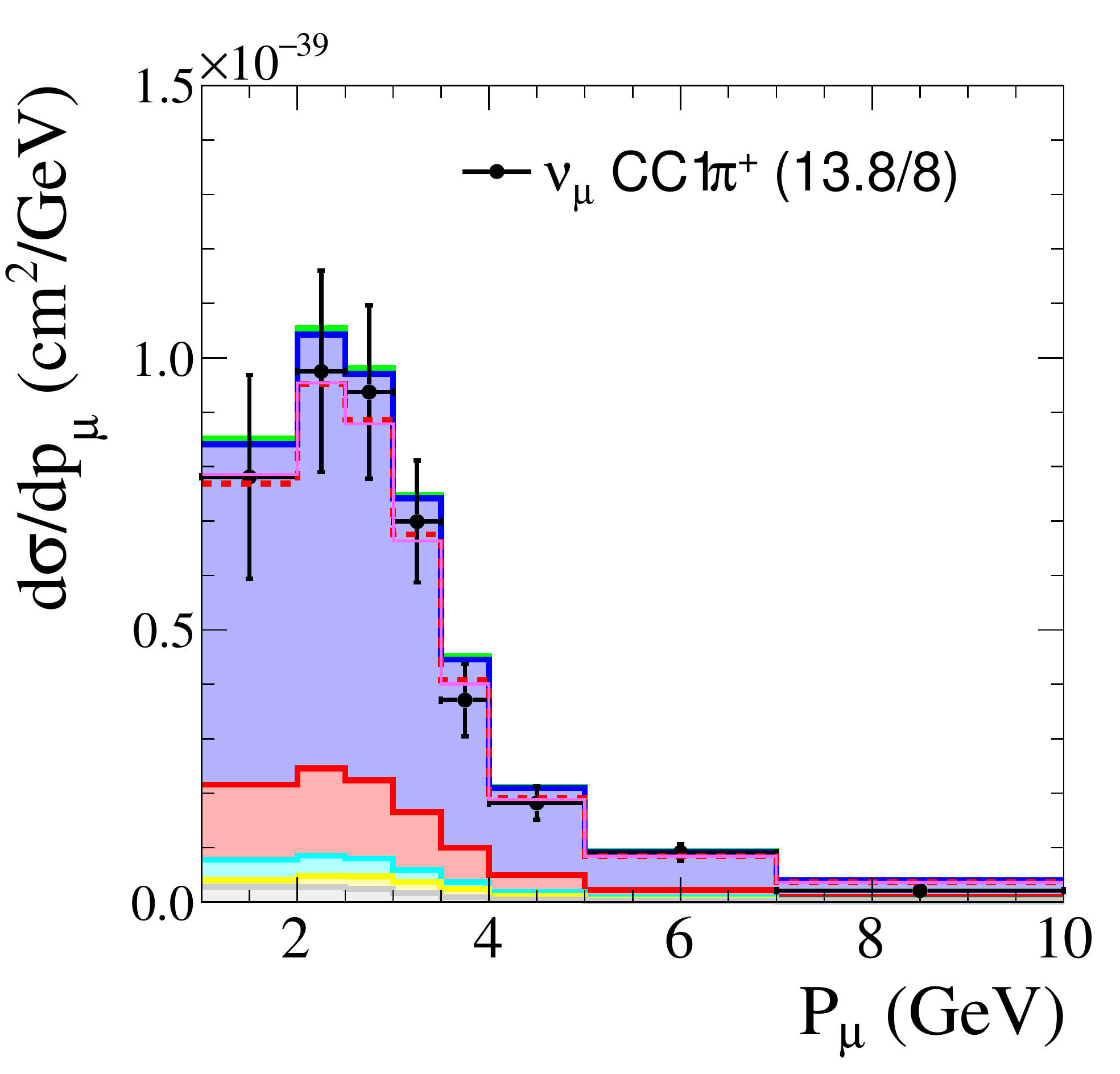}
\includegraphics[width=0.24\textwidth,trim={0mm 2mm 10mm 5mm}, clip]{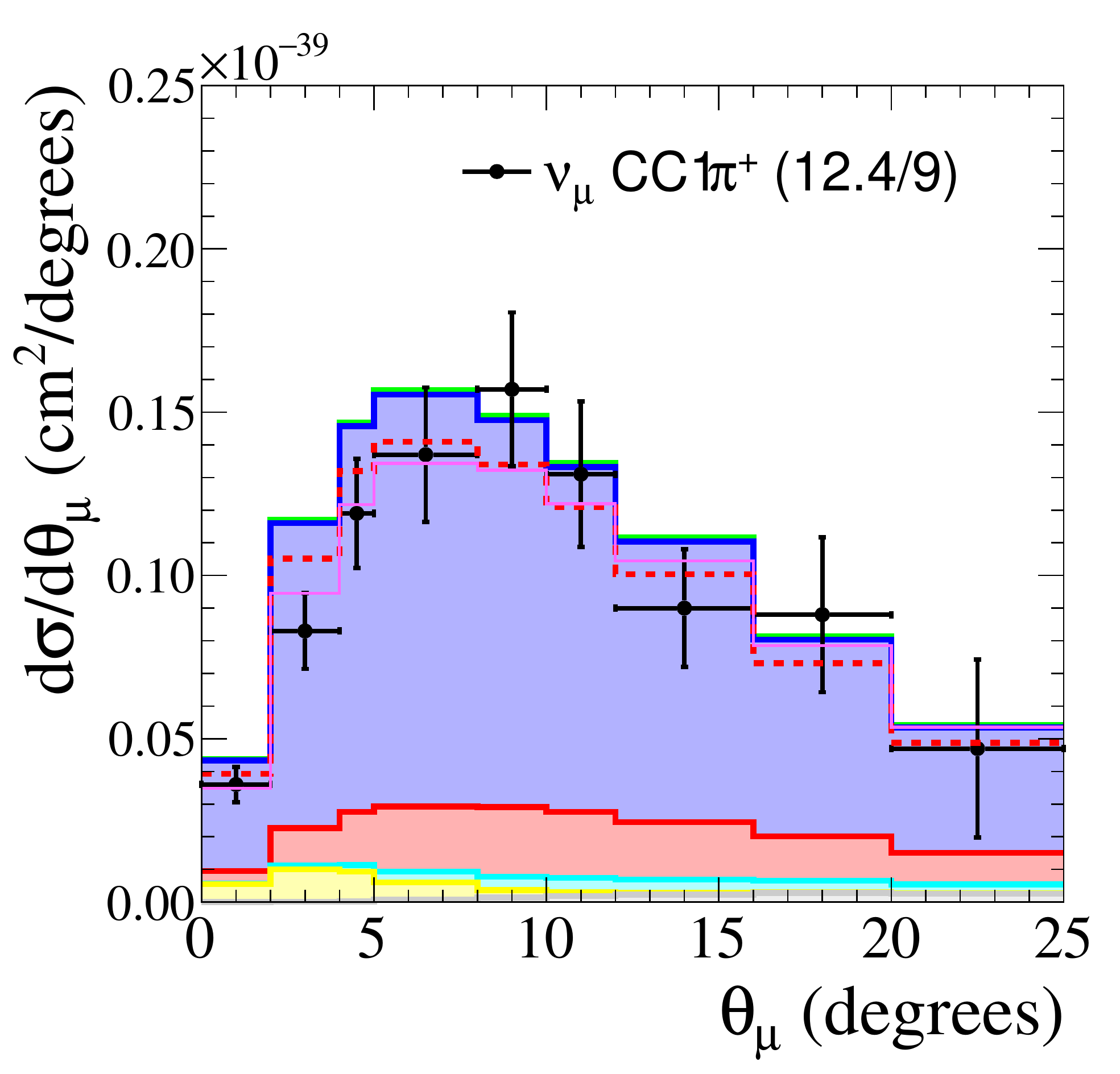}
\includegraphics[width=0.24\textwidth,trim={0mm 2mm 10mm 5mm}, clip]{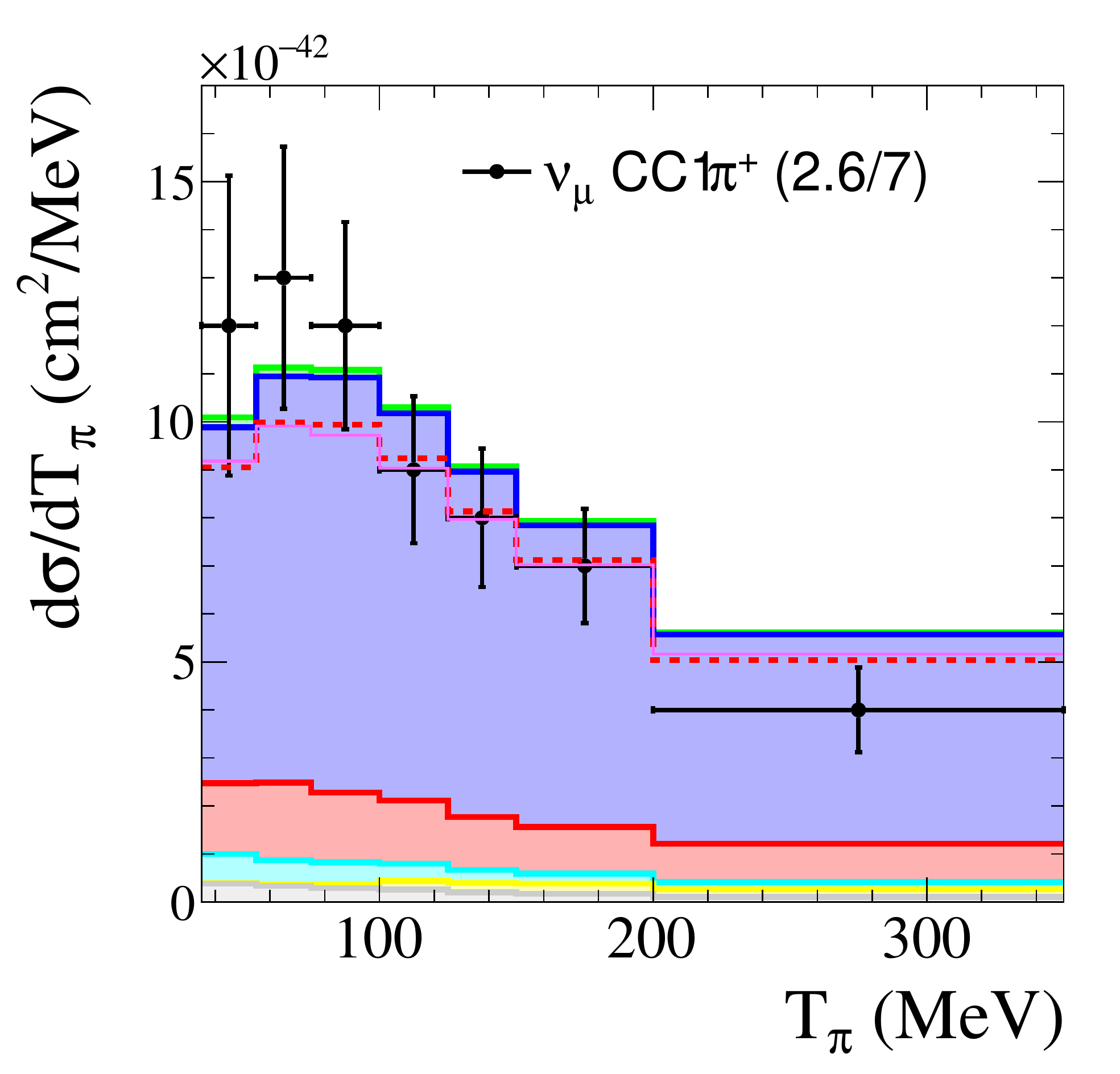}
\includegraphics[width=0.24\textwidth,trim={0mm 2mm 10mm 5mm}, clip]{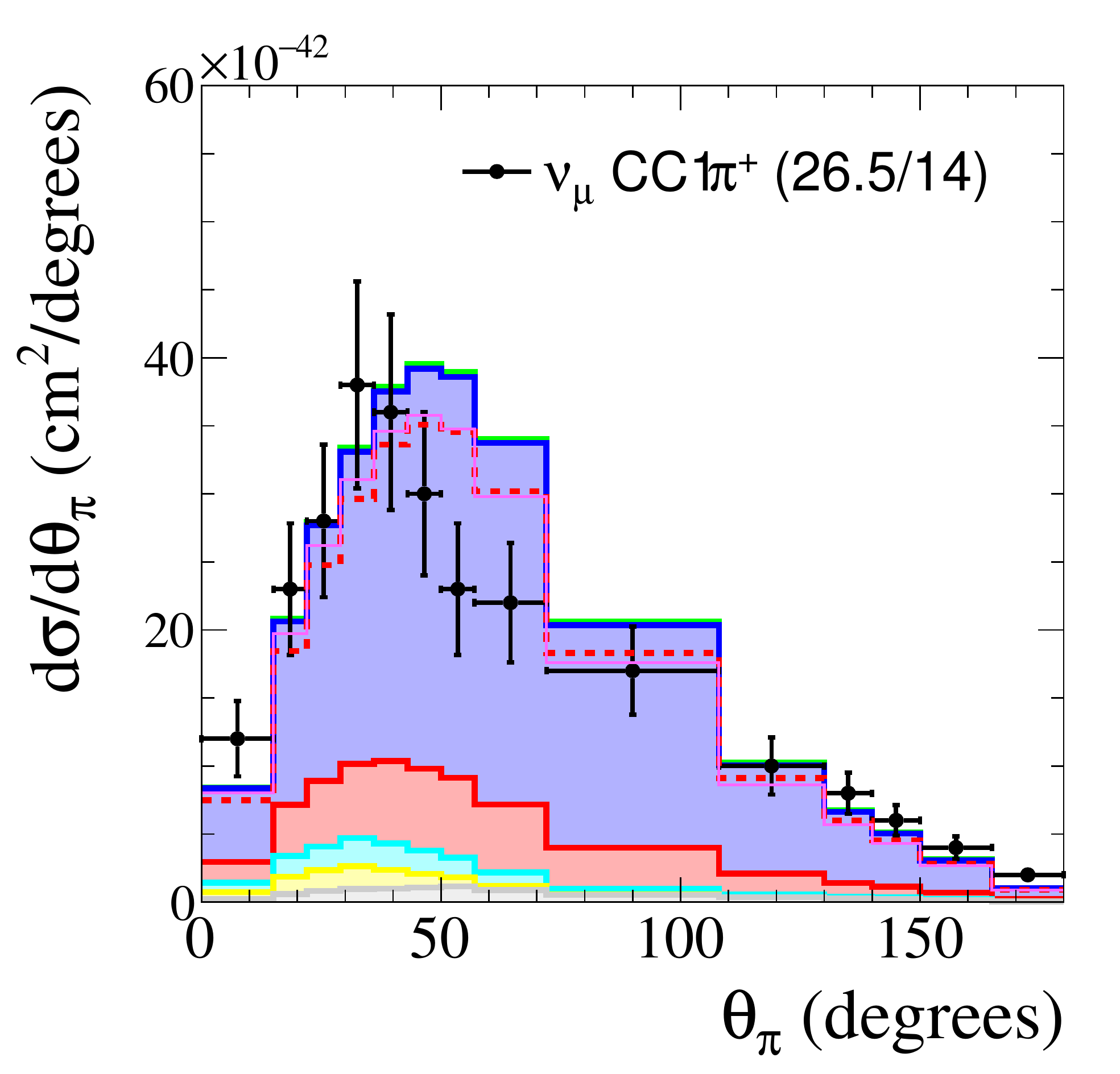} \\
\includegraphics[width=0.24\textwidth,trim={0mm 2mm 10mm 5mm}, clip]{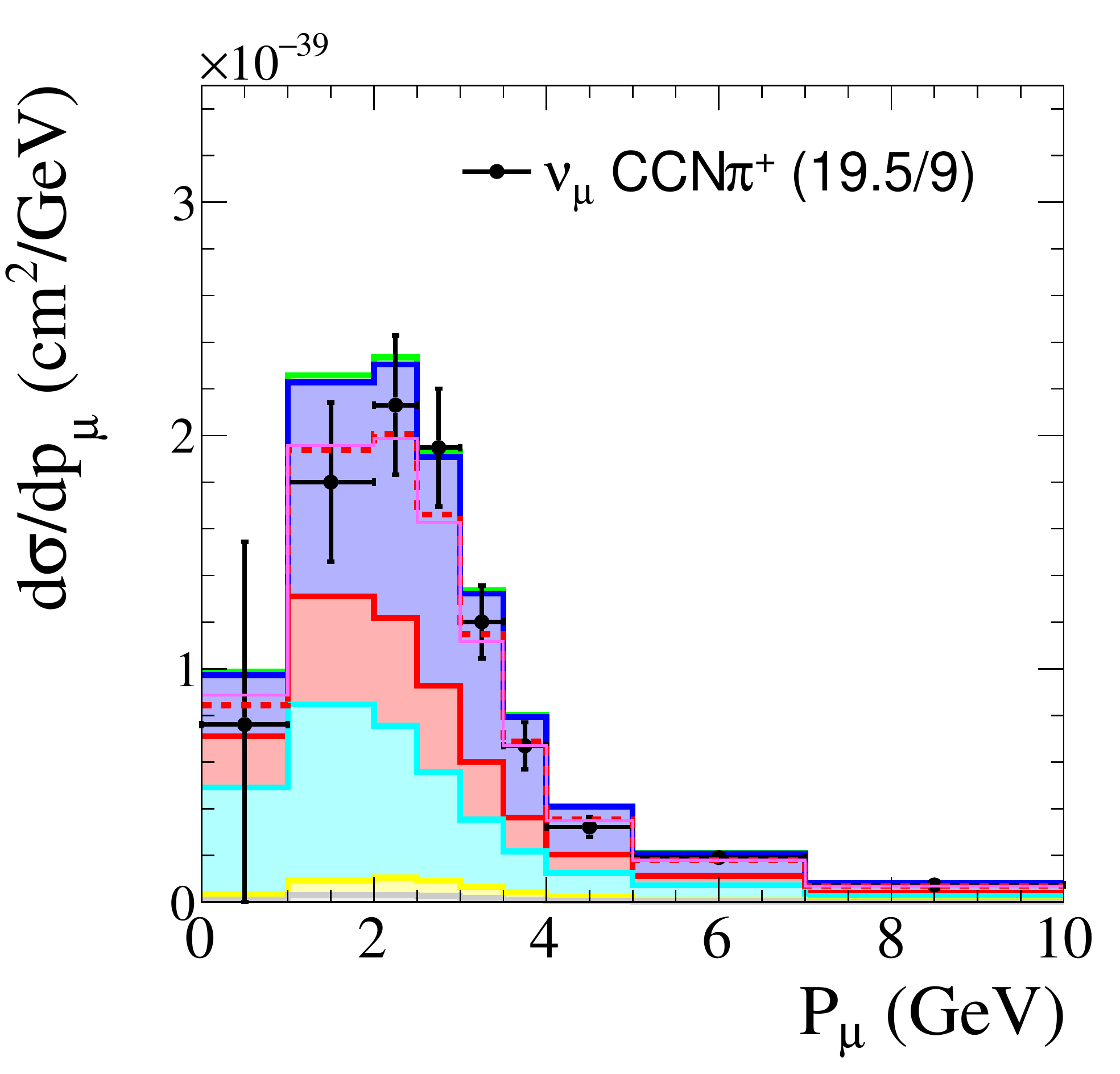}
\includegraphics[width=0.24\textwidth,trim={0mm 2mm 10mm 5mm}, clip]{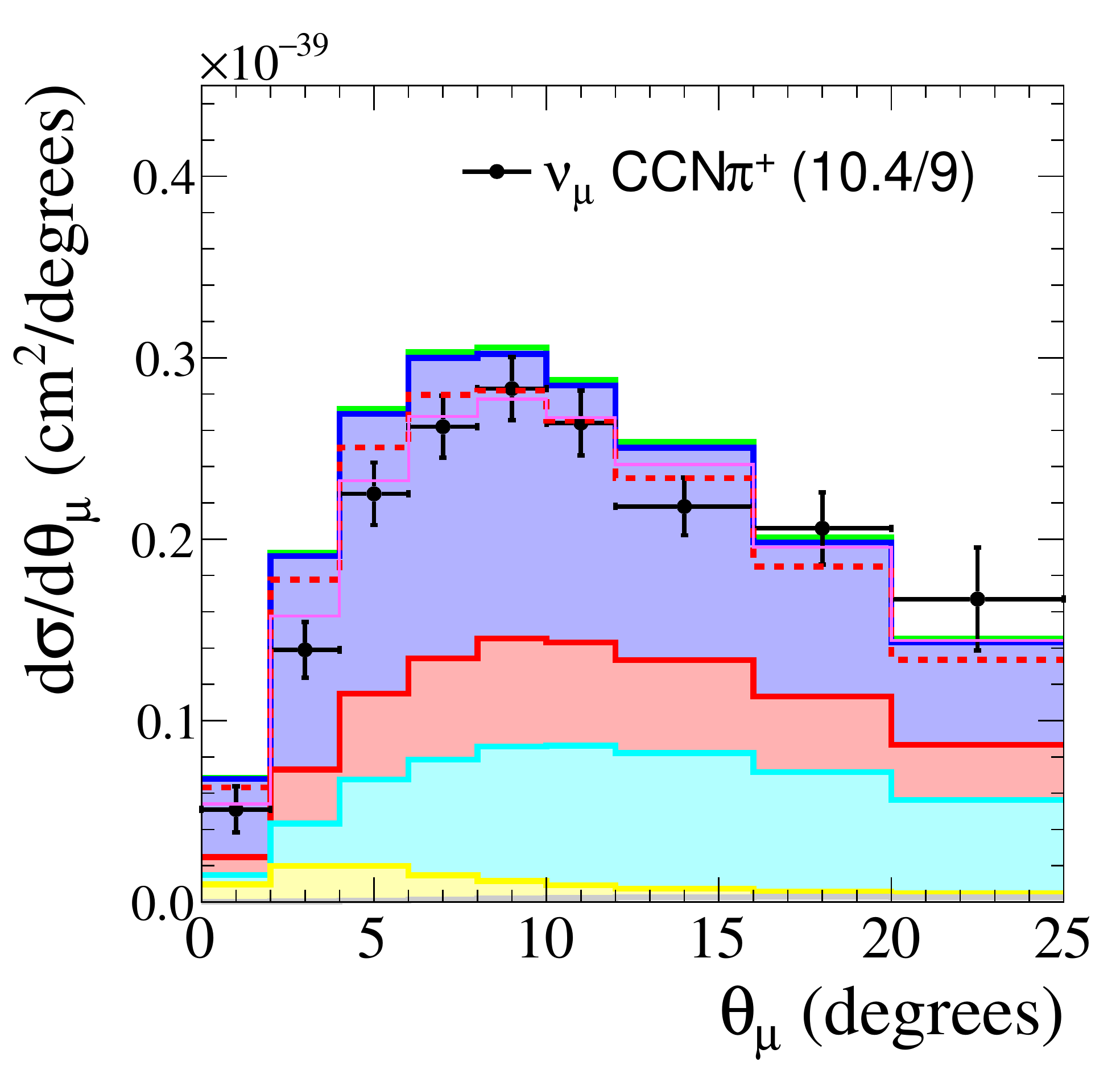}
\includegraphics[width=0.24\textwidth,trim={0mm 2mm 10mm 5mm}, clip]{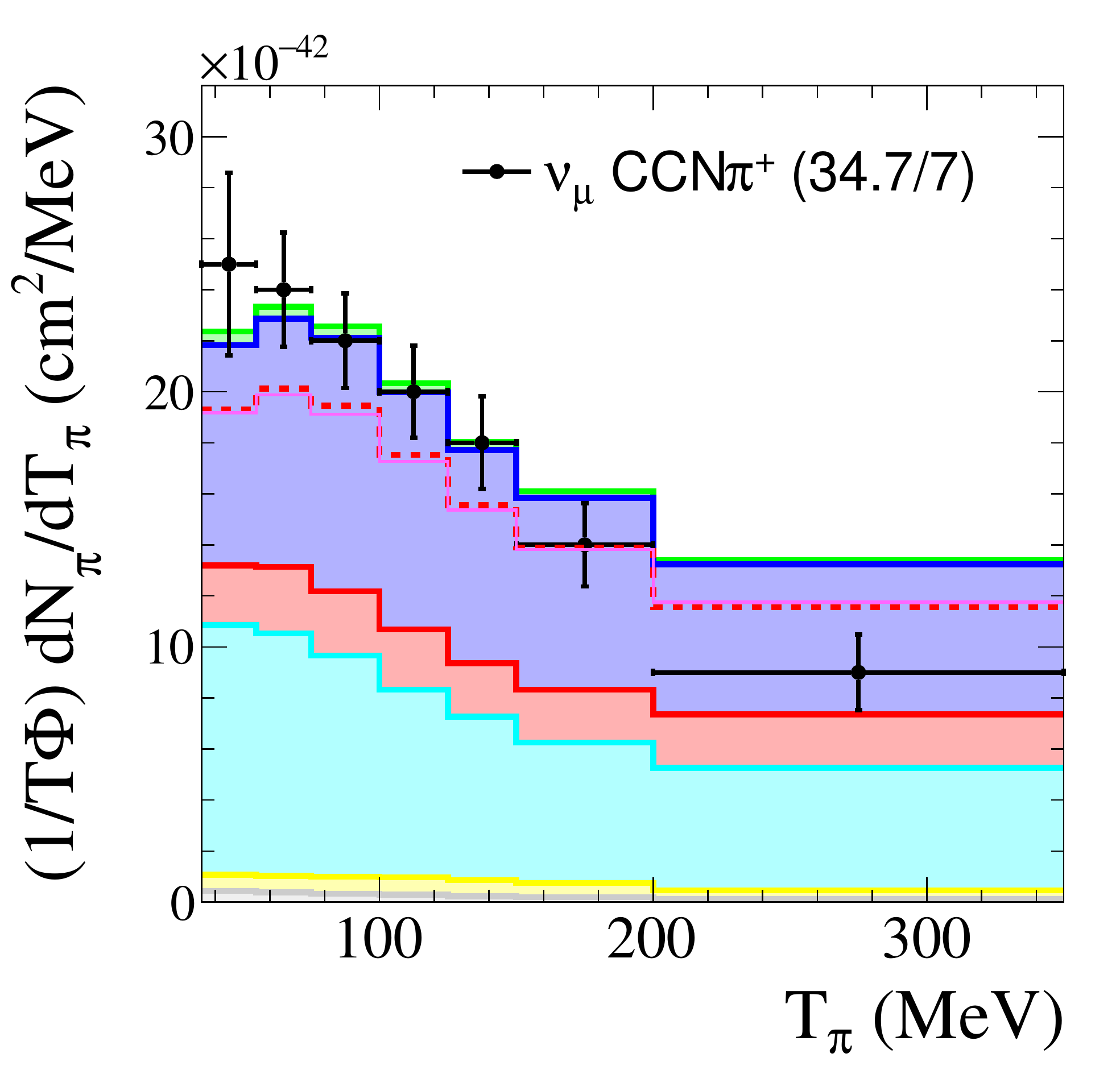}
\includegraphics[width=0.24\textwidth,trim={0mm 2mm 10mm 5mm}, clip]{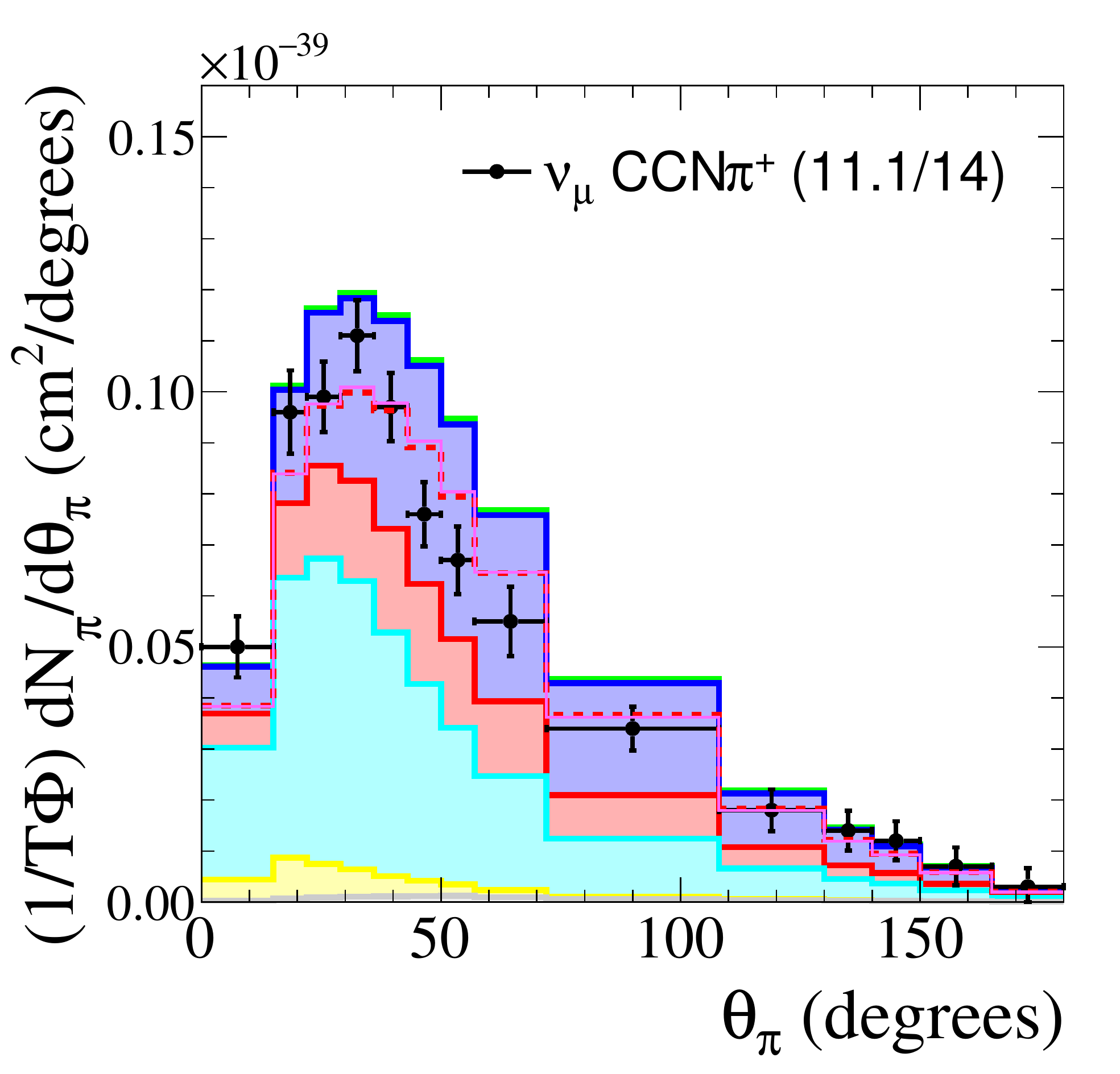} \\
\includegraphics[width=0.24\textwidth,trim={0mm 2mm 10mm 5mm}, clip]{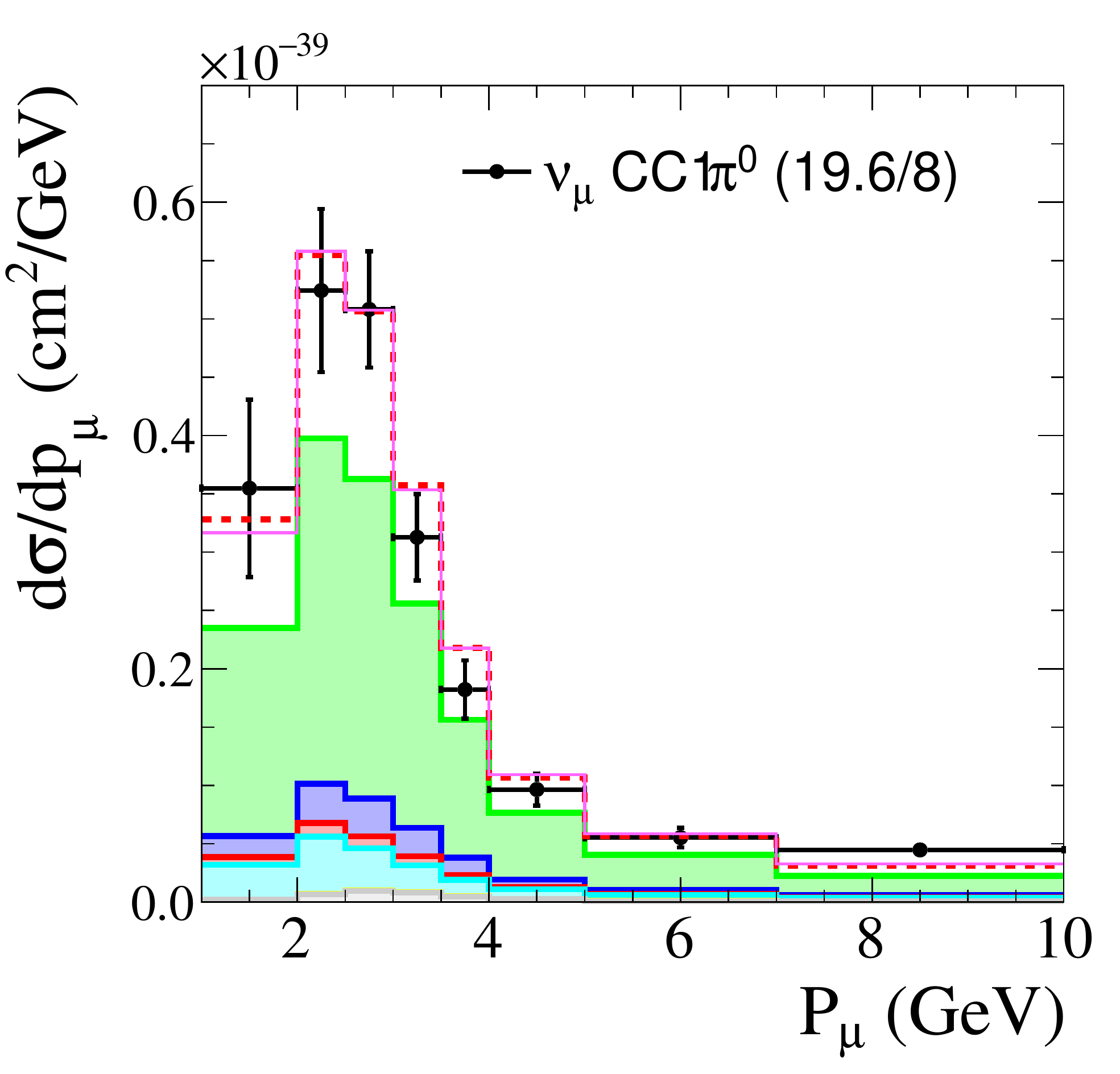}
\includegraphics[width=0.24\textwidth,trim={0mm 2mm 10mm 5mm}, clip]{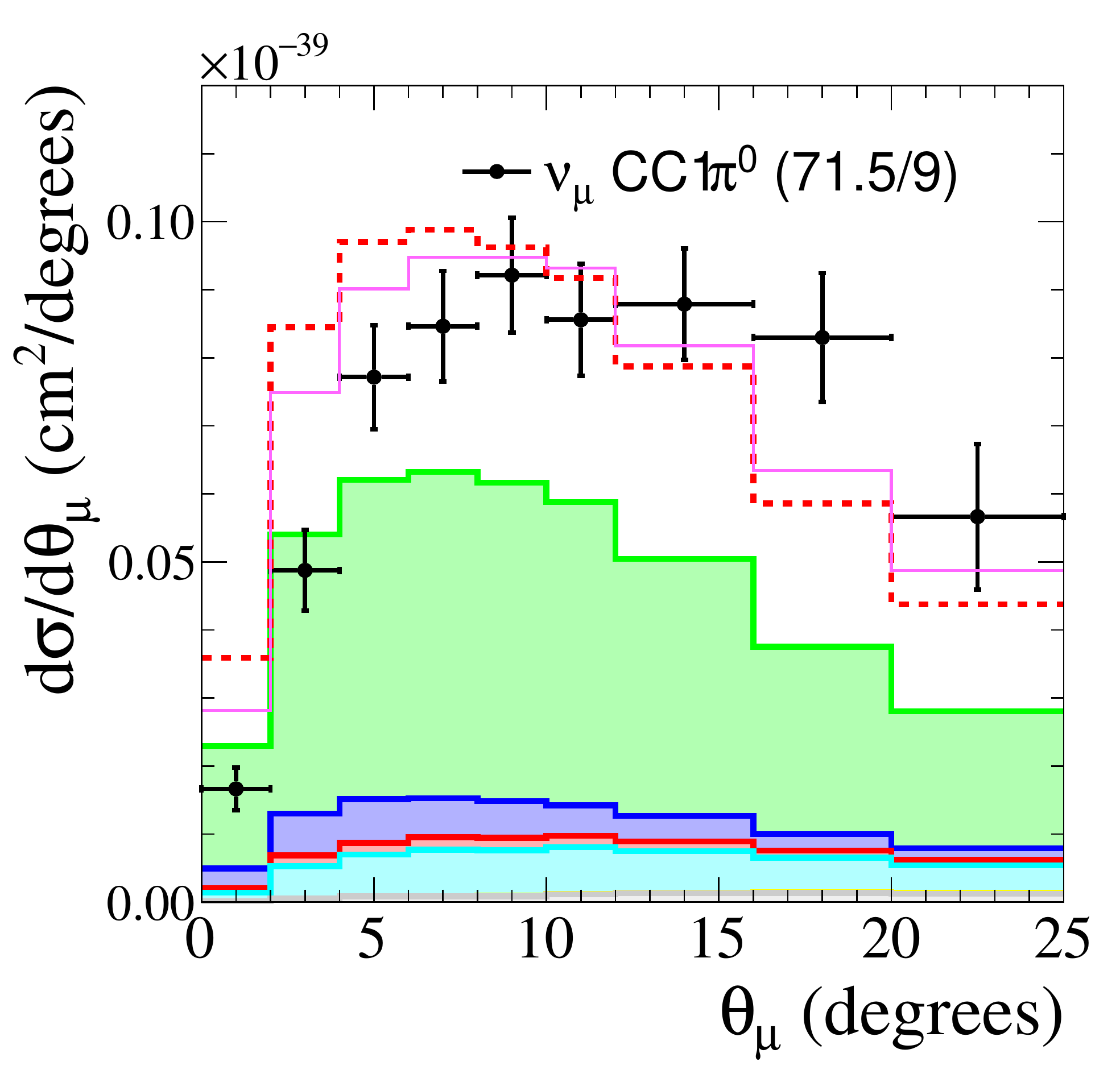}
\includegraphics[width=0.24\textwidth,trim={0mm 2mm 10mm 5mm}, clip]{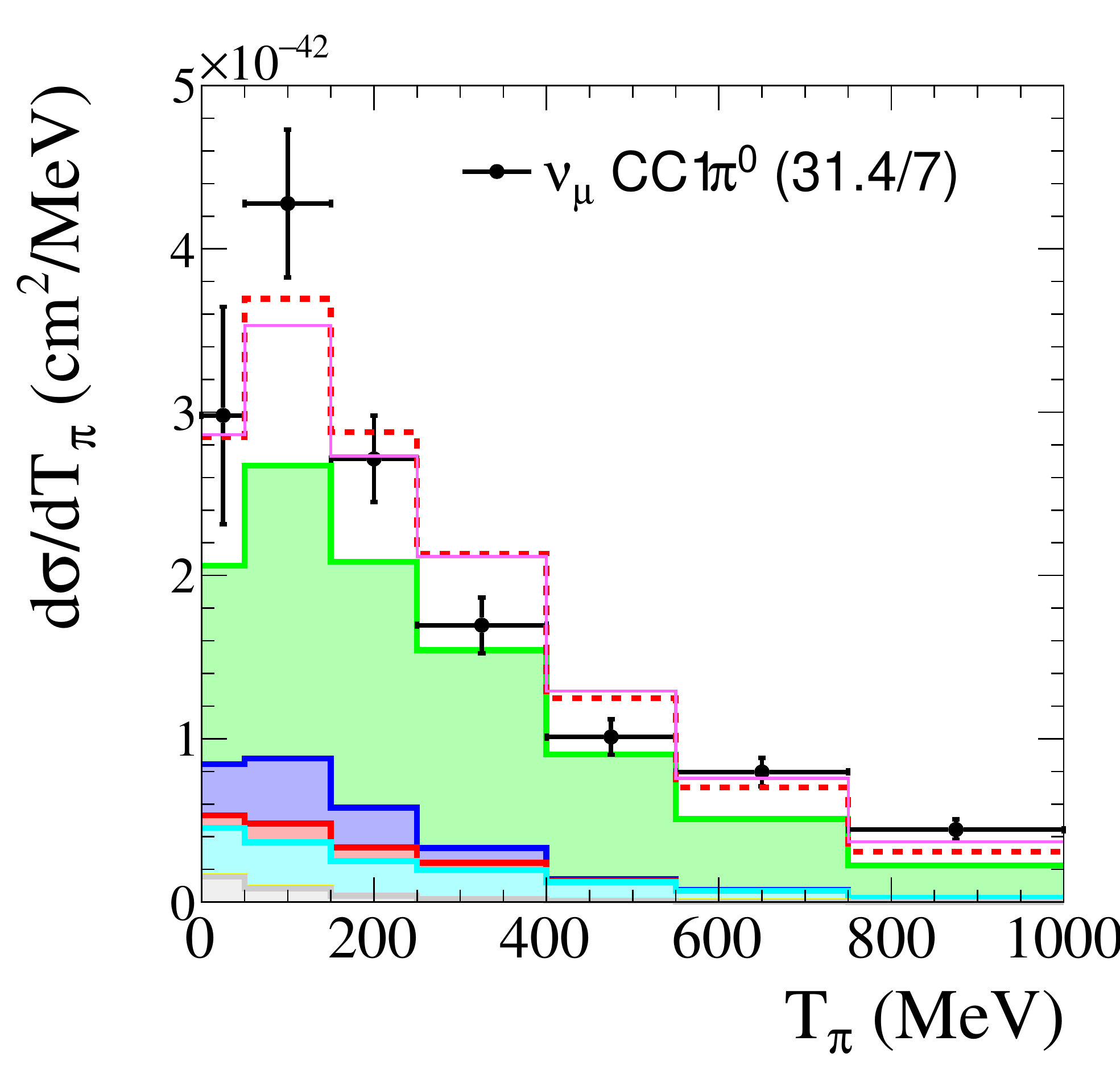}
\includegraphics[width=0.24\textwidth,trim={0mm 2mm 10mm 5mm}, clip]{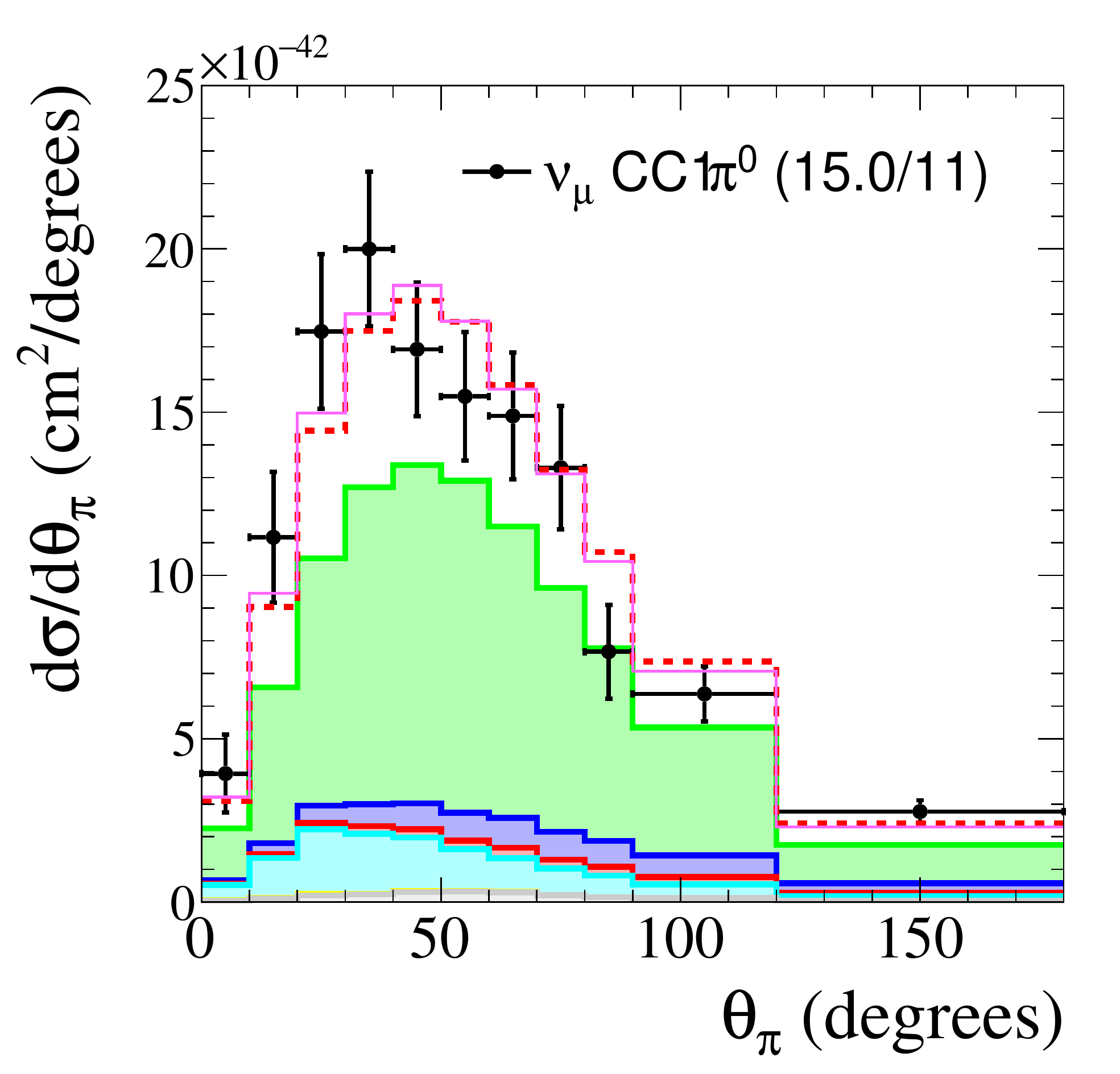} \\
\includegraphics[width=0.24\textwidth,trim={0mm 2mm 10mm 5mm}, clip]{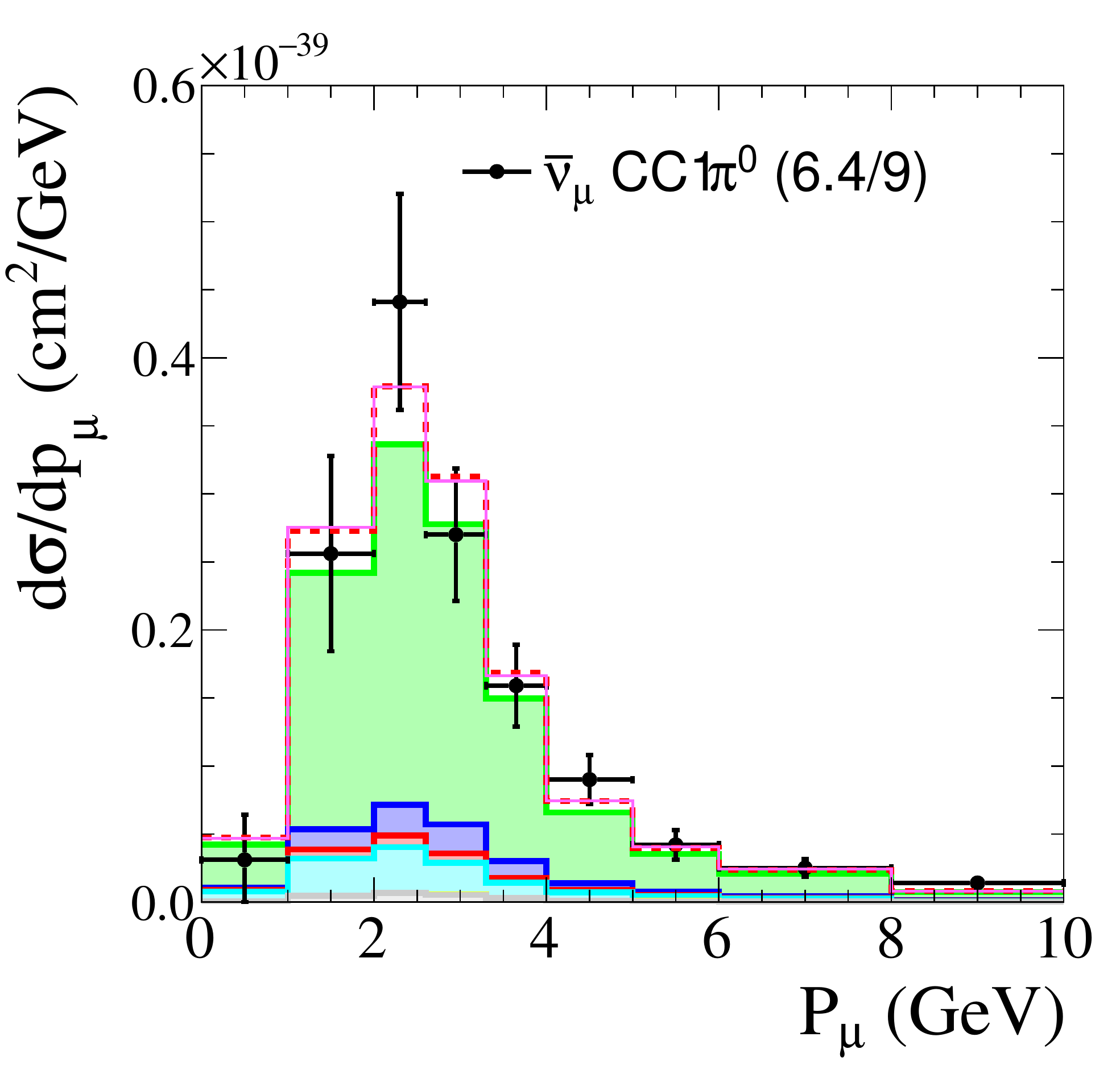}
\includegraphics[width=0.24\textwidth,trim={0mm 2mm 10mm 5mm}, clip]{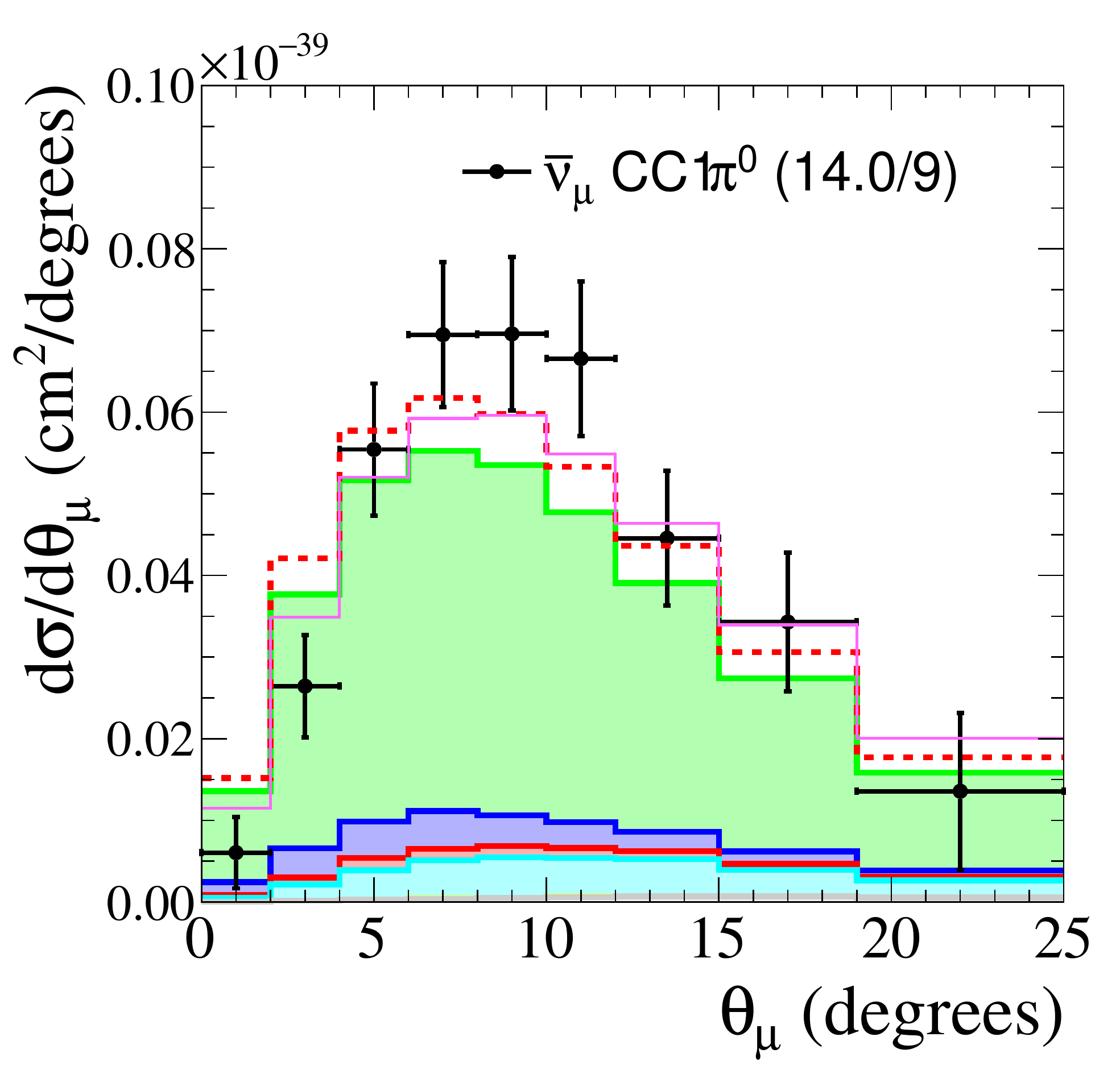}
\includegraphics[width=0.24\textwidth,trim={0mm 2mm 10mm 5mm}, clip]{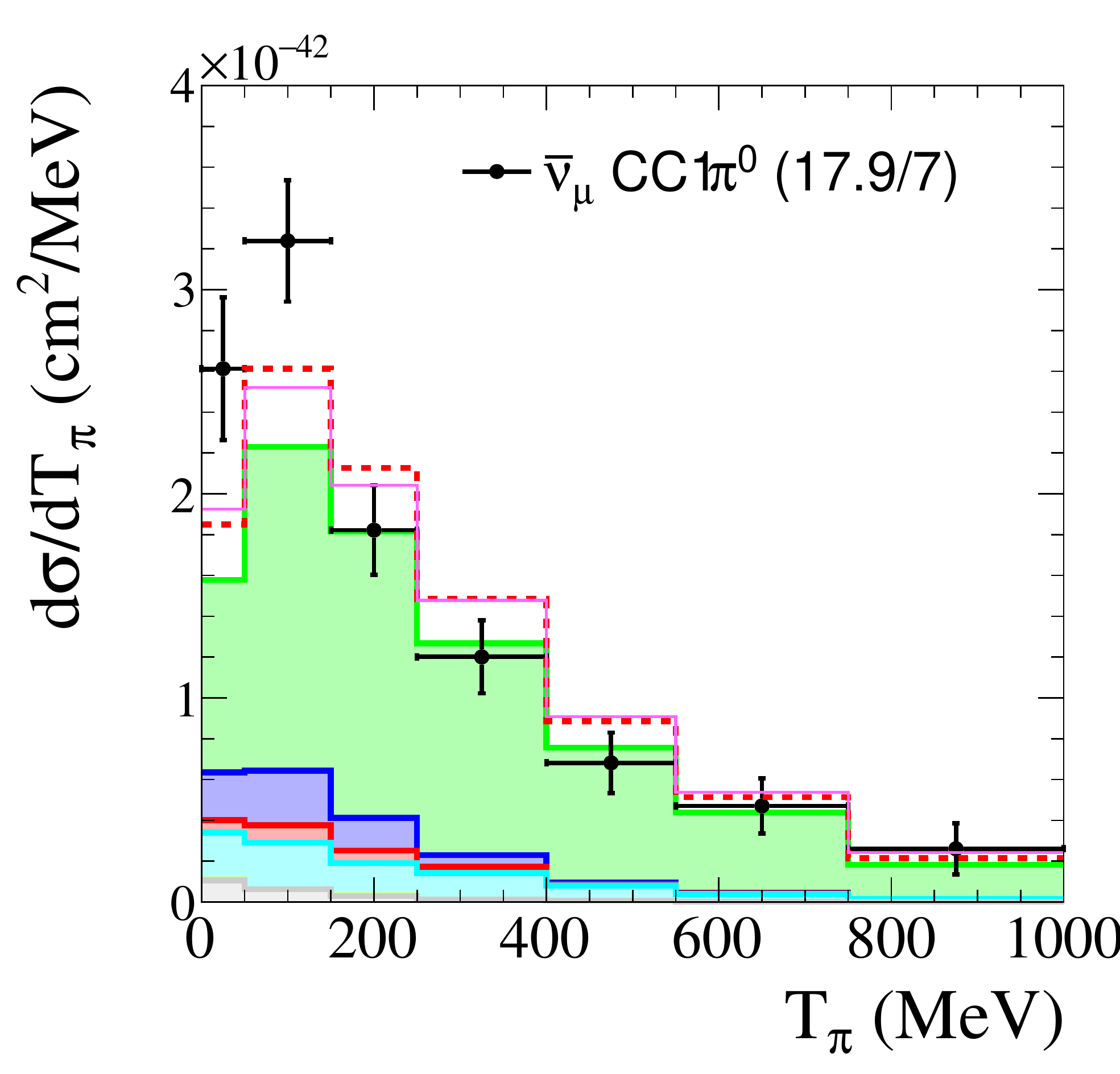}
\includegraphics[width=0.24\textwidth,trim={0mm 2mm 10mm 5mm}, clip]{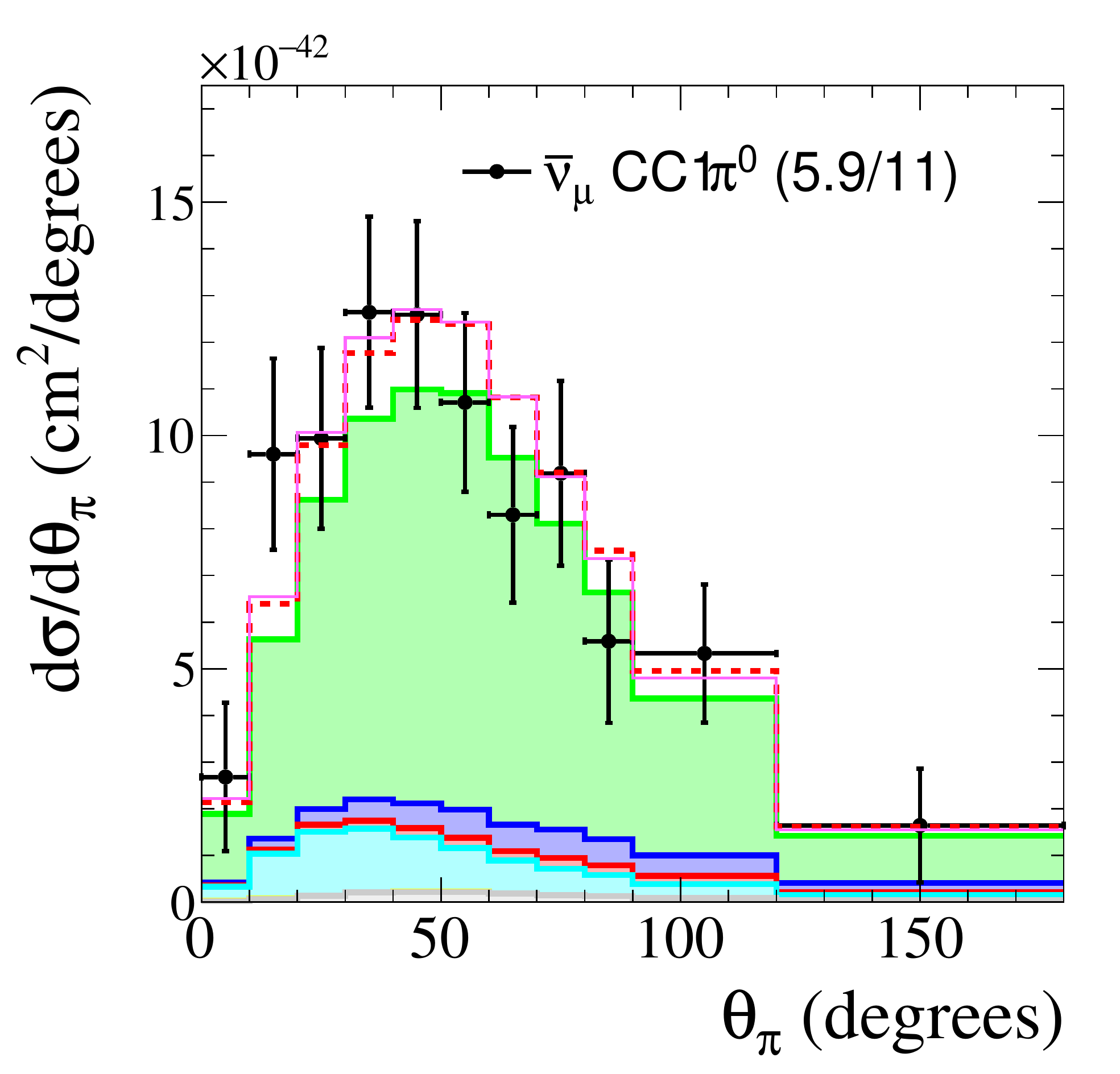}
\caption{\GENIE ANL/BNL single pion tuning model predictions compared to \MINERvA data. The distributions have been weighted to the ANL/BNL tuning parameter set, and have had the coherent pion correction applied.  Colors correspond to particle content at the nucleon interaction.  ``Other'' is dominated by coherent pion production.  ``MC Shape'' shows the total MC prediction after it has been normalized to match the total data normalization.  In the case of the shape-only distributions (\Tmu, \Kpi, \Tpi) the shape-only \chisq/\nbins values are shown.  All cross sections are per nucleon.
\label{fig:geniebctunemodes}}
\end{figure*}

Fig.~\ref{fig:geniebctunemodes} shows \MINERvA data and the predictions of \GENIE when its output has been reweighted to reflect the parameter changes of Table~\ref{tab:callumtuneres}.  The channel-by-channel contributions to the \chisq are given in the fifth column (``ANL/BNL'') of Table~\ref{tab:genienomchi2}. Incorporating the parameter changes improves the total normalization agreement in the \Pmu distributions for \ccpip and \ccnpip.  The \chisq for the \Pmu distribution is also improved in the \ccapin channel, even though the ANL/BNL data is from neutrino interactions only.  The \chisq for the \Pmu distribution in the \ccpin channel is somewhat worse as the parameter tunes reduce the predicted nucleon \ccpin cross section.  The modification of \Mares shifts the \Tmu predictions to lower values, increasing the \chisq contributions. The \Kpi and \Tpi distributions change mostly by normalization, having a smaller effect on the \chisq.  The overall agreement of \GENIE with \MINERvA data is not improved by incorporating the ANL/BNL information.  Indeed, the total \chisq increases, largely because of the \chisq contributions from the \Tmu distributions.

\begin{figure}
\centering
\includegraphics[width=0.48\textwidth,trim={3mm 5mm 30mm 5mm}, clip]{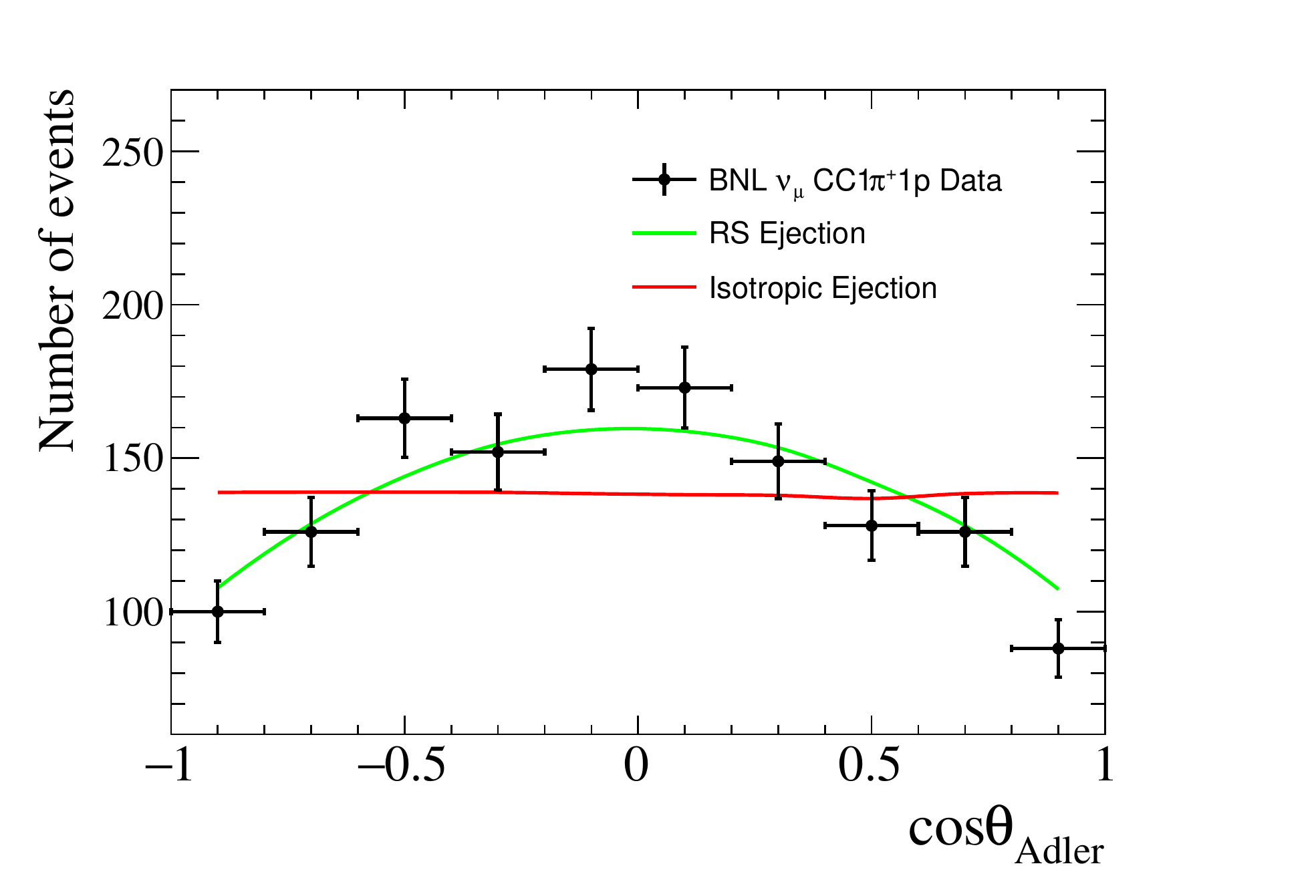}
\includegraphics[width=0.48\textwidth,trim={3mm 5mm 30mm 5mm}, clip]{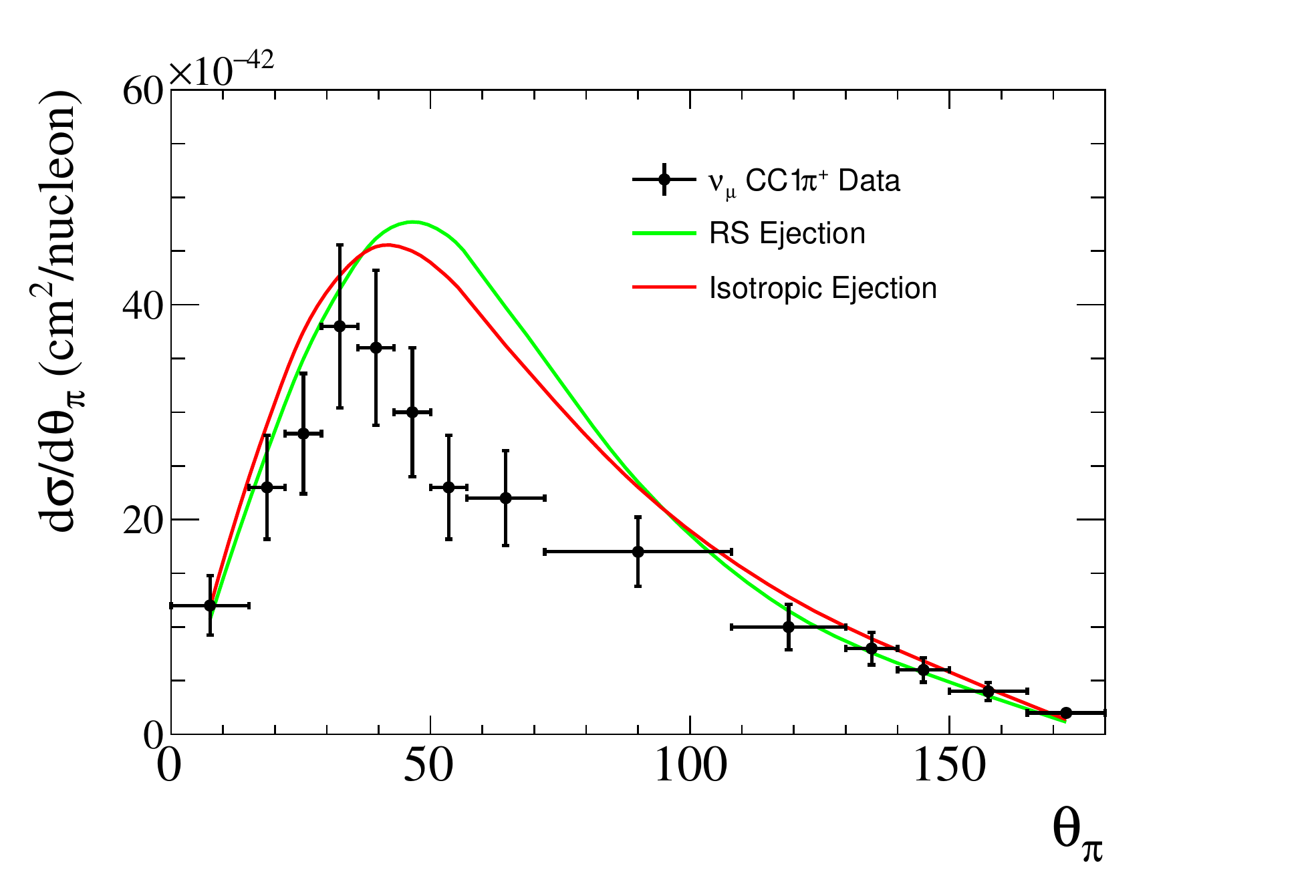}
\caption{Effect of varying the \ThetaPi dial on pion angular distributions for BNL $\nu_\mu$CC1$\pi^+$1p data (top) and \MINERvA $\nu_\mu$CC1$\pi^+$ (bottom).
\label{fig:geniethetavariations}}
\end{figure}

GENIE provides a dial that influences the resonances' decay into the pion-nucleon system in the resonance rest frame, \ThetaPi, and allows events to be reweighted continuously between the default anisotropic distribution (\ThetaPi = 0) and the isotropic distribution (\ThetaPi = 1). The Adler angle\footnote{The Adler angle is the angle between the pion and the three momentum transfer in the resonance rest frame~\cite{ADLER1968189}.} is highly sensitive to the \ThetaPi parameter and has been measured by neutrino induced pion production experiments on single nucleons, such as ANL~\cite{ANL_pion}, BNL~\cite{BNL_pion}, BEBC~\cite{BEBC_pion1, BEBC_pion2} and FNAL~\cite{FNAL_pion}.  Nucleon data strongly prefers an anisotropic process, as shown in Fig.~\ref{fig:geniethetavariations}. Nonetheless, \ThetaPi has some impact, albeit one that does depend on how FSI are modelled, on the shape of \MINERvA \Tpi and \Kpi distributions, seen in the bottom of Fig.~\ref{fig:geniethetavariations}, and was therefore included in this work.

The \GENIE hA model for FSI has uncertainties from the $\pi-A$ cross-section data to which the model was tuned.  The total $\pi-A$ cross section has a stronger constraint than each of the individual interaction cross sections, so \GENIE provides dials to vary the fractional contribution of each component.  The available fractional dials are pion absorption (FrAbs), pion inelastic scattering (FrInel), pion elastic scattering, pion charge exchange and pion production.

\section{Tuning the \GENIE Model}
\label{sec:genie-tuning}
Fig.~\ref{fig:geniebctunemodes} and Table~\ref{tab:genienomchi2} show the unsatisfactory agreement of the \GENIE prediction against \MINERvA data.  The disagreement worsens after incorporating the prior constraint from ANL and BNL data; this correction, based on nucleon data, is inadequate.  This section describes fits that improve the agreement with \MINERvA data.  The parameters \Mares, \Normres and \nonresonepi are included in the fits with a penalty term added to the \chisq from the ANL and BNL data. The penalty term uses the covariance, $M$, shown in Fig.~\ref{fig:callumtunecor}:

\begin{align}
\chisqpen = \sum_{i,j}^{N=3} (x_{i} - f_{i}) \left( M^{-1} \right)_{ij} (x_{j} - f_{j}),
\end{align}

\noindent where $x_{i}$ are the parameter values $i$ at each iteration of the fit, and $f_{i}$ are the parameter values from the fit to ANL and BNL data.  The \GENIE default model is strongly disfavoured with \chisqpen = 299.3, but changing \nonresonepi to 43\% while leaving all the other parameters at their default values reduces the \chisqpen  to 21.8, showing that the largest tension is due to the \nonresonepi parameter.

The \ThetaPi dial is allowed to vary in the range 0--1 in the fit, corresponding to a continuous variation between an RS angular distribution and an isotropic distribution for $\Delta(1232)$ decay.  To avoid the normalization of the \ccnpip measurement pulling parameters in the \ccpip model, the \nonrestwopi dial was allowed to vary between 0-300\% of the nominal value.

When varying one of the five hA pion FSI dials, \GENIE automatically adjusts the remaining parameters to preserve the total pion cross-section and maintain agreement with pion-nucleus scattering data.  This ``cushion'' technique introduced instabilities in the \chisq surface, so it was not possible to include multiple pion FSI parameters in a simultaneous fit. Instead we performed fits with only one of the FSI parameters floating.  No \chisq penalty terms were added for the FSI dial in either tuning: the parameters were driven solely by \MINERvA data.  The charge exchange and pion production dials had small contributions to the overall $\chi^{2}$ for the selected data, forcing the parameters to be inflated beyond +3$\sigma$ of GENIE's recommendation, with large post-fit uncertainties.  Furthermore, the pion elastic scattering parameter is strongly constrained by external data, so its 1$\sigma$ variation has a small impact on the \MINERvA distributions.  The non-FSI fit parameters' (e.g. \Mares) central values and uncertainties all agreed for the five fits. Here we present the results from the FrAbs and FrInel fits.

\begin{figure*}[hbtp]
\centering
\includegraphics[width=0.48\textwidth,trim={12mm 2mm 10mm 8mm}, clip] {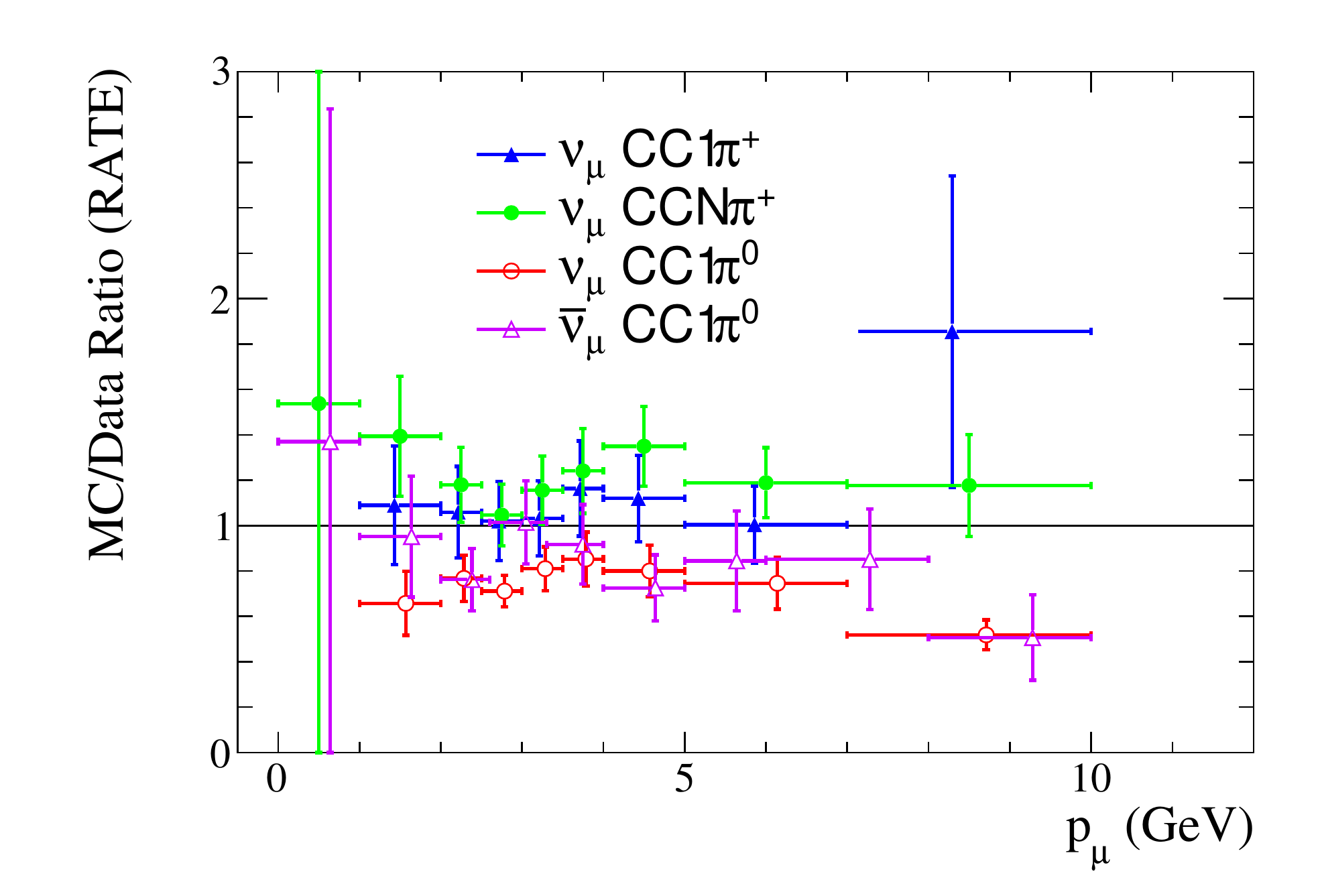}
\includegraphics[width=0.48\textwidth,trim={12mm 2mm 10mm 8mm}, clip] {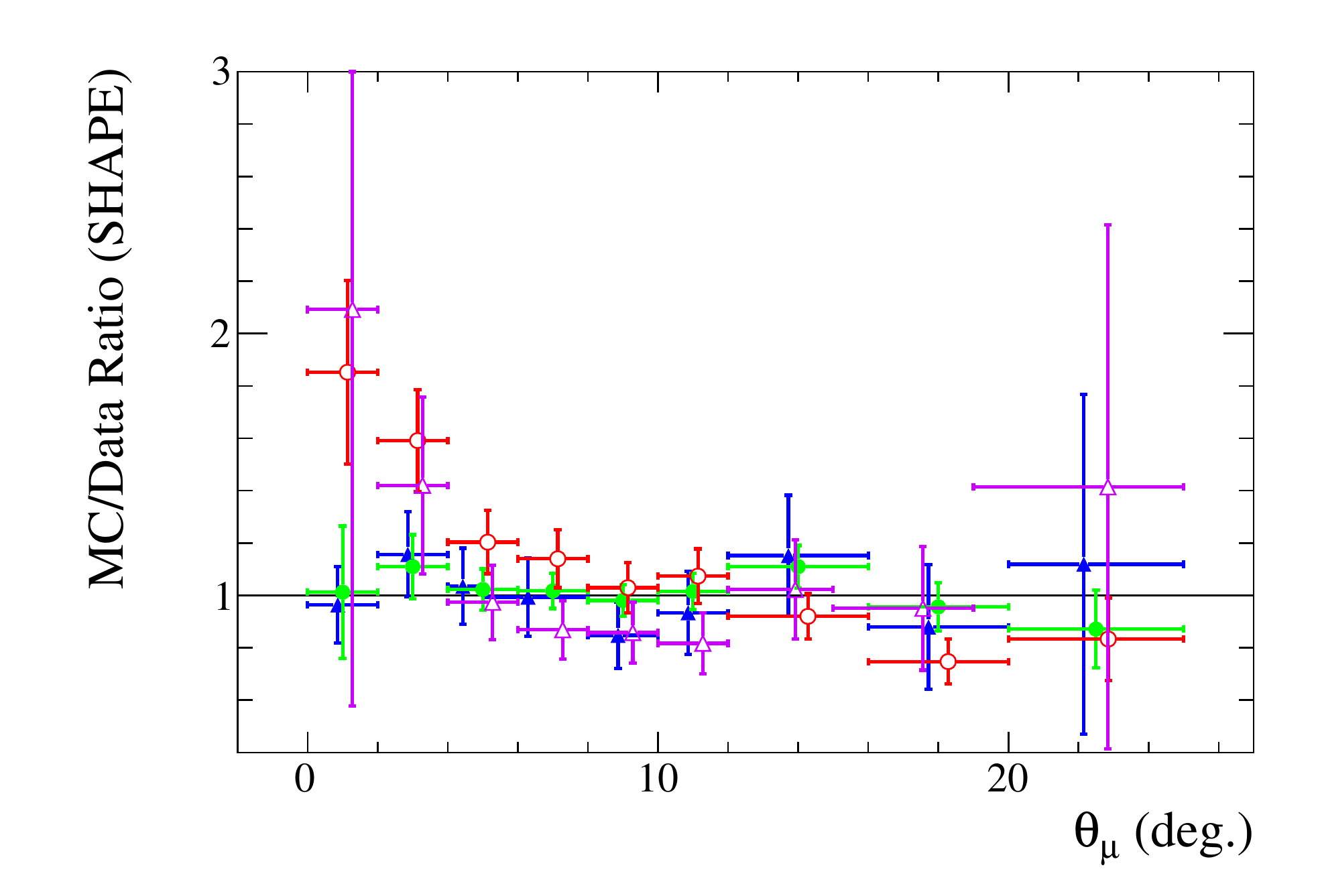}
\includegraphics[width=0.48\textwidth,trim={12mm 2mm 10mm 8mm}, clip] {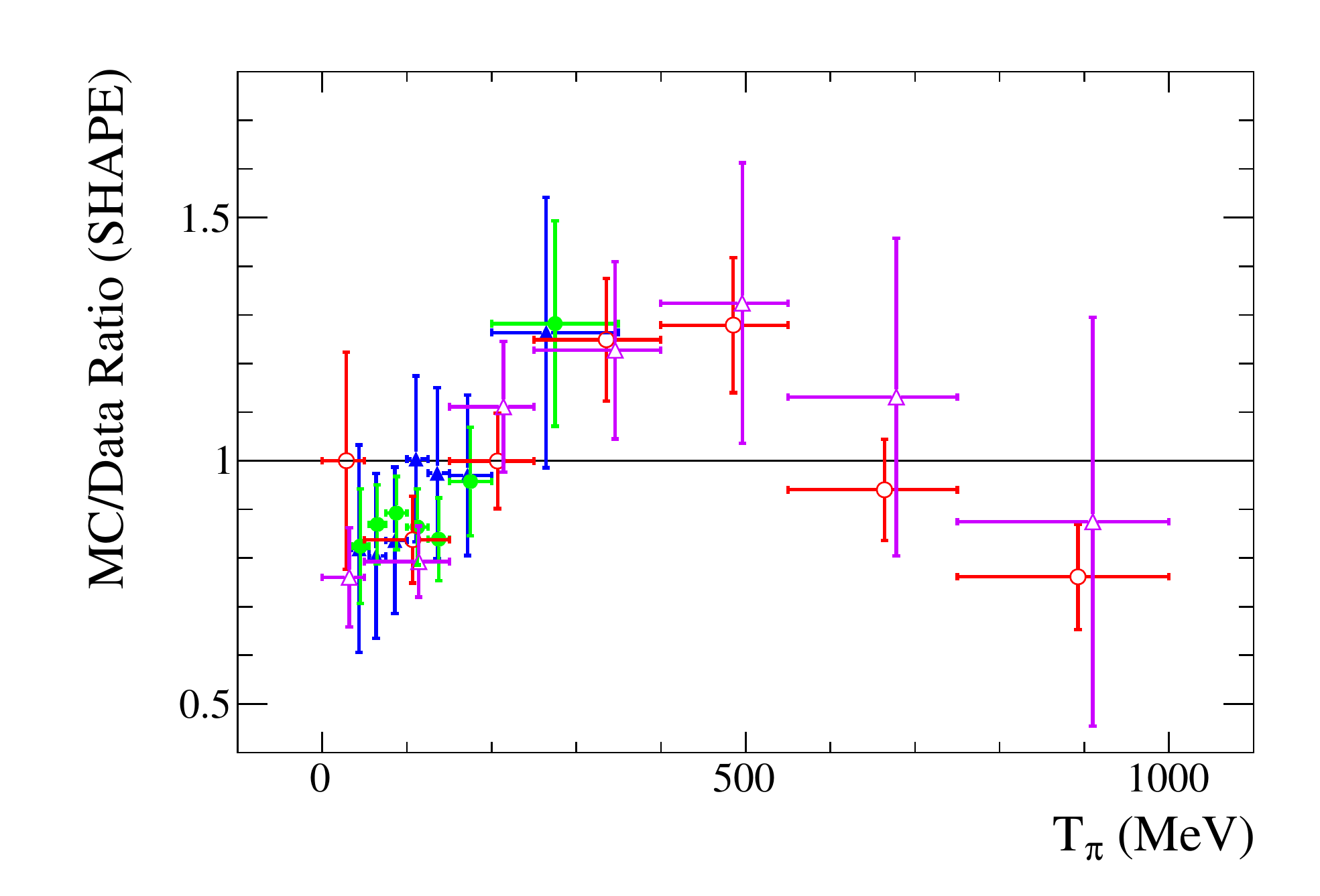}
\includegraphics[width=0.48\textwidth,trim={12mm 2mm 10mm 8mm}, clip] {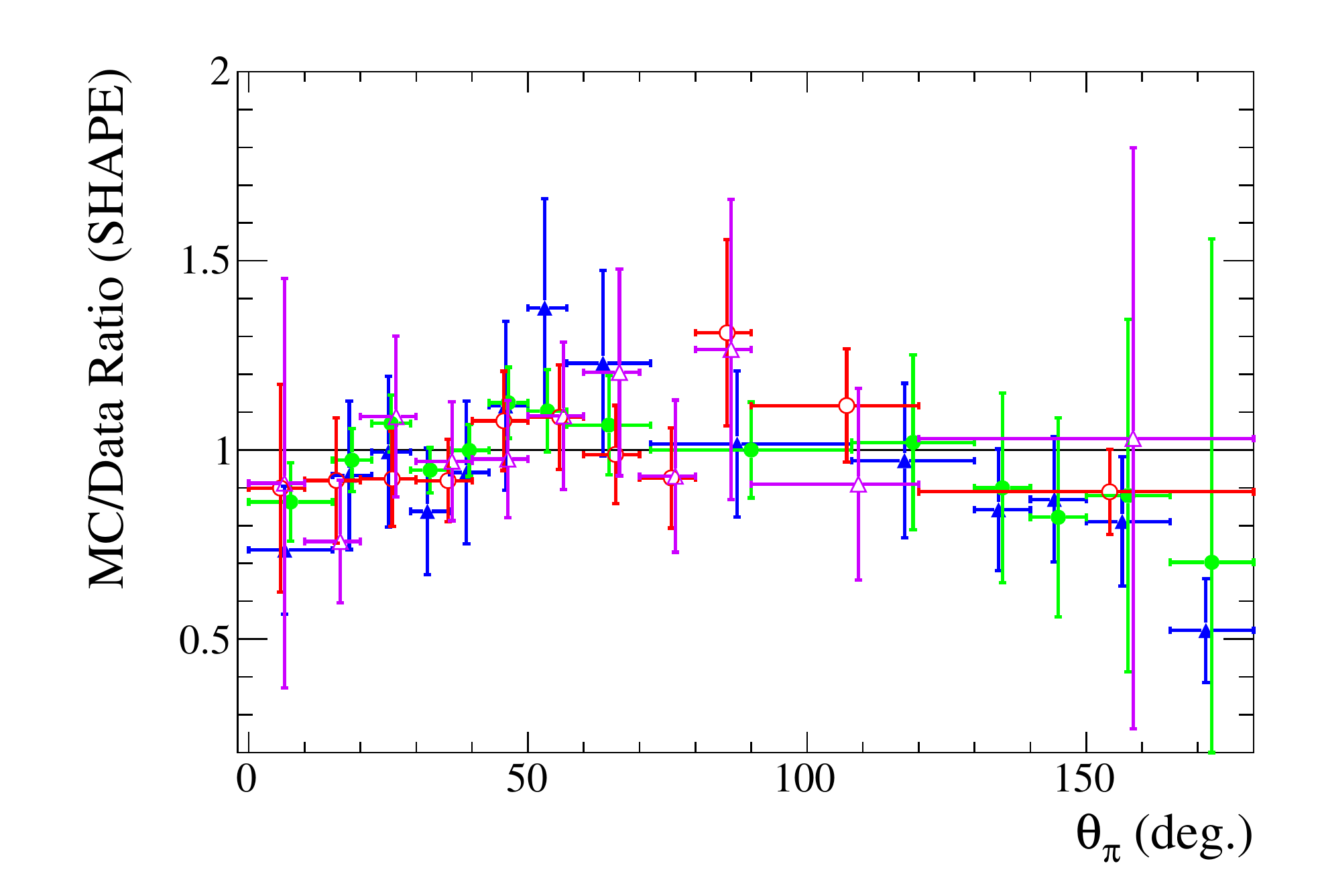}
\caption{MC/data ratios at the best fit point for the FrAbs joint tuning.
\label{fig:frabstuningratio} }
\end{figure*}

\begin{figure*}[hbtp]
\centering
\includegraphics[width=0.48\textwidth,trim={12mm 2mm 10mm 8mm}, clip] {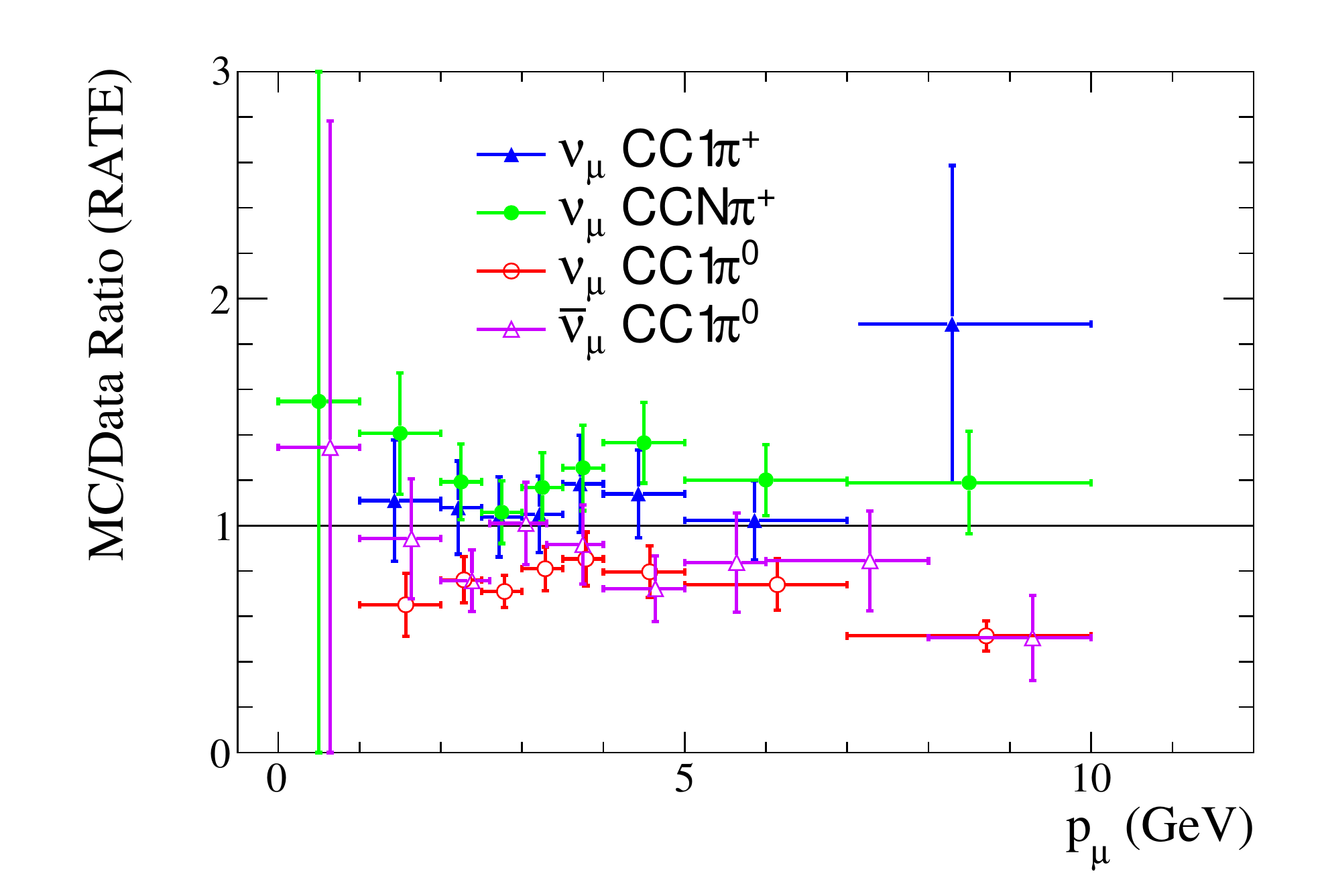}
\includegraphics[width=0.48\textwidth,trim={12mm 2mm 10mm 8mm}, clip] {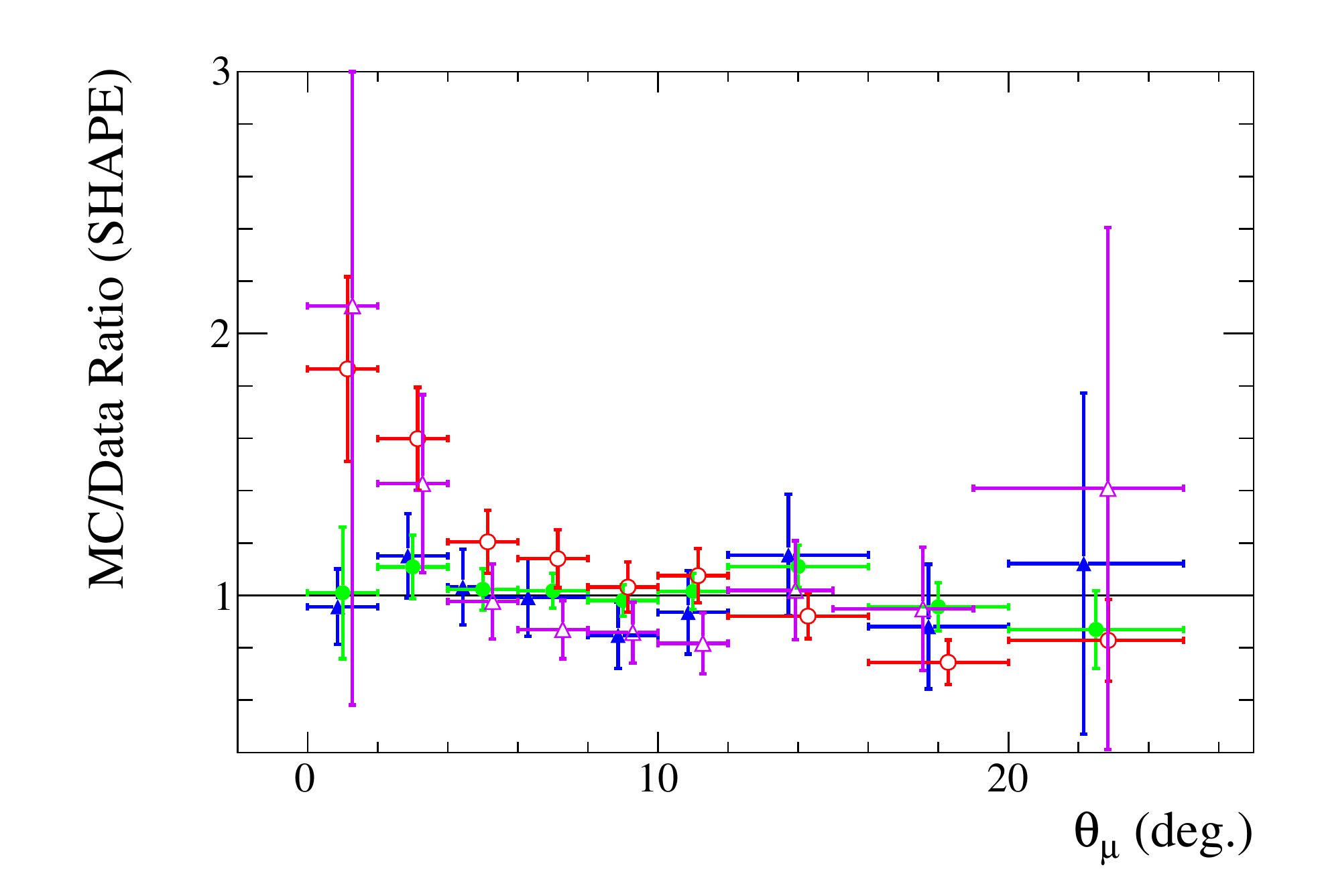}
\includegraphics[width=0.48\textwidth,trim={12mm 2mm 10mm 8mm}, clip] {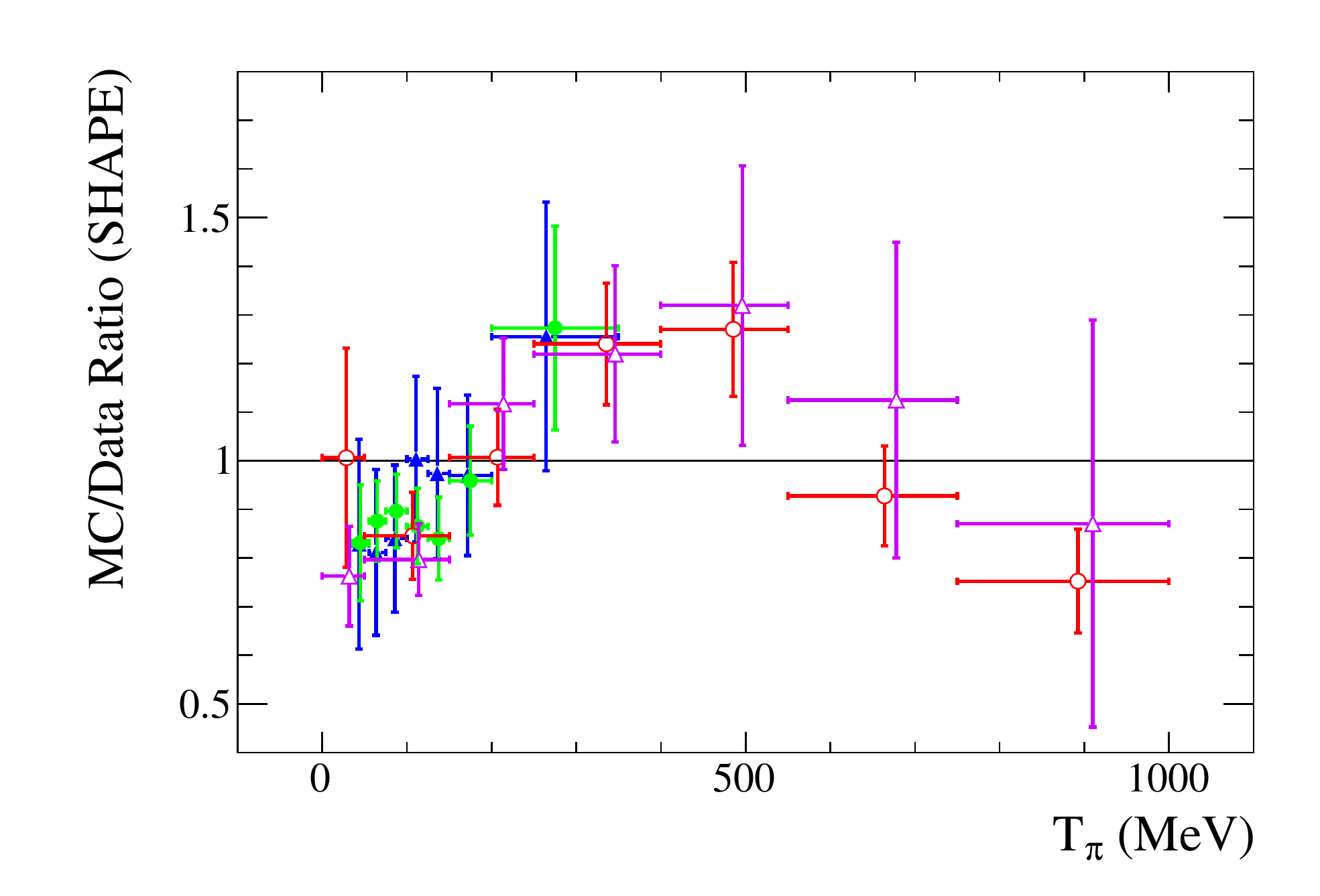}
\includegraphics[width=0.48\textwidth,trim={12mm 2mm 10mm 8mm}, clip] {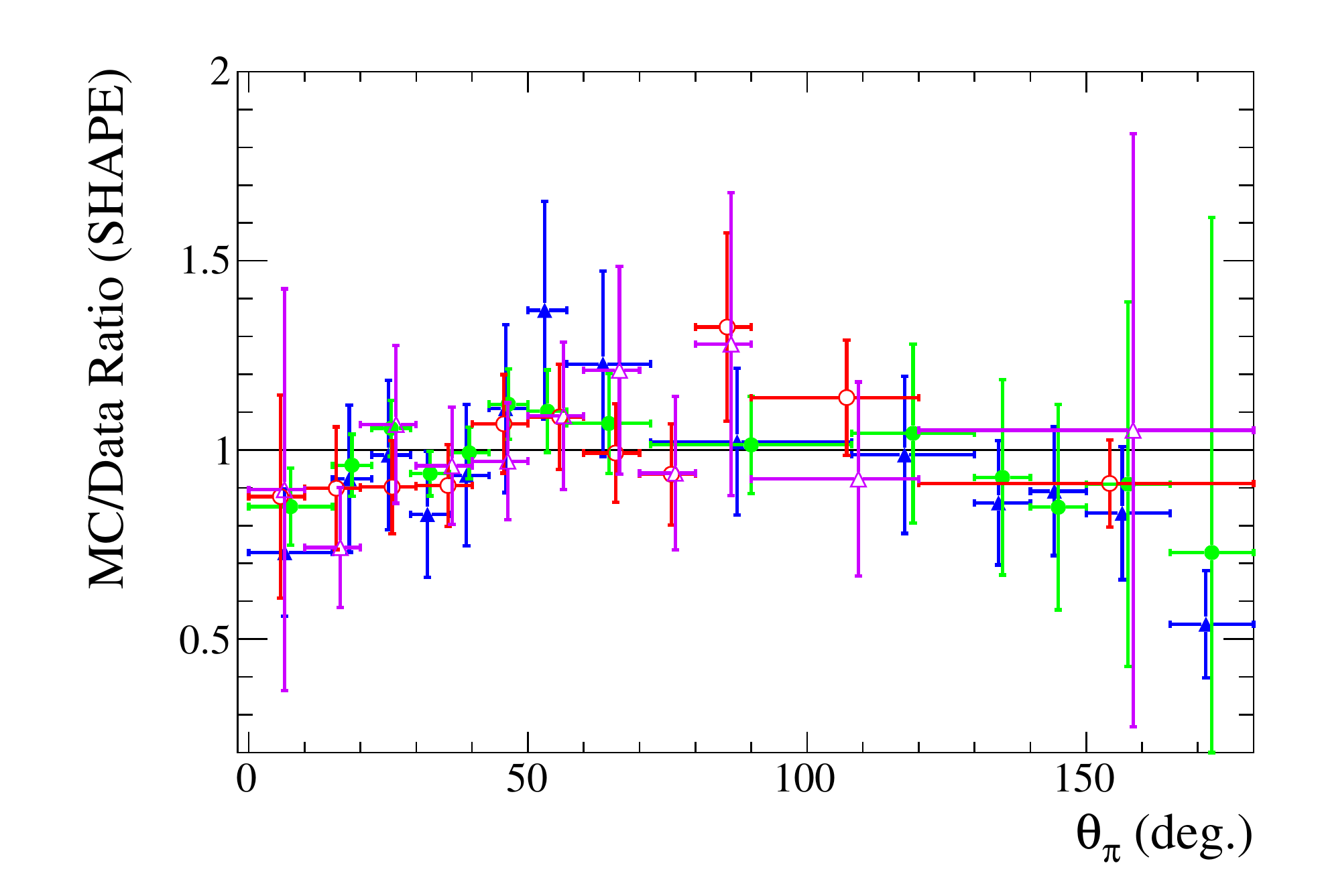}
\caption{MC/data ratios at the best fit point for the FrInel joint tuning.}
\label{fig:frineltuningratio}
\end{figure*}

The \NUISANCE interface to MINUIT2~\cite{James:1975dr} was used to perform the fits. At each iteration, the \GENIE-ReWeight package was used on an event-by-event basis to update the MC predictions before the total \chisq was calculated. The uncertainties in the fitted parameters were determined using the HESSE routine in MINUIT2.  The best fit results from the joint tuning are shown in Table~\ref{tab:combtuningresults}.  Fig.~\ref{fig:frabstuningratio} shows the ratios of the best fit prediction to the data for all four kinematic variables of interest when the pion absorption FSI parameter (FrAbs) is floated in the fit; Fig.~\ref{fig:frineltuningratio} is the same, but when the inelastic scattering FSI parameter (FrInel) is floated. Notably, the two FSI fits are very similar in both minimum $\chi^2$ and best-fit parameter values.

\begin{figure*}[hbtp]
\centering
\includegraphics[width=0.48\textwidth]{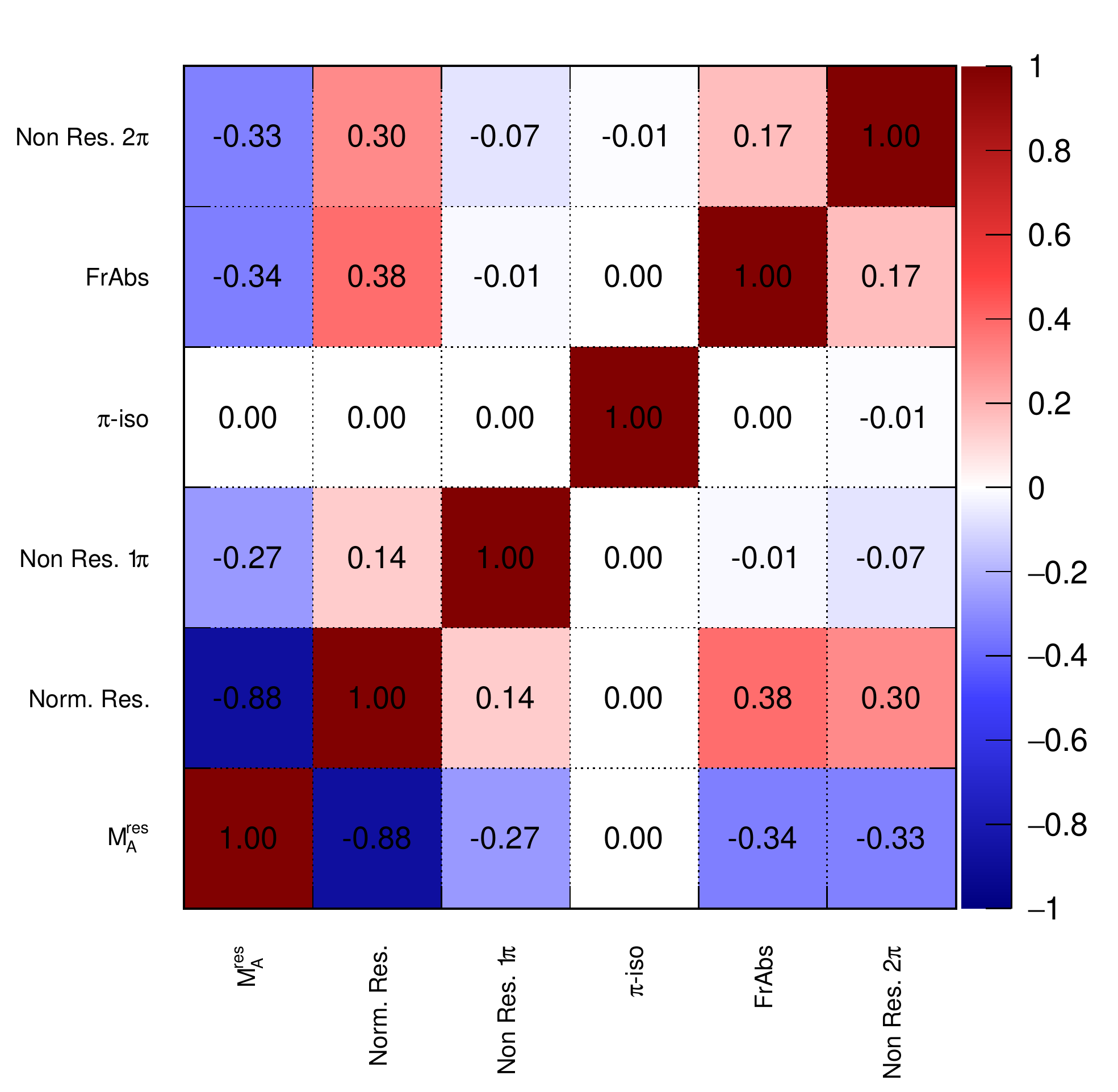}
\includegraphics[width=0.48\textwidth]{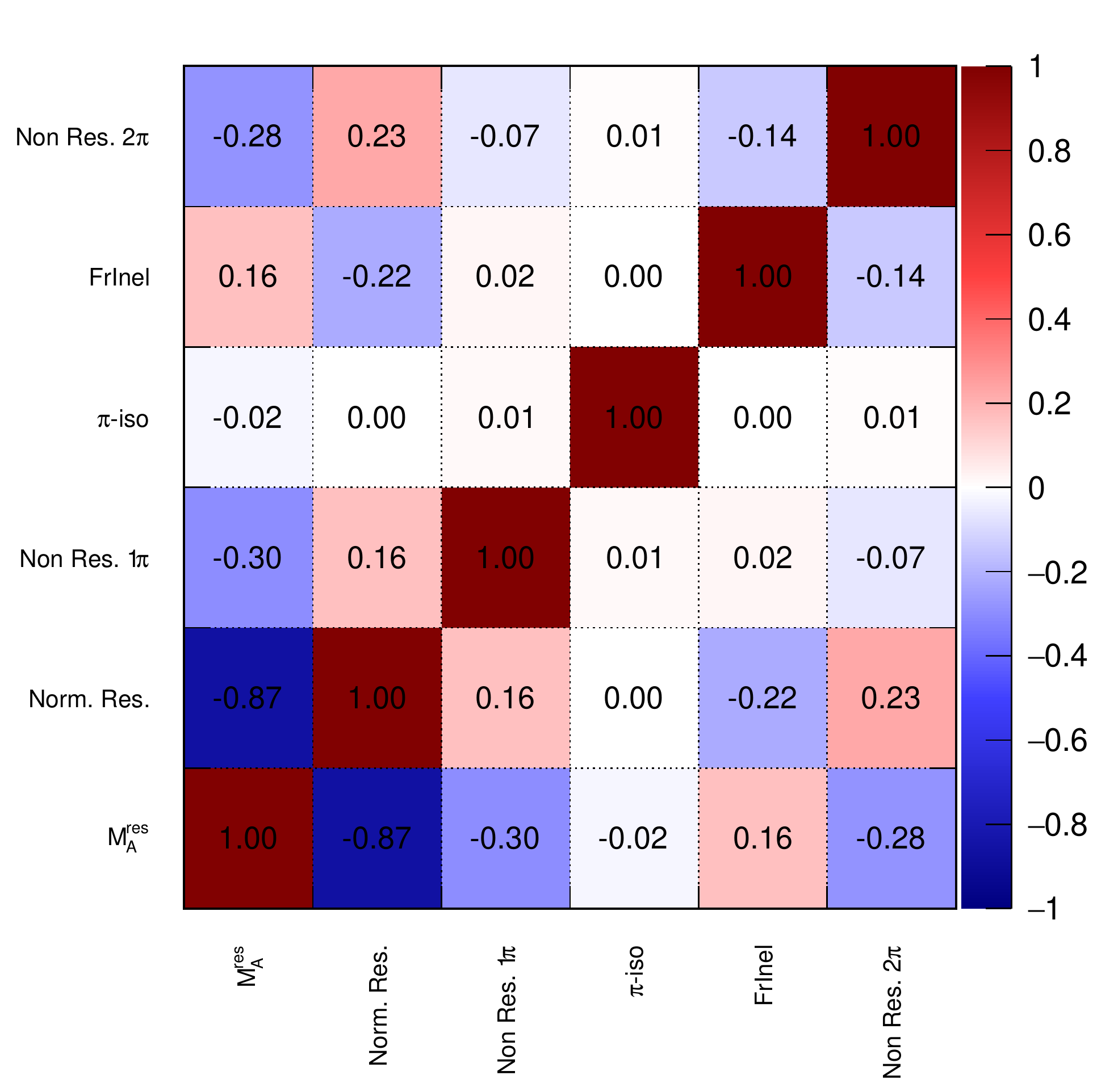}
\caption{Correlation matrix from from tuning \GENIE parameters in \NUISANCE with FrAbs included as a fit parameter (left) and with FrInel included as a fit parameter (right).
\label{fig:postfit_correlations_noq2} }
\end{figure*}

Comparing to the results of the ANL and BNL reanalysis, larger values of \Mares and smaller values of \Normres were found by the fit, pulling the parameters closer to \GENIE nominal.  The \nonresonepi parameter is strongly bound by the bubble chamber data and the \MINERvA data did little to improve on this constraint.  The penalty term contributed to the \chisq by 9.3 for the FrAbs fit and 11.1 for the FrInel fit.  This is a significant improvement over the default, but indicates that there is mild tension between the nucleon and nuclear data. The post-fit correlation matrices are provided in Fig~\ref{fig:postfit_correlations_noq2}.  The ANL/BNL input correlations are largely maintained in our fit.

\begin{table*}[hbtp]
\centering
{\renewcommand{\arraystretch}{1.2}
\begin{tabular}{ccccc}
\hline\hline
Parameter            & Default Value   & ANL/BNL Value & FrAbs Fit Result & FrInel Result \\ 
\hline\hline
\Mares (GeV)         & $1.12 \pm 0.22$ & $0.94\pm0.05$ & $1.07\pm0.04$    & $1.08\pm0.04$ \\
\Normres (\%)        & $100 \pm 20$    & $115\pm7$     & $94\pm6$         & $92\pm6$ \\
\nonresonepi (\%)    & $100 \pm 50$    & $43\pm4$      & $44\pm4$         & $44\pm4$ \\
\nonrestwopi (\%)    & $100 \pm 50$    & -             & $166\pm32$       & $161\pm33$ \\
\ThetaPi             & 0 = RS          & -             & 1 = Iso (limit)  & 1 = Iso (limit) \\
FrAbs (\%)           & 100 $\pm$ 30    & -             & $109\pm16$       & - \\
FrInel (\%)          & 100 $\pm$ 40    & -             & -                & $109\pm24$ \\
\hline\hline
\MINERvA \chisq      & 275.6           & 312.7         & 242.3            & 240.7 \\
\chisqpen            & 299.3           & 0.0           & 9.3              & 11.1 \\
\hline\hline
Total \chisq         & 574.8           & 312.7         & 251.6            & 251.8 \\
\ndof                & 148             & 148           & 145              & 145 \\
\hline\hline
\end{tabular}}
\caption{Fit results from tuning \GENIE parameters in \NUISANCE.  The ``ANL/BNL Value'' column shows the contributions when parameters are fixed at values of Table~\ref{tab:callumtuneres}.
\label{tab:combtuningresults}}
\end{table*}

Tables~\ref{tab:frabsindividual} and~\ref{tab:frinelindividual} show the results when individual \MINERvA data were tuned in separate fits.  Since three of the four channels were removed in these fits the constraint from data is weakened and the total \chisq is steered by the bubble chamber \chisq penalty.  The individual channel fits also found values at the 300\% limit for \nonrestwopi dial, except in the \ccnpip channel, where the result was unchanged by the fit.  Only the \ccnpip channel has a significant contribution from nonresonant 2\pipm production.  In the other fits, the parameter is largely unconstrained and has little impact on the fitted distributions.  The \chisq per degree of freedom is indicative of a poor fit in the \ccnpip and \ccpin channels, but not in the \ccpip or \ccapin channels.  Furthermore, the \ccpin shows the strongest \chisq penalty, indicating tension with the ANL/BNL prior.  Given the different kinematic regions covered by the channels (see Table~\ref{tab:datareleases}) and the different physics (\eg fraction of coherent pion production) it is difficult to infer what combination of effects are at work. Isotropic emission was preferred in all fits, driven by the \Tpi distributions. Disagreements in the \Tpi spectrum are clearly seen in the data/MC ratios of Fig.~\ref{fig:frabstuningratio} and~\ref{fig:frineltuningratio}, and the large \chisq values observed for the \ccnpip and \ccpin channels.

The individual \chisq contributions in the joint tuning best fit, shown in sixth and seventh columns (``FrAbs Tune'' and ``FrInel Tune'') of  Table~\ref{tab:genienomchi2}, show that not all distributions in all channels benefit from the model variations, as the default \GENIE fits have a better \chisq for some distributions.  In particular, the \ccpin channel distributions have worse agreement after the tuning, with only the \Tpi distribution improving in \chisq, whereas all channels benefit from the shift to isotropic emission.  While there is an overall improvement over the ANL/BNL tune when comparing the combined \chisq results, Figs.~\ref{fig:frabstuningratio} and~\ref{fig:frineltuningratio} show that there are still unresolved shape disagreements in both the \Kpi and \Tmu kinematics.

\begin{table*}[hbtp]
\centering
{\renewcommand{\arraystretch}{1.2}
\begin{tabular}{ccccc}
\hline\hline
Parameter            & \ccpip          & \ccnpip         & \ccpin          & \ccapin \\
\hline\hline
\Mares (GeV)         & 0.97 $\pm$ 0.05 & $0.97\pm0.05$   & $1.02\pm0.05$   & $0.96\pm0.05$ \\
\Normres (\%)        & 110 $\pm$ 7     & $110\pm7$       & $104\pm7$       & $111\pm7$ \\
\nonresonepi (\%)    & 43 $\pm$ 4      & $42\pm4$        & $44\pm4$        & $43\pm4$ \\
\nonrestwopi (\%)    & $300$ (limit)   & 99$\pm$30       & $300$ (limit)   & $300$ (limit) \\
\ThetaPi             & 1 = Iso (limit) & 1 = Iso (limit) & 1 = Iso (limit) & 1 = Iso (limit) \\
FrAbs (\%)           & 156 $\pm$ 53    & $128\pm34$      & $126\pm17$      & $82\pm31$ \\
\hline\hline
\MINERvA \chisq      & 36.6            & 64.1            & 92.3            & 34.6 \\
\chisqpen            & 0.5             & 0.7             & 3.2             & 0.3 \\
\hline\hline
Total \chisq         & 37.1            & 64.8            & 95.5            & 34.9 \\
\ndof                & 35              & 36              & 32              & 33 \\
\hline\hline
\end{tabular}}
\caption{Individual channel tuning results when the FrAbs dial is treated as the free FSI parameter.
\label{tab:frabsindividual}}
\end{table*}

\begin{table*}[hbtp]
\centering
{\renewcommand{\arraystretch}{1.2}
\begin{tabular}{ccccc}
\hline\hline
Parameter            & \ccpip          & \ccnpip         & \ccpin          & \ccapin \\
\hline\hline
\Mares (GeV)         & 0.97 $\pm$ 0.05 & $0.97\pm0.05$   & $1.03\pm0.05$   & $0.96\pm0.05$ \\
\Normres (\%)        & 109 $\pm$ 7     & $108\pm7$       & $103\pm7$       & $112\pm7$ \\
\nonresonepi (\%)    & 42 $\pm$ 4      & $42\pm4$        & $43\pm4$        & $43\pm4$ \\
\nonrestwopi (\%)    & $300$ (limit)   & 110$\pm$30      & $300$ (limit)   & $300$ (limit) \\
\ThetaPi             & 1 = Iso (limit) & 1 = Iso (limit) & 1 = Iso (limit) & 1 = Iso (limit) \\
FrInel (\%)          & 117 $\pm$ 54    & $127\pm33$      & 0 (limit)       & $80\pm59$ \\
\hline\hline
\MINERvA \chisq      & 37.1            & 63.4            & 86.9            & 34.9 \\
\chisqpen            & 0.7             & 1.3             & 3.4             & 0.2 \\
\hline\hline
Total \chisq         & 37.8            & 64.7            & 90.3            & 35.1 \\
\ndof                & 35              & 36              & 32              & 33 \\
\hline\hline
\end{tabular}}
\caption{Individual channel tuning results when the FrInel dial is treated as the free FSI parameter.
\label{tab:frinelindividual}}
\end{table*}

The tension between \MINERvA's nuclear data and the constraints from ANL and BNL nucleon data is difficult to confidently pinpoint; the lack of lepton mass effects~\cite{Berger:2007rq}, modification to the resonance propagator in the nucleus~\cite{singh,oset}, missing diagrams describing the non-resonant background contributions~\cite{Kabirnezhad:2017jmf}, dynamical coupled channels~\cite{dcc}, interactions on correlated initial states, and the pion FSI model~\cite{Mosel:2017nzk} are all part of an incomplete list of possible culprits.

\section{Ad hoc \qq Supression}
\label{sec:q2-corrections}
\begin{figure*}[hbtp]
\centering
\includegraphics[width=0.48\textwidth,page=3,trim={0mm 10mm 0mm 10mm},clip]{RPAFunctions/q2_correction.pdf}
\includegraphics[width=0.48\textwidth,page=4,trim={0mm 10mm 0mm 10mm},clip]{RPAFunctions/q2_correction.pdf}
\includegraphics[width=0.48\textwidth,page=1,trim={0mm 10mm 0mm 10mm},clip]{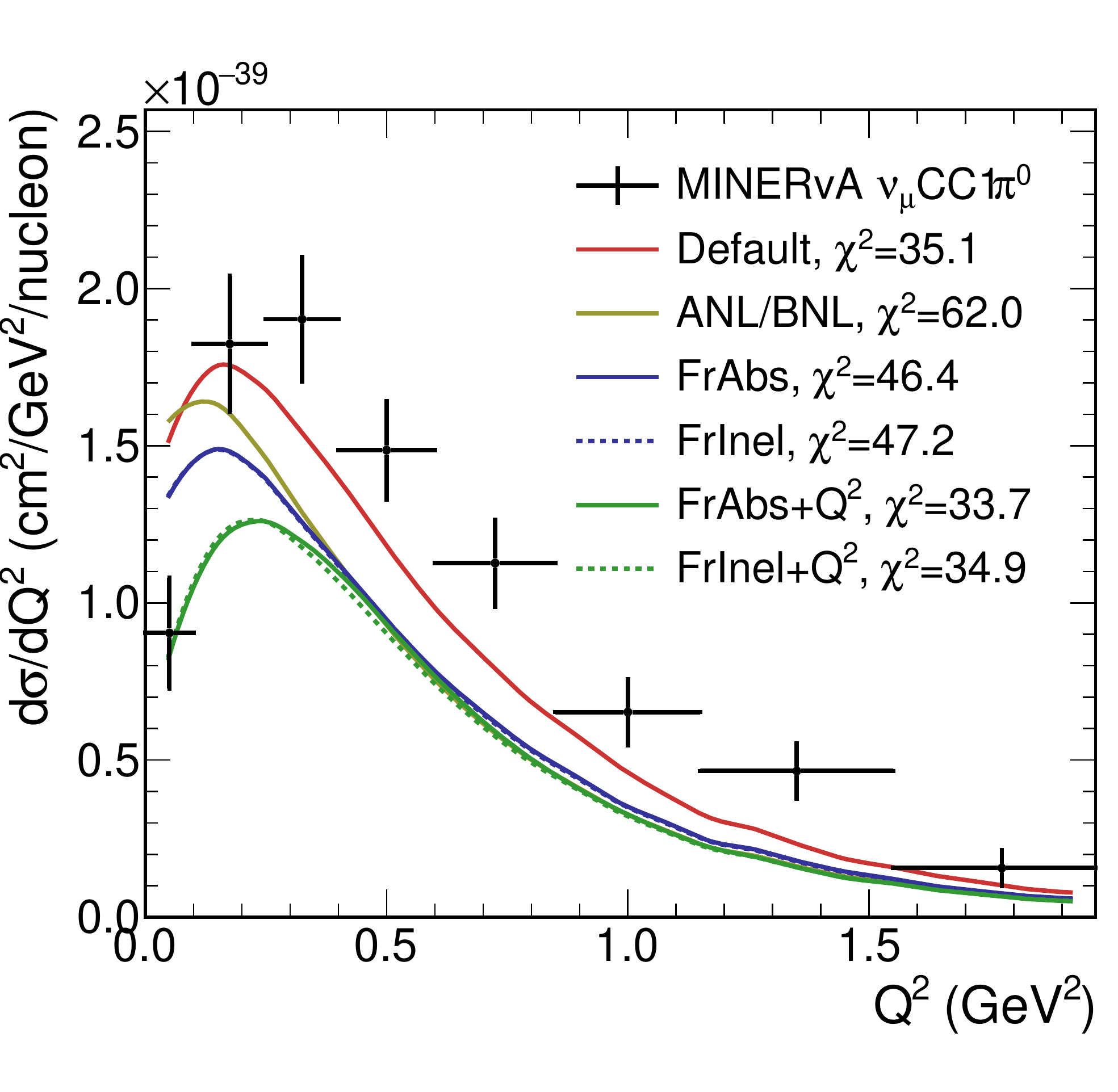}
\includegraphics[width=0.48\textwidth,page=2,trim={0mm 10mm 0mm 10mm},clip]{RPAFunctions/q2_correction.pdf}
\caption{Comparisons of the nominal and tuned models to \MINERvA \ccpip (left top), \ccnpip (right top), \ccpin (left bottom) and \ccapin (right bottom) distributions in \qq. The \chisq is computed using the full covariance matrices. The distributions were not explicitly used in the tuning procedure. \label{fig:q2difatminerva}}
\end{figure*}

Further modifications beyond the standard \GENIE dials are required to resolve the observed tensions.  Fig.~\ref{fig:q2difatminerva} (not used for any tuning) shows the \qq distributions observed at \MINERvA in for our tunes.  The data is below the predictions of the tunes of Section~\ref{sec:genie-tuning} at low values of \qq.  There are also also differences at low \Tmu, as shown in Fig.~\ref{fig:frabstuningratio} and~\ref{fig:frineltuningratio}.  Measurements of \ccpip and \ccpin interactions on mineral oil at \MiniBooNE have shown a data/MC shape discrepancy for the RS implementation in the \NUANCE model~\cite{Casper:2002sd, Przewlocki:2009zz} in both \qq and $\cos\Tmu$ distributions~\cite{AguilarArevalo:2010bm, AguilarArevalo:2010xt}.  In the \MINOS quasielastic analysis~\cite{Adamson:2014pgc} on iron, which used NeuGen~\cite{Gallagher:2002}, a similar disagreement was observed when studying pion production dominated sidebands.  Indeed, concerns about low-\qq modeling date back almost a decade~\cite{Graczyk:2008zz}.  The data from MINOS and \MiniBooNE experiments and the \MINERvA data on CH studied herein suggest that the RS implementation common to each of the generators needs to be suppressed at low \qq. Collective effects, which are usually modeled in the random phase approximation, are known to affect the \qq distribution of neutrino-nucleus reactions at low \qq. Motivated by these considerations, we attempted to improve the \Tmu modeling by introducing a \qq-dependent correction to the model. 

 The \MINOS collaboration suppression was expressed as
\begin{align}
R = \frac{A}{ 1 + \exp\{ 1 - \sqrt{Q^{2}}/Q_{0} \}}  \label{eq:minosrpa},
\end{align}
where the free parameters $A=1.010$ and $Q_{0}=0.156\units{GeV}$ were empirically extracted from bin-by-bin fits in \qq to the data, and a hard cut-off at $Q^2 < 0.7\units{GeV}^2$ was imposed.

We chose an empirical function so that the shape of the suppression preferred by each of the \MINERvA channels could be extracted.   The empirical correction function is applied to events with a resonance decay inside the nucleus giving rise to a pion. Our suppression term is defined by choosing 3 points $(x_i, R_i)_{i=1,2,3}$ between $0.0 < x < 1.0$ and $0.0 < R < 1.0$, where $x \equiv Q^2$.  Motivated the ANL/BNL curves in Fig.~\ref{fig:q2difatminerva}, the correction is assumed to approach unity as \qq approaches $0.7\units{GeV}^2$, providing the constraint $(x_3, R_3)=(0.7\units{GeV}^2,1.0)$.  Lagrange interpolation is used to derive a curvature from $R_{2}$ by assuming a simple interpolation between the points $(x_{1},0.0)$, $(x_2, R_2)$, and $(0.7\units{GeV}^2,1.0)$:
\begin{align}
R(\qq<x_3) & = \frac{R_{2} (\qq-x_1)(\qq-x_3)}{(x_2-x_1)(x_2-x_3)} \notag\\
           & + \frac{(\qq-x_1)(\qq-x_2)}{(x_3-x_1)(x_3-x_2)}  \label{eq:ourrpa1}.
\end{align}
\noindent This interpolation function is then used to calculate the correction for each event as 
\begin{align}
w(Q^{2}) = 1-(1-R_{1})(1-R(Q^{2}))^{2} \label{eq:ourrpa2}.
\end{align}
\noindent where $R_{1}$ defines the magnitude of the correction function at the intercept, $x_{1}=0.0$. $x_{2}$ is chosen to be $Q^2=0.35\units{GeV}^2$ so that $R_{2}$ describes the curvature at the centre point of the correction.  Expressing the weights with Equations~\ref{eq:ourrpa1} and \ref{eq:ourrpa2} ensures that the magnitude at $x_{2}$ always lies between $R_{1}$ and 1.0, avoiding parameter sets with large unphysical peaks in the correction function.  Additionally, the squared term in Equation~\ref{eq:ourrpa2} ensures that $w(Q^{2})\rightarrow 1.0$ as $x \rightarrow x_{3}$, avoiding discontinuous steps in the weighting function at $x_{3}$. The fitted parameters $R_1$ and $R_2$ were limited to $0.0<R_1<1.0$ and $0.5<R_2<1.0$ to avoid extraneous solutions, \eg double peaks.

\begin{figure*}[hbtp]
\centering
\includegraphics[width=0.48\textwidth,trim={5mm 5mm 30mm 16mm}, clip]{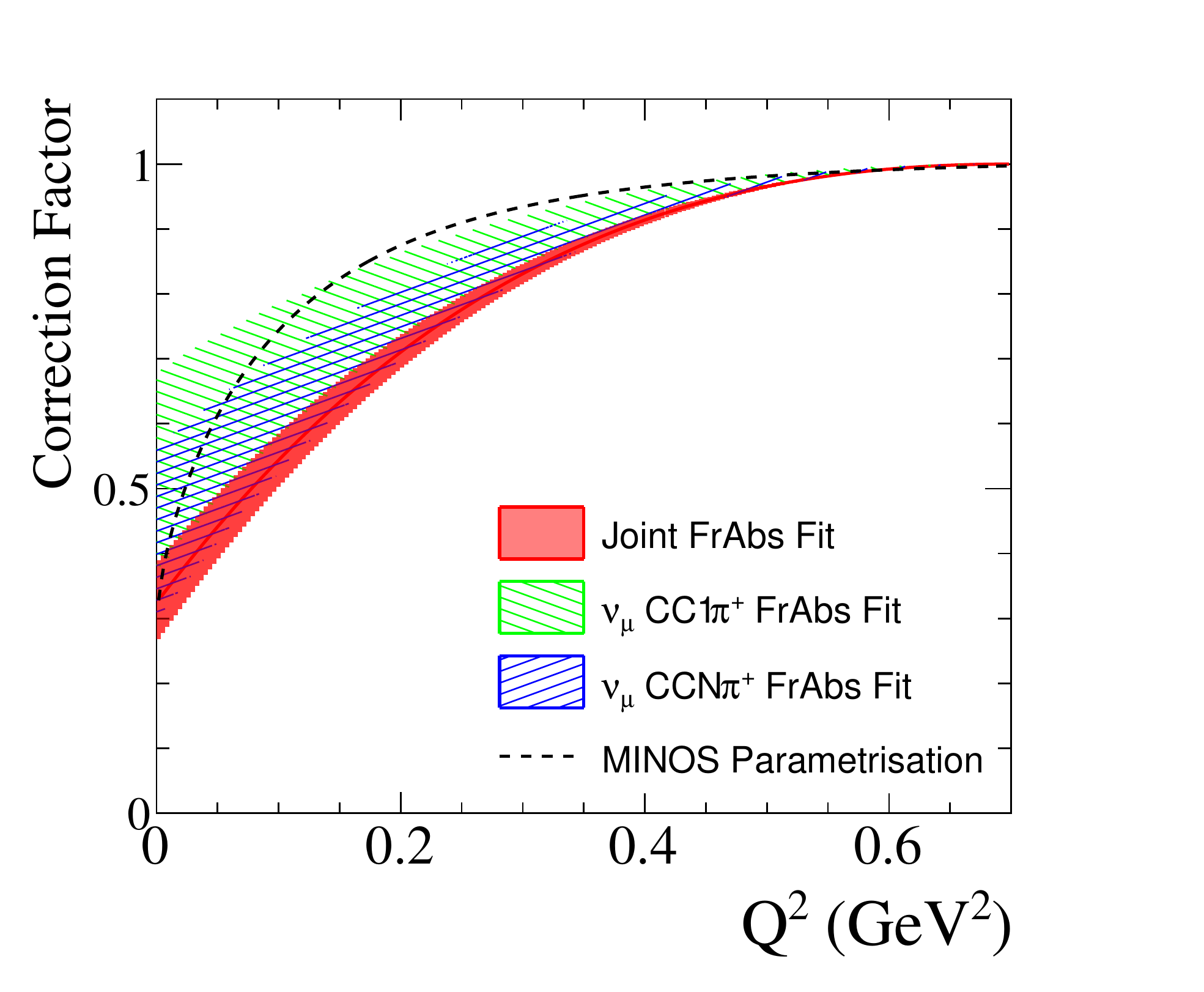}
\includegraphics[width=0.48\textwidth,trim={5mm 5mm 30mm 16mm}, clip]{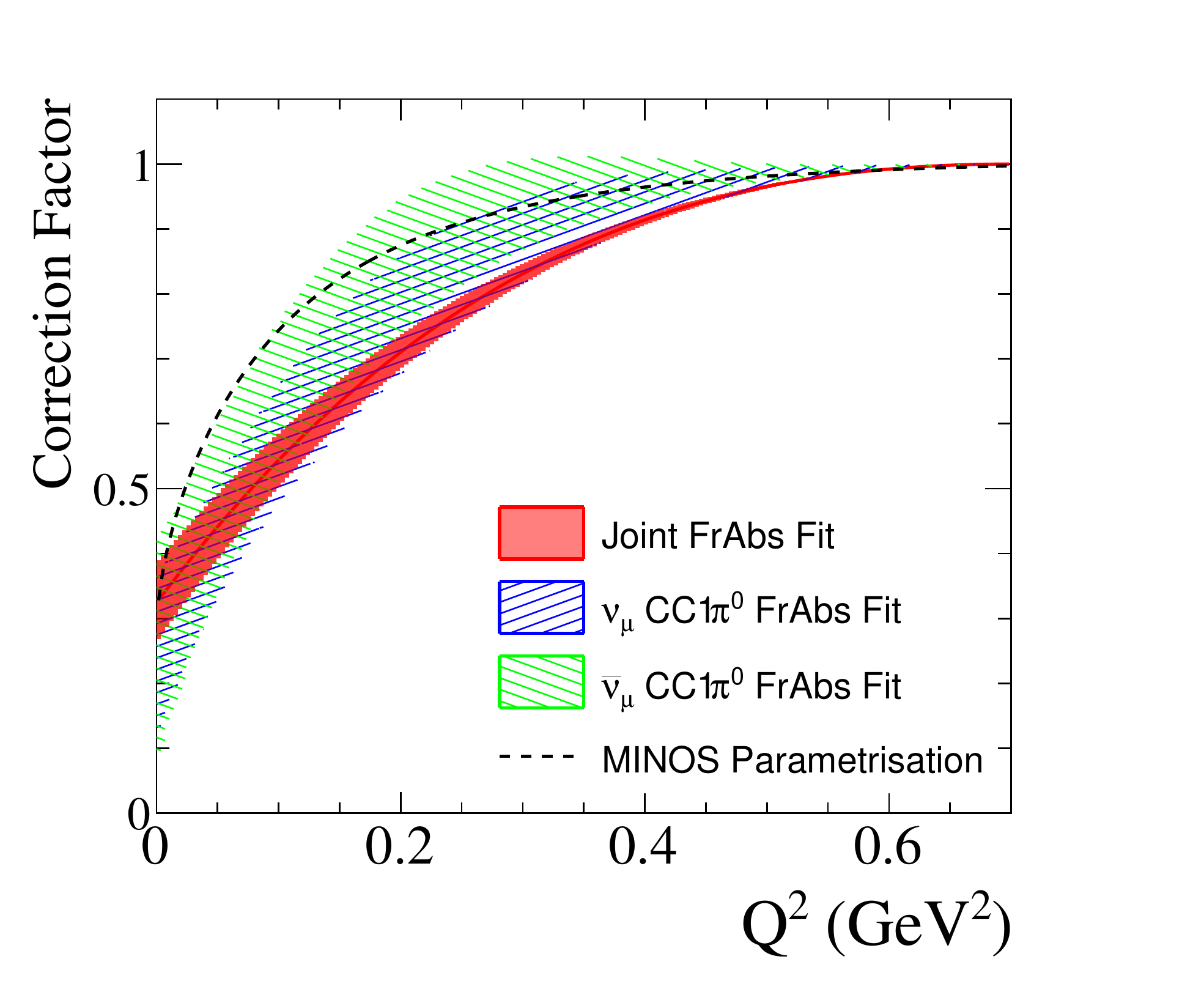}
\caption{Extracted low-\qq suppression factors from the FrAbs + low-\qq tuning to each channel. The left and right plots compare the results for the charged and neutral pion production channels respectively. Shown in red  is the uncertainty band extracted from the joint fit to all 4 channels simultaneously.
\label{fig:jointfitlagrangian}}
\end{figure*}

The fit results are shown in Table~\ref{tab:rpajointfits}.  The correction from the fit with FrAbs taken as a free parameter are compared to the \MINOS low-\qq correction in Fig.~\ref{fig:jointfitlagrangian}.  Our fits obtain a suppression factor that is similar to the \MINOS one, with almost identical suppression at \qq = 0, albeit with less curvature, particularly in the \ccpip and \ccnpip channels. The correction factors from the fit with FrInel or FrAbs as free parameters give similar results.

\begin{figure*}[hbtp]
\centering
\includegraphics[width=0.48\textwidth]{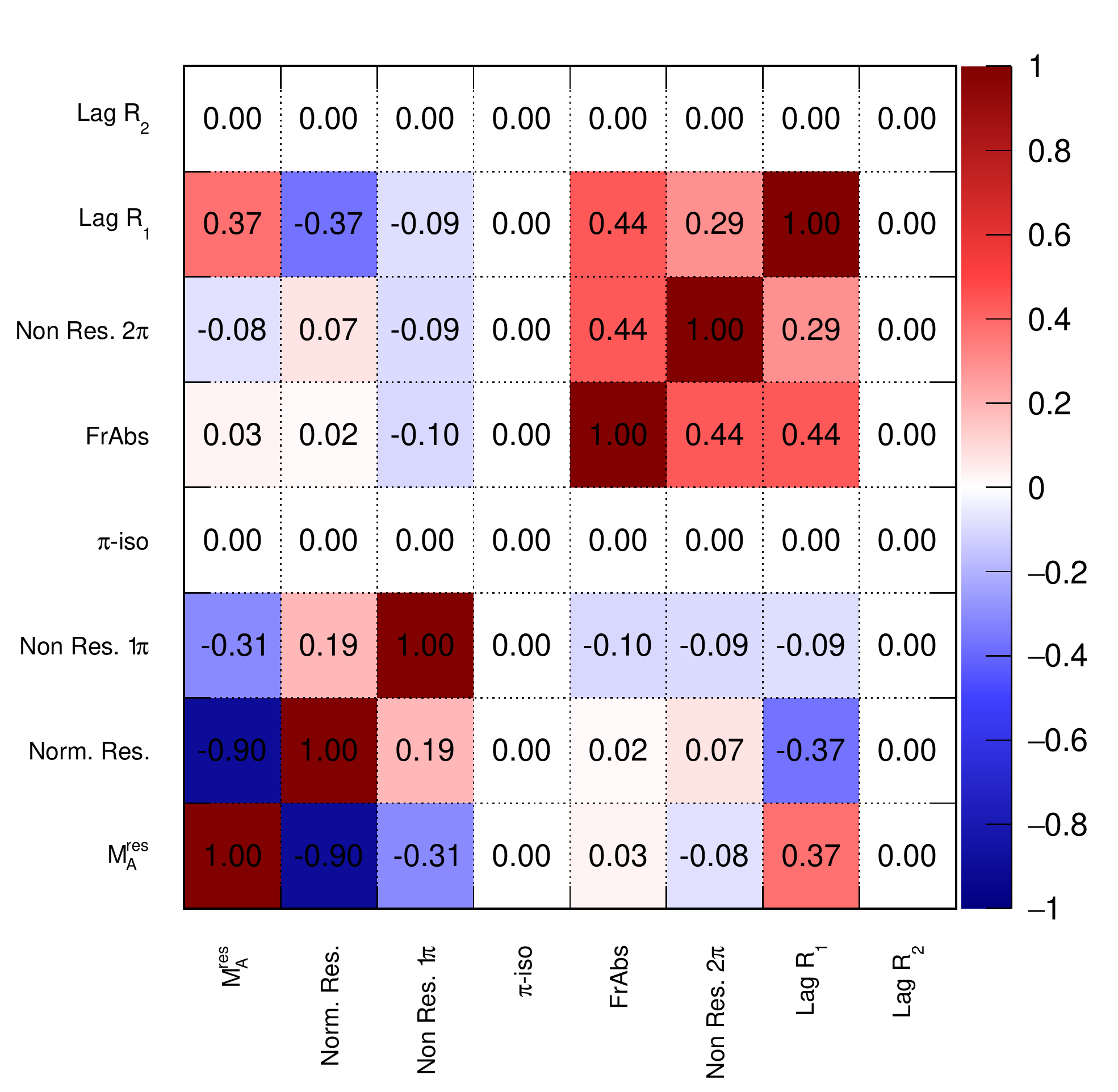}
\includegraphics[width=0.48\textwidth]{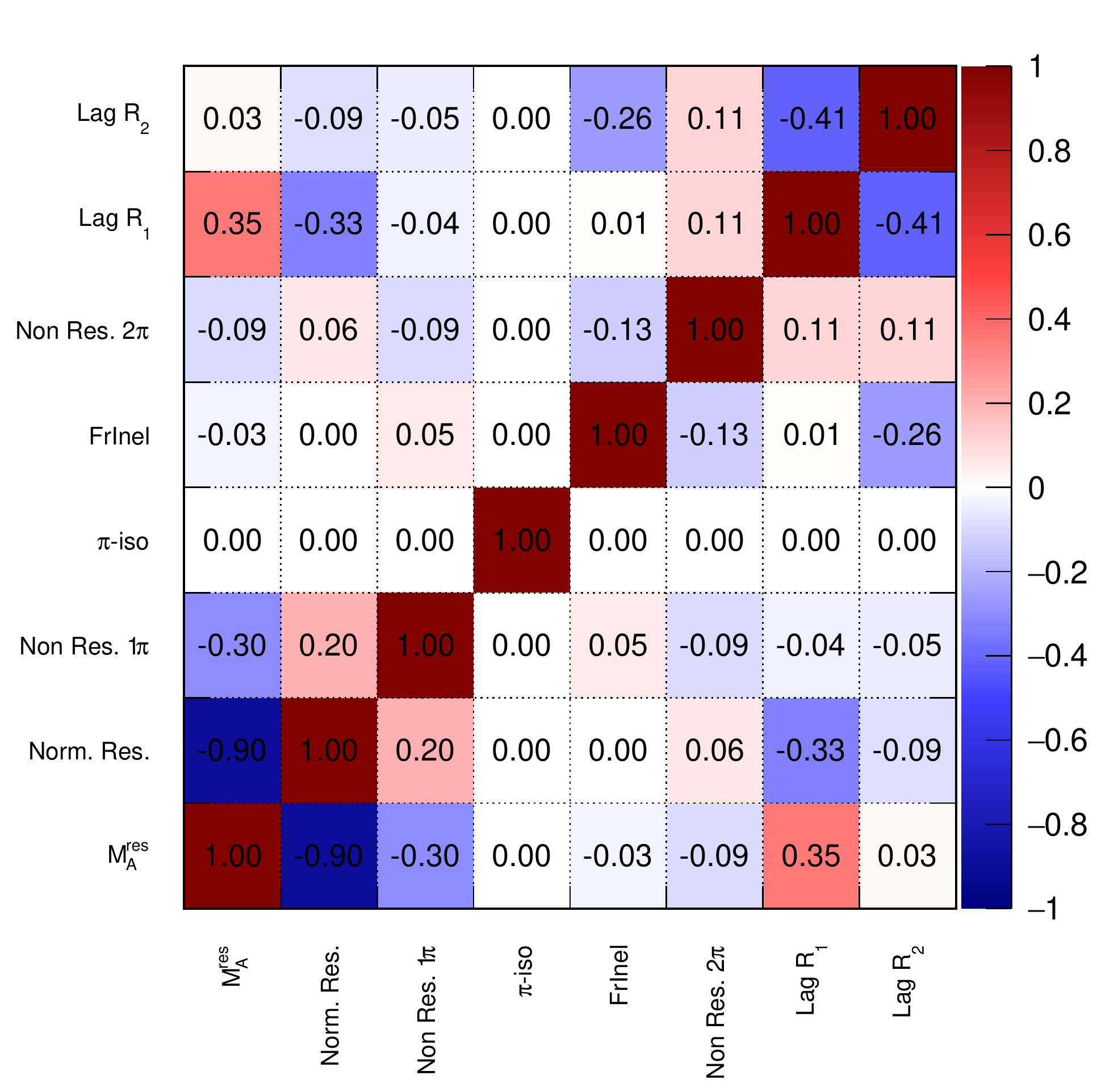}
\caption{Correlation matrix from tuning \GENIE parameters with an ad hoc low-\qq supression with FrAbs included as a fit parameter (left) and with FrInel included as a fit parameter (right).}
\label{fig:postfit_correlations_q2}
\end{figure*}

The correlation matrices for the fits including a $Q^2$ dependent suppression are provided in Fig~\ref{fig:postfit_correlations_q2}.  Again, the ANL/BNL input prior covariance is maintained.  The parameters largely correlate in the same way for the FrAbs and FrInel fit, and for the FrInel fit the $R_1$ and $R_2$ parameters are negatively correlated.

\begin{table*}[hbtp]
\centering
{\renewcommand{\arraystretch}{1.2}
\begin{tabular}{ccccc}
\hline\hline
Parameter              & FrAbs Tune     &  FrAbs + low-\qq Tune & FrInel Tune   &  FrInel + low-\qq Tune \\
\hline\hline
\Mares $(GeV)$         & $1.07\pm0.04$  & $0.92\pm0.02$         & $1.08\pm0.04$ & $0.93\pm0.05$ \\
\Normres (\%)          & $94\pm6$       & $116\pm3$             & $92\pm6$      & $116\pm7$ \\
\nonresonepi (\%)      & $43\pm4$       & $46\pm4$              & $44\pm4$      & $46\pm4$ \\
\nonrestwopi (\%)      & $166\pm32$     & 99$\pm$31             & $161\pm33$    & $120\pm32$ \\
\ThetaPi               & $1.0$ (limit)  & 1.0 (limit)           & 1.0 (limit)   & 1.0 (limit) \\
FrAbs (\%)             & $109\pm16$     & $48\pm21$             & -             & - \\
FrInel (\%)            & -              & -                     & $109\pm24$    & $132\pm27$ \\
Lag. $R_1$             & -              & $0.32\pm0.06$         & -             & $0.37\pm0.09$ \\
Lag. $R_2$             & -              & $0.5$ (limit)         & -             & $0.60\pm0.16$  \\
\hline\hline
\MINERvA \chisq        & 242.3          & 212.2                 & 240.7         & 215.7 \\
\chisqpen              & 9.3            & 0.7                   & 11.1          & 0.5 \\
\hline\hline
Total\chisq            & 251.6          & 212.9                 & 251.8         & 216.2 \\
\ndof                  & 145            & 143                   & 145           & 143 \\
\hline\hline
\end{tabular}}
\caption{Ad hoc low-\qq suppression model tuning results compared to the tuning results without the low-\qq suppression. \label{tab:rpajointfits}}
\end{table*}

\begin{figure*}[hbtp]
\centering
\includegraphics[width=0.48\textwidth,trim={12mm 2mm 10mm 8mm}, clip]{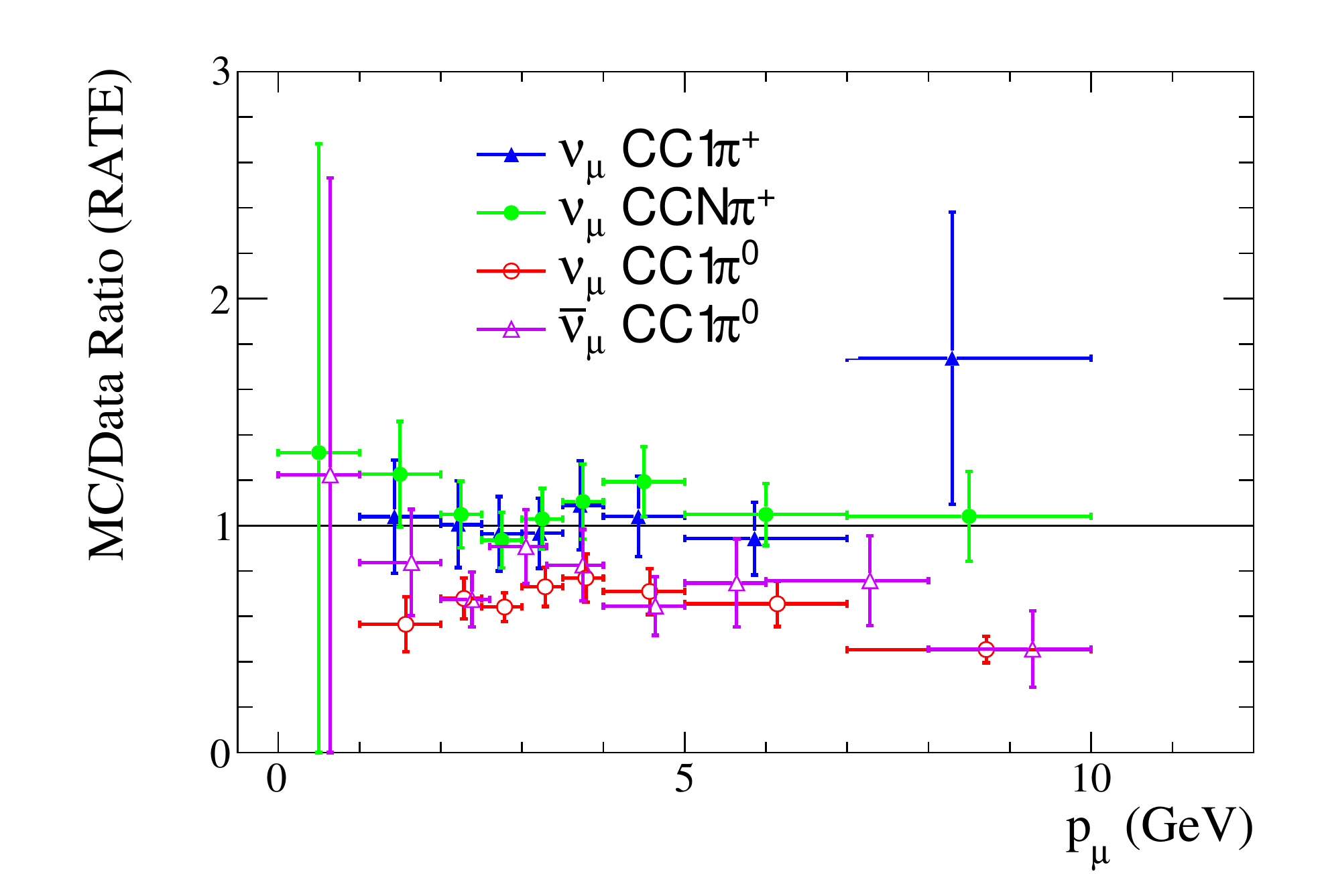}
\includegraphics[width=0.48\textwidth,trim={12mm 2mm 10mm 8mm}, clip]{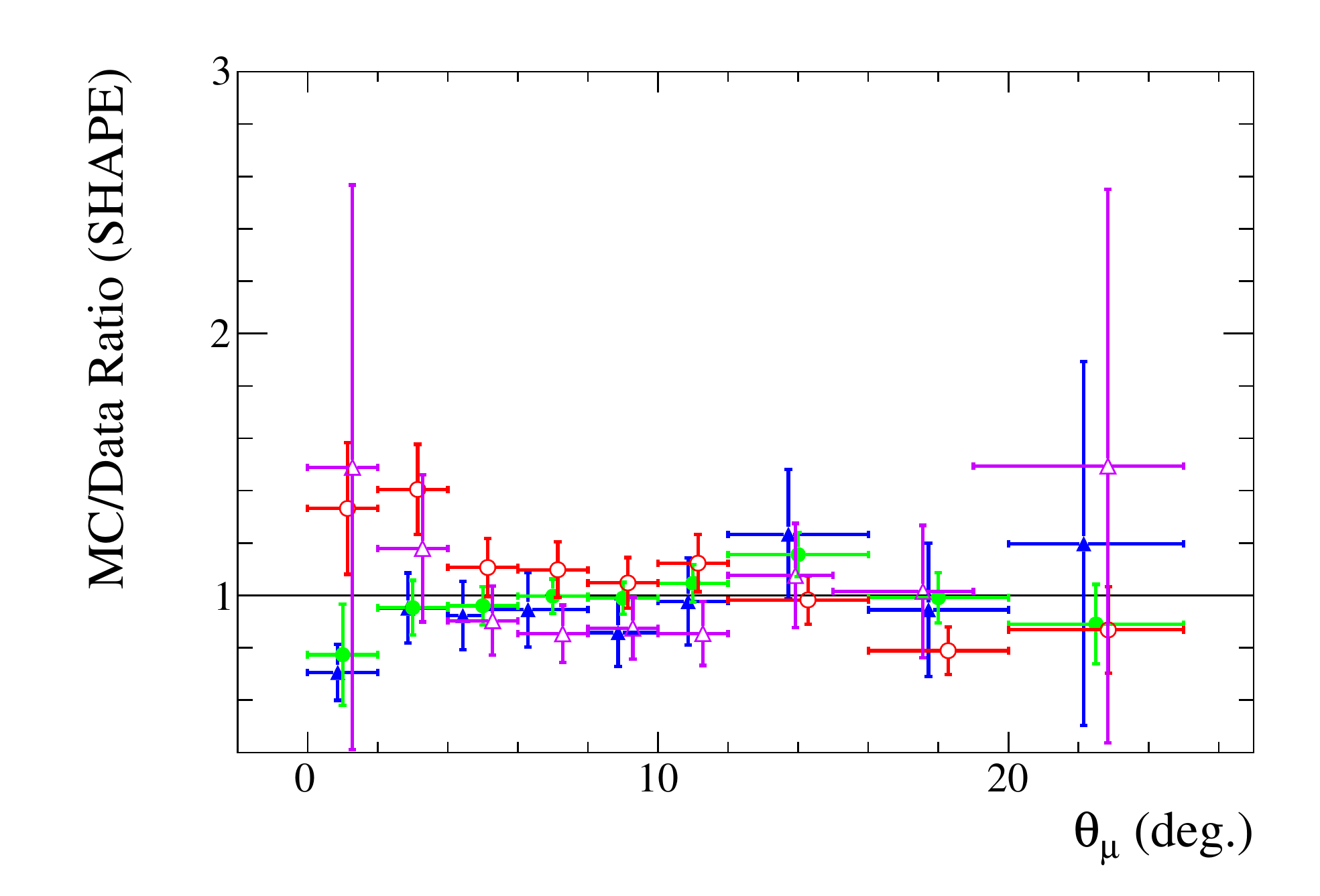}
\includegraphics[width=0.48\textwidth,trim={12mm 2mm 10mm 8mm}, clip]{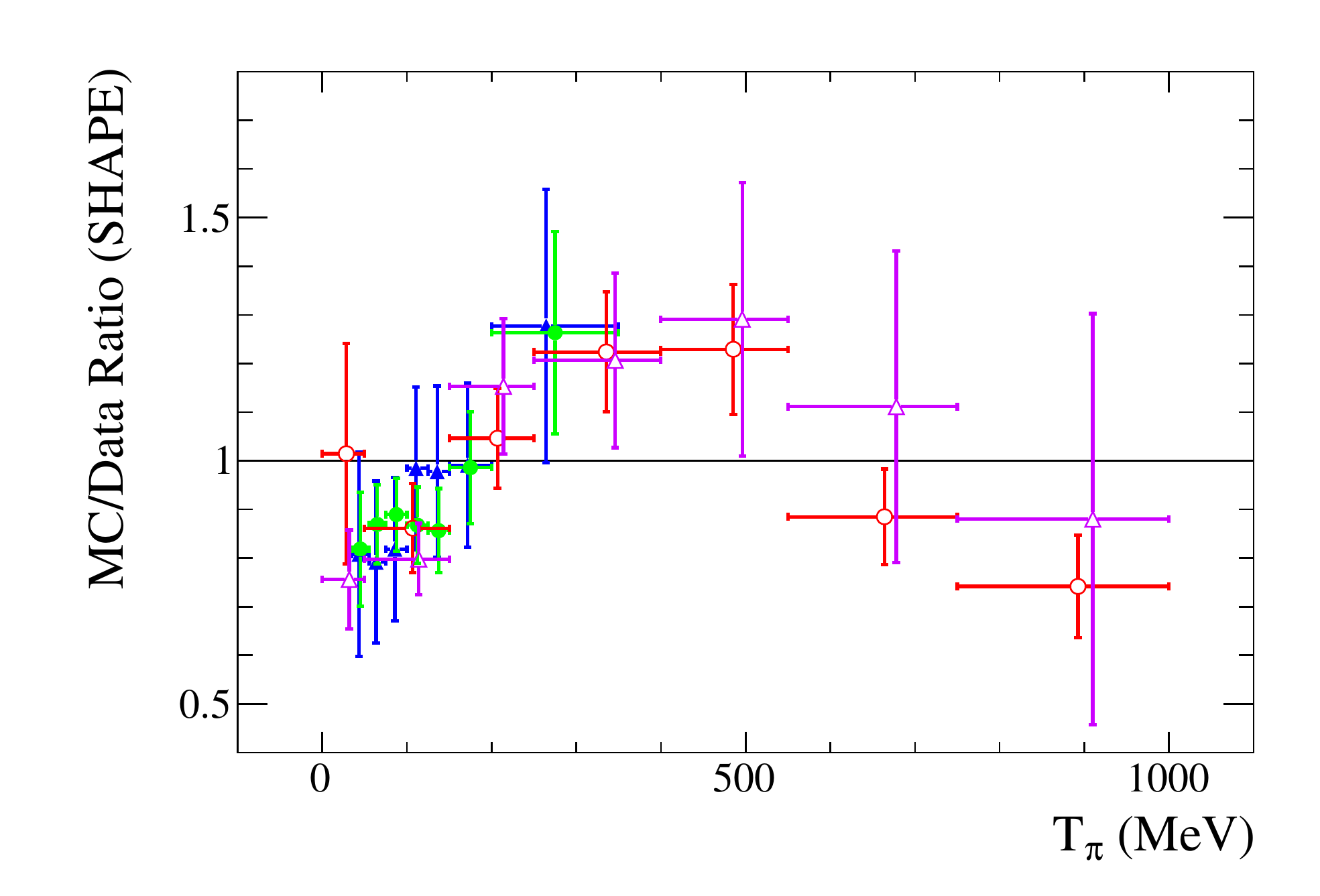}
\includegraphics[width=0.48\textwidth,trim={12mm 2mm 10mm 8mm}, clip]{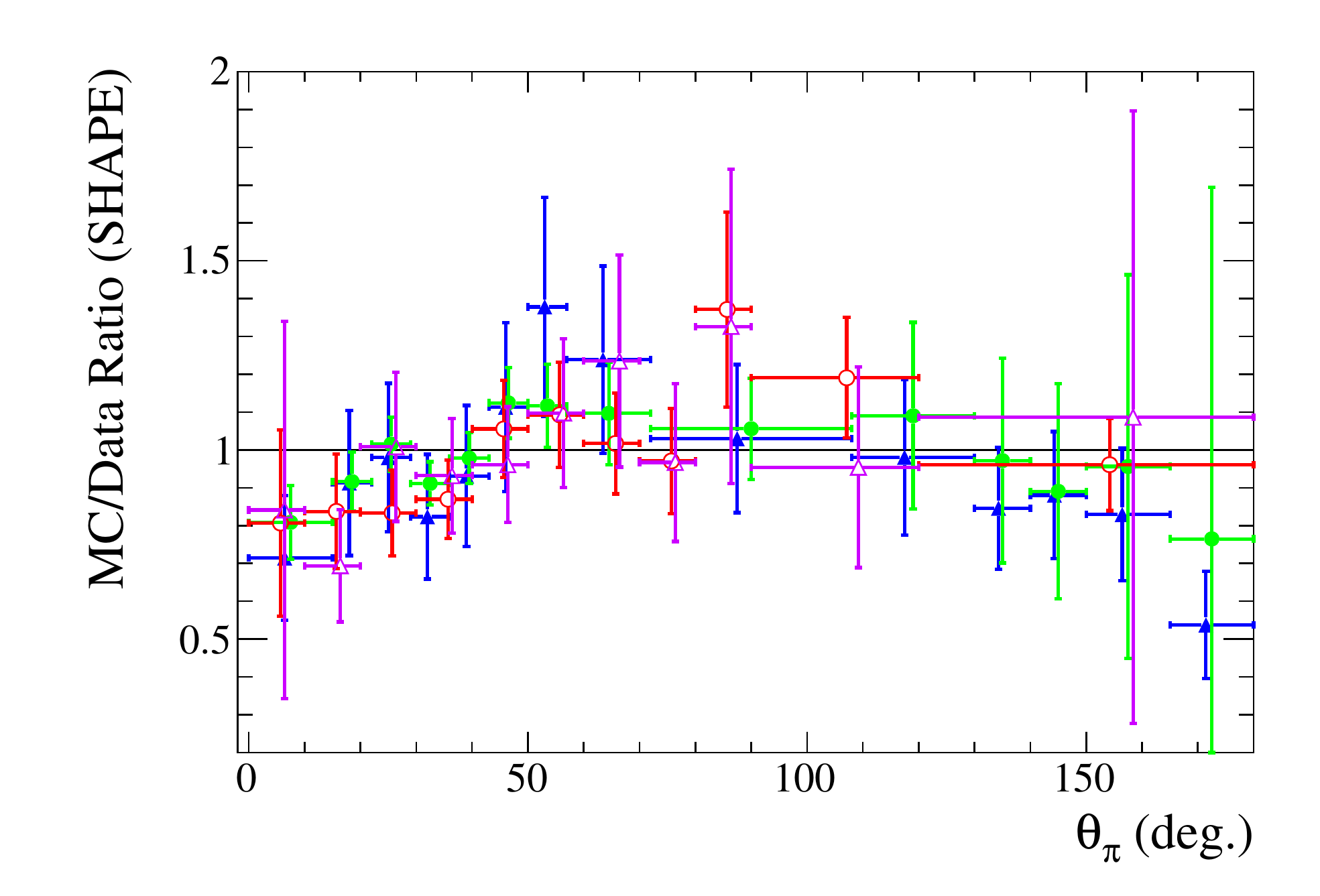}
\caption{MC/data ratios at the best fit points from the FrAbs tuning with low-\qq suppression included.}
\label{fig:frabsLagtuneratios}
\end{figure*}

\begin{figure*}[hbtp]
\centering
\includegraphics[width=0.48\textwidth,trim={12mm 2mm 10mm 8mm}, clip]{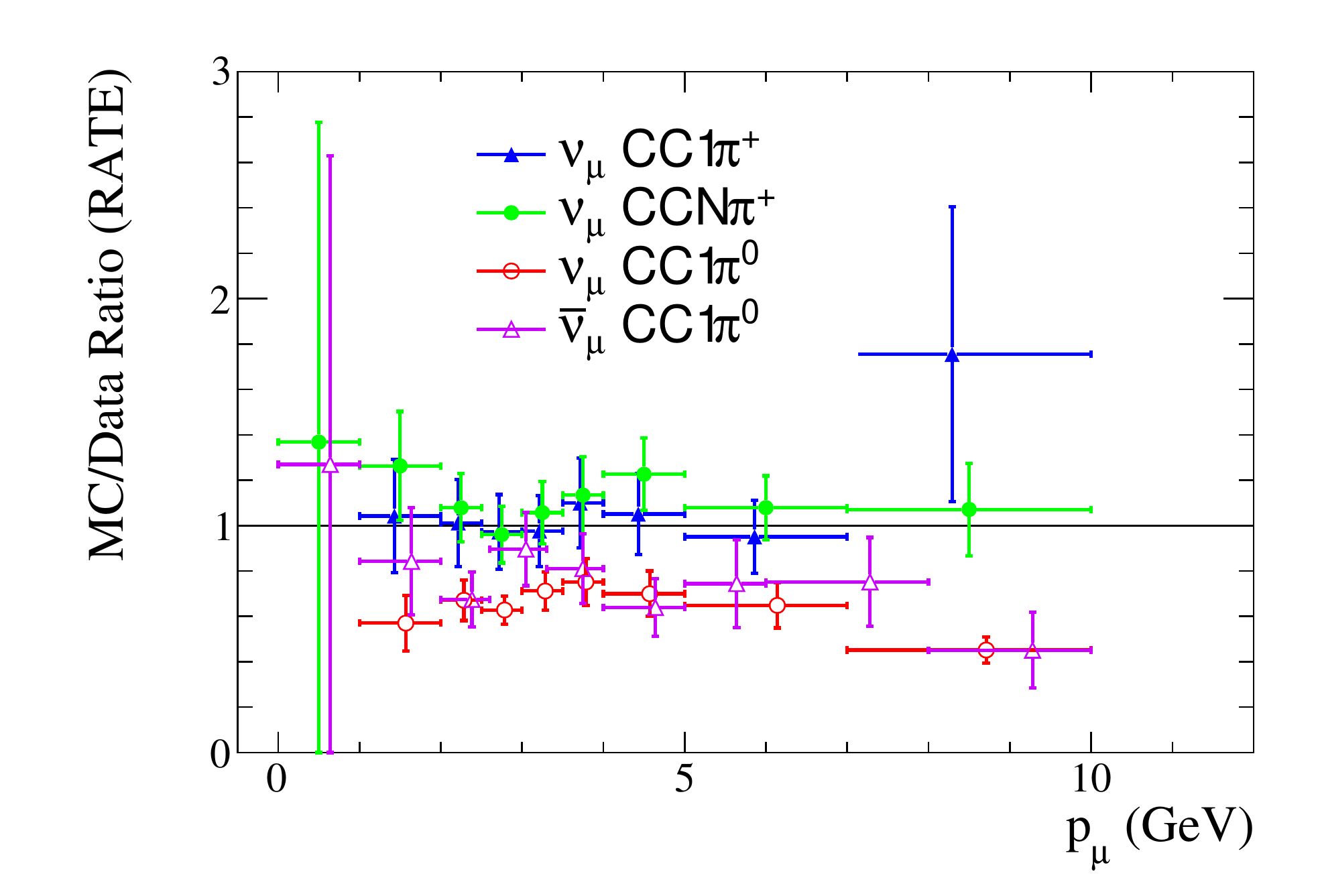}
\includegraphics[width=0.48\textwidth,trim={12mm 2mm 10mm 8mm}, clip]{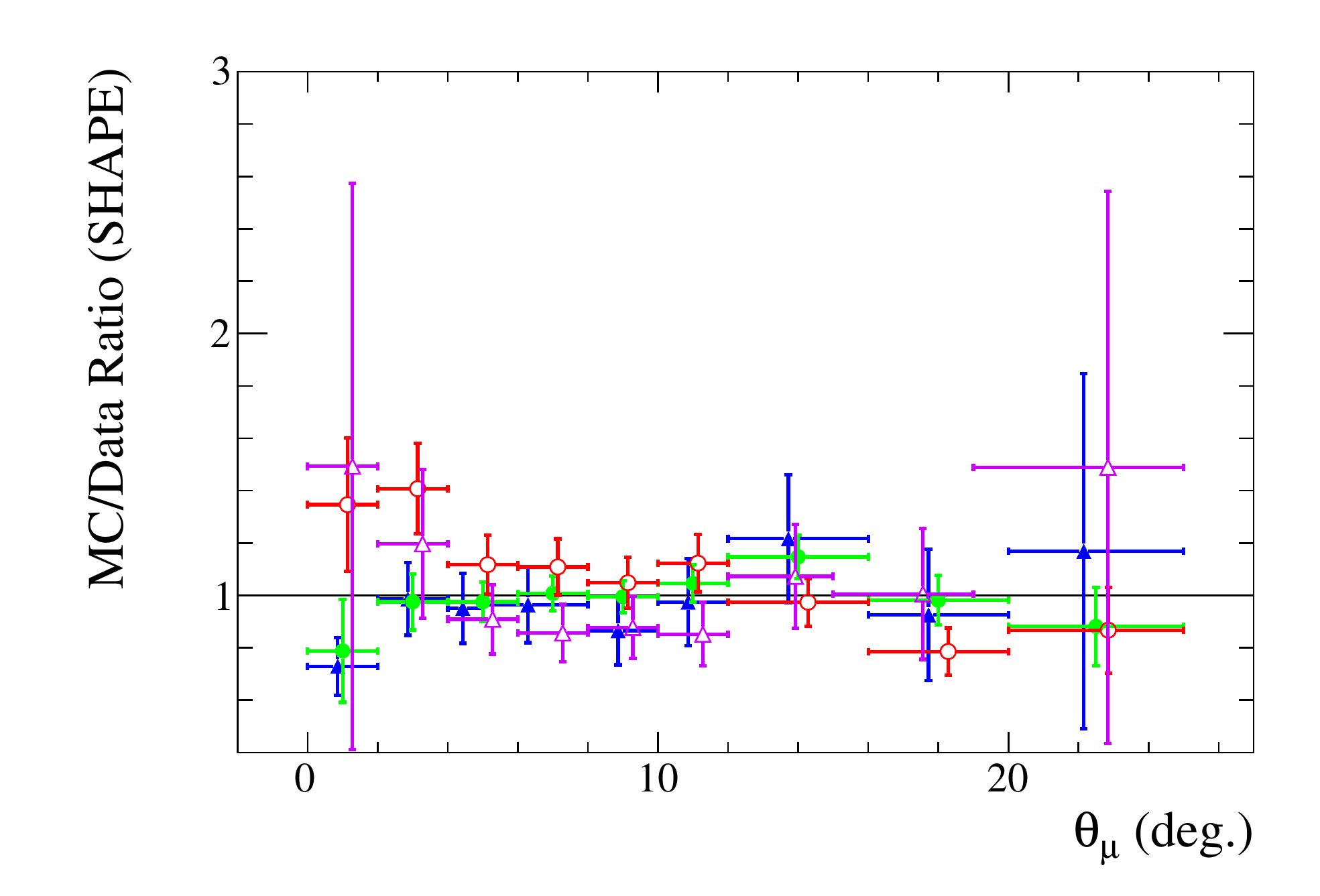}
\includegraphics[width=0.48\textwidth,trim={12mm 2mm 10mm 8mm}, clip]{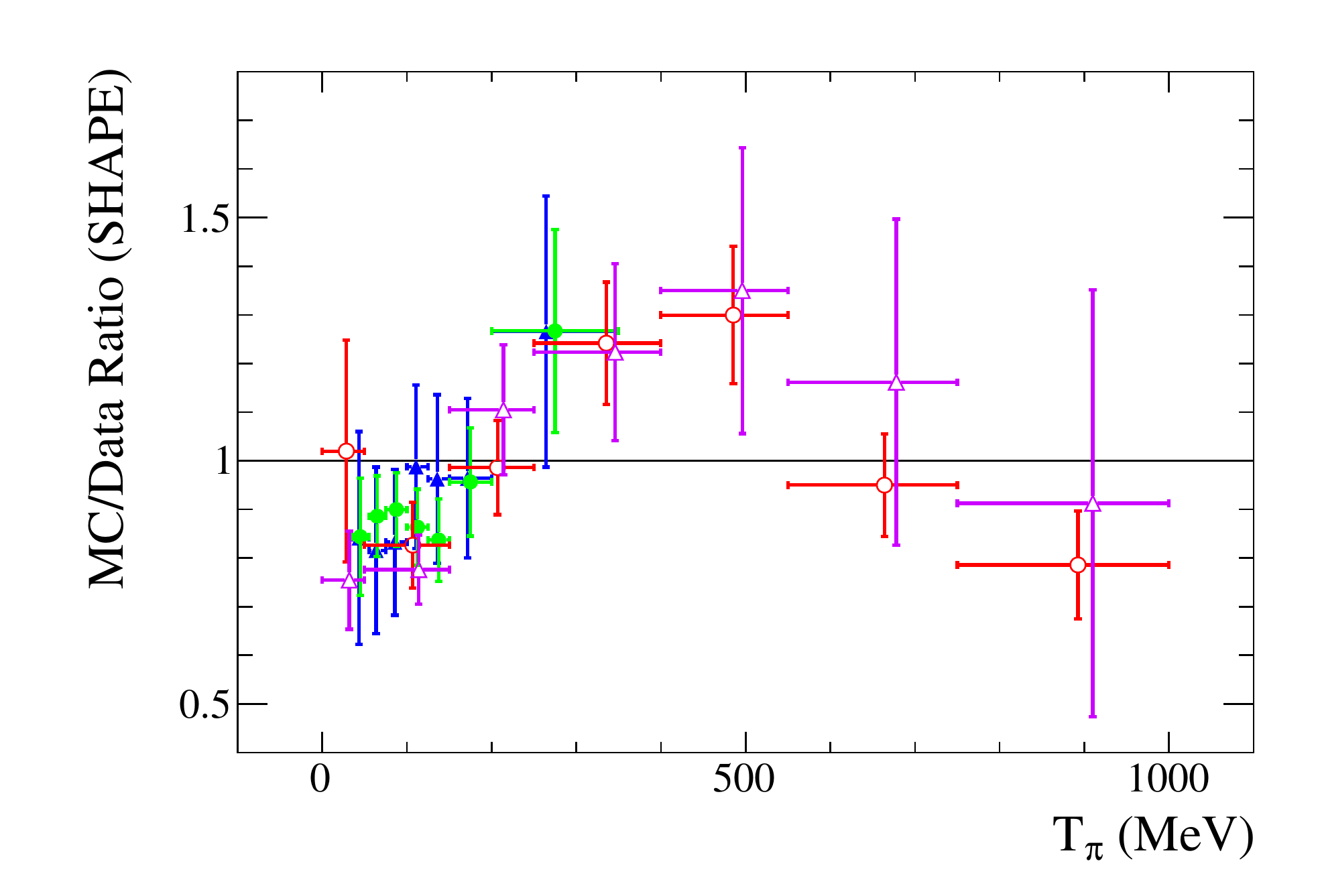}
\includegraphics[width=0.48\textwidth,trim={12mm 2mm 10mm 8mm}, clip]{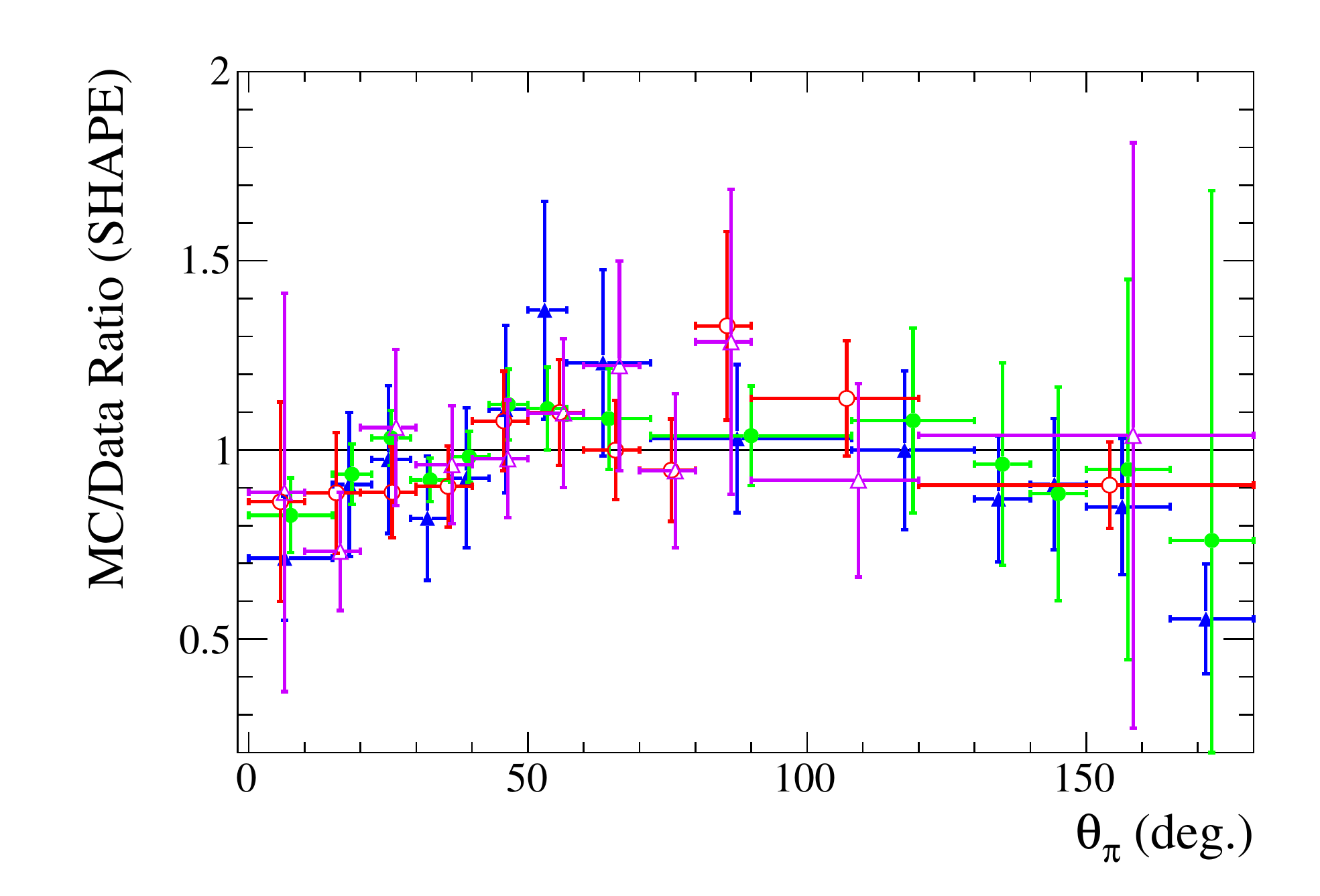}
\caption{MC/data ratios at the best fit points from the FrInel tuning with low-\qq suppression included.}
\label{fig:frinelLagtuneratios}
\end{figure*}

Figure~\ref{fig:frabsLagtuneratios} (Fig.~\ref{fig:frinelLagtuneratios}) shows the ratio of the resulting fits to the \MINERvA data when FrAbs (FrInel) is taken as a free parameter. As anticipated, the predictions now have better agreement with the data in regards to the \Tmu distribution, and the \chisq values are improved by the introduction of our ad hoc low-\qq correction.  Other fit parameters are for the most part unchanged by the introduction of the low-\qq correction. Furthermore, \Mares and \Normres are closer to their values when fitting ANL and BNL data, indicating the \qq correction alleviates the tension between nucleon and nuclear modeling.  Fig.~\ref{fig:q2difatminerva} shows the comparison of all our models directly against \MINERvA data in \qq.  Although the tuning sees improvement in the \chisq for the \ccpin and \ccapin distributions, the \ccpip and \ccnpip distributions get worse, hinting at tensions in the charged and neutral pion production channels.

Tables~\ref{tab:frabsrpaindividual} and~\ref{tab:frinelrpaindividual} show the results of the fits to individual channels, and Table~\ref{tab:adhocchi2} shows the breakdown of contributions to the \chisq from the individual channels.  The best fit \chisq value was significantly improved for each channel tuning when using a low-\qq suppression and the extracted parameters were consistent with the ANL/BNL tunings. Pion kinematic distributions are not improved, and in some cases are slightly worse, as a result of including the low-\qq suppression. It is clear from Table~\ref{tab:rpajointfits} (or by comparing Tables~\ref{tab:frabsindividual} and~\ref{tab:frabsrpaindividual}) that the low-\qq suppression has a similar effect in the fit to the FrAbs parameter. When the low-\qq suppression is introduced, FrAbs tends to consistently lower values. It is also clear that the \ccapin channel favors stronger low-\qq suppression than the other channels.

\begin{table*}[hbtp]
\centering
{\renewcommand{\arraystretch}{1.2}
\begin{tabular}{ccccc}
\hline\hline
Parameter            & \ccpip          & \ccnpip         & \ccpin          & \ccapin \\
\hline\hline
\Mares (GeV)         & $0.93\pm0.02$   & $0.92\pm0.02$   & $0.96\pm0.05$   & $0.94\pm0.05$ \\
\Normres (\%)        & $115\pm3$       & $117\pm3$       & $114\pm7$       & $115\pm7$ \\
\nonresonepi (\%)    & $43\pm4$        & $43\pm4$        & $45\pm4$        & $43\pm4$ \\
\nonrestwopi (\%)    & $300$ (limit)   & $70\pm28$       & $300$ (limit)   & $300$ (limit) \\
\ThetaPi             & 1 = Iso (limit) & 1 = Iso (limit) & 1 = Iso (limit) & 1 = Iso (limit) \\
FrAbs (\%)           & 92 $\pm$ 65     & $79\pm40$       & $74\pm22$       & $34\pm35$ \\
Lag. $R_1$           & $0.53\pm0.16$   & $0.43\pm0.13$   & $0.21\pm0.14$   & $0.14\pm0.22$ \\
Lag. $R_2$           & $0.50$ (limit)  & $0.50$ (limit)  & $0.63\pm0.31$   & $1.00$ (limit) \\
\hline\hline
\MINERvA \chisq      & 32.2            & 55.7            & 71.2            & 27.7 \\
\chisqpen            & 0.1             & 0.4             & 0.5             & 0.0 \\
\hline\hline
Total \chisq         & 32.3            & 56.1            & 71.7            & 27.7 \\
\ndof                & 33              & 34              & 30              & 31 \\
\hline\hline
\end{tabular}}
\caption{Individual channel FrAbs + low-\qq tuning results.
\label{tab:frabsrpaindividual}}
\end{table*}

\begin{table*}[hbtp]
\centering
{\renewcommand{\arraystretch}{1.2}
\begin{tabular}{ccccc}
\hline\hline
Parameter            & \ccpip          & \ccnpip         & \ccpin          & \ccapin \\
\hline\hline
\Mares (GeV)         & $0.93\pm0.02$   & $0.91\pm0.02$   & $0.95\pm0.05$   & $0.94\pm0.05$ \\
\Normres (\%)        & $116\pm3$       & $117\pm3$       & $114\pm7$       & $115\pm7$ \\
\nonresonepi (\%)    & $43\pm4$        & $43\pm4$        & $44\pm4$        & $43\pm4$ \\
\nonrestwopi (\%)    & $300$ (limit)   & $78\pm28$       & $300$ (limit)   & 300 (limit) \\
\ThetaPi             & 1 = Iso (limit) & 1 = Iso (limit) & 1 = Iso (limit) & 1 = Iso (limit) \\
FrInel (\%)          & $179\pm63$      & $173\pm37$      & $8\pm125$       & $103\pm57$ \\
Lag. $R_1$           & $0.49\pm0.14$   & $0.38\pm0.13$   & $0.25\pm0.17$   & $0.31\pm0.26$ \\
Lag. $R_2$           & $0.50$ (limit)  & $0.50$ (limit)  & $0.76\pm0.37$   & $1.00$ (limit) \\
\hline\hline
\MINERvA \chisq      & 30.8            & 52.1            & 69.5            & 30.9 \\
\chisqpen            & 0.1             & 0.6             & 0.2             & 0.0 \\
\hline\hline
Total \chisq         & 30.9            & 52.7            & 69.7            & 30.9 \\
\ndof                & 33              & 34              & 30              & 31 \\
\hline\hline
\end{tabular}}
\caption{Individual channel FrInel + low-\qq tuning results.
\label{tab:frinelrpaindividual}}
\end{table*}

\begin{table*}[hbtp]
\centering
\small
{\renewcommand{\arraystretch}{1.2}
\begin{tabular}{ccccccc}
\hline\hline
Distribution & Channel & \nbins & FrAbs Tune & FrAbs + low-\qq Tune & FrInel Tune & FrInel + low-\qq Tune \\
\hline\hline
\Pmu (Rate)  & \ccpip  & 8      & 12.0       & 10.8                 & 12.3        & 10.9 \\
             & \ccnpip & 9      & 26.1       & 16.2                 & 26.8        & 17.9 \\
             & \ccpin  & 8      & 19.0       & 26.2                 & 19.3        & 26.9 \\
             & \ccapin & 9      & 6.2        & 7.1                  & 6.3         & 7.2 \\
\hline\hline
\Tmu (Shape) & \ccpip  & 9      & 7.5        & 7.4                  & 7.4         & 7.1 \\
             & \ccnpip & 9      & 4.0        & 6.3                  & 4.1         & 5.6 \\
             & \ccpin  & 9      & 44.5       & 20.0                 & 45.6        & 20.5 \\
             & \ccapin & 9      & 10.2       & 7.0                  & 10.3        & 6.9 \\
\hline\hline
\Kpi (Shape) & \ccpip  & 7      & 2.5        & 2.5                  & 2.3         & 2.4 \\
             & \ccnpip & 7      & 31.2       & 28.9                 & 29.4        & 27.7 \\
             & \ccpin  & 7      & 30.9       & 27.1                 & 29.9        & 32.0 \\
             & \ccapin & 7      & 16.6       & 15.7                 & 16.0        & 18.7 \\
\hline\hline
\Tpi (Shape) & \ccpip  & 14     & 13.0       & 13.4                 & 12.6        & 12.6 \\
             & \ccnpip & 14     & 6.9        & 7.0                  & 6.2         & 6.3 \\
             & \ccpin  & 11     & 8.3        & 12.2                 & 8.9         & 9.4 \\
             & \ccapin & 11     & 3.4        & 4.4                  & 3.5         & 3.7 \\
\hline\hline
Total \chisq &         & 148    & 242.3      & 212.2                &  240.7      & 215.7 \\
\hline\hline
\end{tabular}}
\caption{Channel by channel contributions to the \chisq at for the \GENIE tunings with and without the low-\qq correction included.
\label{tab:adhocchi2} }
\end{table*}

\section{Concluding Remarks}
\label{sec:conclusions}
We have adjusted the parameters of the \GENIE model that are important for pion production to match \MINERvA data in the \ccpip, \ccnpip, \ccpin and \ccapin channels, using the \NUISANCE framework. We incorporate existing results which informs the \GENIE model using ANL and BNL bubble chamber data from scattering off protons and deuterons.  Fits of selected \GENIE model parameters were done using the kinematic distributions \Pmu, \Tmu, \Kpi and \Tpi.  Parameter fits were performed with either the fraction of pions absorbed or the fraction of pions inelastically scattered in FSI as a floating parameter, with broadly similar conclusions for the two cases.

The results of the fit (see Table~\ref{tab:combtuningresults}) show that the tuning improves the \GENIE pion production model significantly, but tensions remain.  The pull on the ANL/BNL prior demonstrates a tension between \MINERvA nuclear target data and the light-target bubble-chamber data sets used to make the prior, indicating a deficiency in the \GENIE nuclear model which cannot be fixed by modifying the available reweighting dials.   Additionally, fitting to individual \MINERvA pion production channels produces different best-fit parameters, demonstrating that \GENIE cannot describe the different exclusive channels in a consistent manner with the available dials (shown in Tables~\ref{tab:frabsindividual} and~\ref{tab:frinelindividual}).  Because the four channels cover different kinematic regions (see Table~\ref{tab:datareleases}) and contain different physics (\eg different coherent pion production contributions or nonresonant processes), it is difficult to pinpoint the origin of the discrepancy between the model and the different \MINERvA data sets.

Following experimental hints of discrepancies at low-\qq for a variety of cross-section measurements on nuclear targets, an additional empirical low-\qq suppression was introduced and the fits were repeated.  Although the data showed a preference for a strong suppression at low-\qq and the agreement improved for \Tmu and \qq distributions, tensions remain.  In particular, fits to individual \MINERvA channels still produced different results, and favor different parameter values for the low-\qq suppression.

The main conclusion of this work is that current neutrino experiments operating in the few--GeV region should think critically about single pion production models and uncertainties, as the Monte Carlo models which are currently widely used in the field are unable to explain multiple data sets, even when they are from a single experiment.

A key strength of this analysis is its development within the \NUISANCE framework, allowing it to be easily repeated with alternate model assumptions, neutrino interaction generators, and different data.  The developments presented here will be used in future iterations of this work, as the \MINERvA collaboration works towards a \GENIE model that provides a good description of all their available data, and can be easily applied to other measurements and experiments.

\section*{Acknowledgements}
P.S., L.P., and C.V.C.W. would like to thank the UK Science and Technology Facilities Council (STFC) for Ph.D. funding support. P.S. acknowledges the Fermilab Neutrino Physics Center for the scholarship that funded this work and thanks Fermilab and the \MINERvA collaboration for their hospitality during this work. C.W. acknowledges the support of the Swiss National Science Foundation and SERI.

This document was prepared by members of the MINERvA Collaboration using the resources of the Fermi National Accelerator Laboratory (Fermilab), a U.S. Department of Energy, Office of Science, HEP User Facility.  Fermilab is managed by Fermi Research Alliance, LLC (FRA), acting under Contract No. DE-AC02-07CH11359.  These resources included support for the \MINERvA construction project, and support for construction also was granted by the United States National Science Foundation under Award No. PHY-0619727 and by the University of Rochester.  Support for participating scientists was provided by NSF and DOE (USA); by CAPES and CNPq (Brazil); by CoNaCyT (Mexico); by Proyecto Basal FB 0821, CONICYT PIA ACT1413, Fondecyt 3170845 and 11130133 (Chile); by CONCYTEC, DGI-PUCP, and IDI/IGI-UNI (Peru); and by the Latin American Center for Physics (CLAF); NCN Opus Grant No. 2016/21/B/ST2/01092 (Poland).  We thank the MINOS Collaboration for use of its near detector data. Finally, we thank the staff of Fermilab for support of the beam line, the detector, and computing infrastructure.

\bibliography{bib}

\end{document}